\documentclass[journal]{IEEEtran}
\ifCLASSINFOpdf
\else
\fi
\usepackage{graphicx} 
\usepackage{multirow}
\usepackage{array}
\usepackage{rotating} 
\usepackage{amssymb}
\usepackage{pifont}
\usepackage{longtable}
\usepackage{xltabular}
\usepackage{makecell}
\usepackage[justification=centering]{caption}
\usepackage{threeparttable}
\usepackage{tikz}
\usepackage[edges]{forest}
\usetikzlibrary{arrows.meta}
\newcommand{\tikzred}{\tikz\draw[red,fill=red] (0,0) circle (.5ex);}
\newcommand{\tikzgreen}{\tikz\draw[green,fill=green] (0,0) circle (.5ex);}
\usepackage{caption}
\usepackage{subcaption}
\begin{document}
%
\title{Cooperative Localization for Autonomous Underwater Vehicles - a comprehensive review}
%
%
%

\author{Milind~Fernandes,
	Soumya~Ranjan~Sahoo,
	and~Mangal~Kothari
\thanks{M. Fernandes and S. R. Sahoo are with the Department of Electrical Engineering,
	Indian Institute of Technology Kanpur, Kanpur 208016, India (e-mail:
	milindf@iitk.ac.in; srsahoo@iitk.ac.in).}
\thanks{M. Kothari is with the Department of Aerospace Engineering,
	Indian Institute of Technology Kanpur, Kanpur 208016, India (e-mail:
	mangal@iitk.ac.in).}
}

%
%

\markboth{}%
{Shell \MakeLowercase{\textit{et al.}}: Bare Demo of IEEEtran.cls for IEEE Journals}
%



\maketitle

\begin{abstract}
Cooperative localization is an important technique in environments devoid of GPS-based localization, more so in underwater scenarios, where none of the terrestrial localization techniques based on radio frequency or optics are suitable due to severe attenuation. Given the large swaths of oceans and seas where autonomous underwater vehicles (AUVs) operate, traditional acoustic positioning systems fall short on many counts. Cooperative localization (CL), which involves sharing mutual information amongst the vehicles, has thus emerged as a viable option in the past decade. This paper assimilates the research carried out in AUV cooperative localization and presents a qualitative overview. The cooperative localization approaches are categorized by their cooperation and localization strategies, while the algorithms employed are reviewed on the various challenges posed by the underwater acoustic channel and environment. Furthermore, existing problems and future scope in the domain of underwater cooperative localization are discussed.

\end{abstract}


%
\IEEEpeerreviewmaketitle

\section{Introduction}
%
%
%
%
\IEEEPARstart{I}{t} has been said that we know more about the surface of the moon than the ocean floor. A large swath of our oceans remains unexplored and unmapped, owing mainly to its massive size and the cost of operation for any survey activity. A modern solution to this age-old problem is unmanned underwater vehicles or UUV's instead of human-crewed ships. While the UUV's generally consist of remotely operated vehicles (ROV) and autonomous underwater vehicles (AUV), it's the latter that is best suited for long-duration and long-range survey missions in the oceans. An AUV is capable of autonomously carrying out pre-planned surveys, opportunistic seeking missions, and target tracking without human intervention and control. However, even though the AUV's have increased our capabilities to survey the ocean depths, they do have a few shortcomings. First, an AUV needs a support vehicle in the form of a human-crewed ship for deployment and recovery, which is costly \cite{6380737}. Second, since the AUV works underwater, localization is a huge challenge.%

In a terrestrial setting, the localization problem is solved mainly by relying on GPS. But for underwater environments, no such large-scale system exists. This is because electromagnetic signals attenuate very rapidly in water and do not propagate useful distances \cite{775301}, \cite{tan_survey_2011}. In the absence of localization references such as GPS, it is common for an autonomous vehicle to rely on its onboard sensors for dead reckoning. But despite the advances in accuracy and resolutions of sensors such as accelerometers, gyroscopes, compass, etc., the dead reckoning approach still suffers significant drift from true location, especially in large distance surveys \cite{4099086}, \cite{emami_taban_2018}. Thus, the location of the AUV becomes increasingly uncertain. The cost of these high-accuracy sensors renders them prohibitively expensive in missions utilizing large teams of AUVs. A Doppler velocity log (DVL), which provides velocity measurements, can be used to bound the growth rate of location error to some extent, as was shown in \cite{whitcomb1999combined}. However, the ocean floor needs to be constantly in the range of DVL for it to be useful. This is not possible if the vehicle operates far above the ocean floor in the Pelagic zone. This necessitates external reference systems to minimize or bound the uncertainty in the position within a specified range.

One of the simplest solutions is for the AUV itself to periodically surface for a GPS fix. But this is not a very elegant solution because significant energy and mission time is wasted for surfacing and then heading back down. Traditionally underwater localization has relied on acoustic localization systems such as the long baseline (LBL), GPS intelligent buoys (GIB), and ship/surface vehicle-based short or ultra-short baseline systems (SBL/USBL) \cite{kinsey_survey_2006}, \cite{kebkal_auv_2017}. Other methods, such as those based on Simultaneous Localization and Mapping (SLAM), geophysical features obtained using camera or SONAR imagery, magnetic field maps, and bathymetric maps, have elicited interest in the recent past. The reader can find an excellent review of AUV navigation technologies and techniques in \cite{6678293}, \cite{gonzalez-garcia_autonomous_2020}. 

The traditional localization systems, such as LBL and GIB, suffer from installation/deployment and maintenance/recovery issues, whereas SBL and USBL suffer from lower precision and operating range with bearing measurements further affected by the surface vehicles' roll and pitch. Even with LBL, the operating range is limited to a few square kms. Underwater SLAM, on the other hand, suffers from a limited set of available underwater features, whereas optical solutions need clear waters and fail in turbid environments. They also have high computational requirements and a high monetary cost that escalates with the number of AUVs for a given mission. Another option that has gained significant attention in recent times is the range-only single beacon-based localization due to its inherent simplicity and cost-effectiveness \cite{8744517}. The beacon could be either static or moving. While a moving beacon falls in the realm of cooperative localization (CL), as will be seen in the sequel, the static beacon-based localization requires the AUV to perform fast turning or encircling maneuvers for the system to be observable. While this is acceptable in target tracking (where the beacon is the target), if the AUV has a specific mission trajectory, this method cannot be used. 

All the above issues have contributed to the increasing interest in cooperative localization-based approaches, especially in AUV teams working together. At the very least, CL requires a set of sensors that will, by necessity, be present on each of the vehicles, such as the Inertial Navigation System (INS), acoustic modem, and DVL. Some CL approaches can even accommodate vehicles with lower accuracy sensors. The CL-based approach's primary requirement is the mutual information exchange between vehicles, using which each vehicle can improve its respective localization accuracy.

With the growing interest in this area, the literature on underwater CL has reached a critical mass. However, to the best of the authors' knowledge, no publication has yet classified and categorized this trove of information. Thus, this paper aims to review and classify the existing literature in this domain qualitatively. With this in mind, the contributions of this paper are as follows:

\begin{itemize}
	\item We present an exhaustive review of underwater cooperative localization literature for AUVs up to date.
	\item We classify the CL approaches and bring forth their salient features.
	\item We identify and discuss the open problems in underwater cooperative localization.
\end{itemize}

The paper is organized as follows. Section II gives the requisite background that underlines the operational performance of underwater cooperative localization strategies. Section III discusses the different categories of cooperative localization strategies and provides a detailed comparison of various performance parameters. In section IV, a discussion on the current shortcomings and future directions is presented.  Section V concludes the paper.

\section{BACKGROUND}
In this section, we put forth some of the considerations relevant to the underwater environment and cooperative localization algorithms. These include the underwater acoustic channel, state estimation techniques, measurements for state estimators, among others. This section also introduces some of the criteria used to compare the current state-of-the-art cooperative localization strategies for AUVs. 
\subsection{Underwater acoustic channel}
A water body as a communication channel is quite challenging, especially considering the severe attenuation of electromagnetic signals, be it radio frequencies or light. This has led to widespread adoption and advancements in acoustic communication technologies for underwater use. Still, the underwater acoustic channel has a fair share of issues that need to be dealt with and kept in mind while developing algorithms that use acoustically transmitted information, as highlighted below.

\begin{table*}[!t]
	\renewcommand{\arraystretch}{2}
	\caption{COMMON STATE ESTIMATORS IN UNDERWATER CL}
	\label{table_1}
	\centering
	\begin{tabular} {p{0.05\textwidth}p{0.13\textwidth}p{0.13\textwidth}p{0.13\textwidth}p{0.13\textwidth}p{0.13\textwidth}p{0.13\textwidth}}
		\hline
				\hline
		\textbf{Sr. No.} & \multicolumn{1}{c}{\textbf{State Estimator}} & \multicolumn{1}{c}{\textbf{Belief}}       & \multicolumn{1}{c}{\textbf{Model}}       & \multicolumn{1}{c}{\textbf{Advantages}}                  & \multicolumn{1}{c}{\textbf{Disadvantages}}       & \multicolumn{1}{c}{\textbf{Remarks}}            \\ \hline
		\hline
		1                & Kalman~Filter~(KF)                            & Gaussian (Mean and Variance)             & Linear                                    & Fast, Optimal                                             & Not applicable for Non-Linear models            & -                                                \\ \hline
		2                & Extended~Kalman~Filter (EKF)                & Gaussian (Mean and Variance)             & Linearized NL (1st order Taylor Series approx.) & Fast, NL models                                           & Linearization errors, Requires matrix inversion & Unsatisfactory performance with outliers       \\ \hline
		3                & Unscented Kalman Filter (UKF)               & Gaussian (Mean and Variance)             & Non-Linear (Sigma points)                 & Lower NL errors                                           & Higher computational cost                         & Refines only current state                     \\ \hline
		4                & Extended~Information~Filter~(EIF)           & Gaussian (Information   matrix and vector) & Linearized NL (1st order Taylor Series approx.) & Fast measurement update                                   & Slow prediction step                              & Canonical Gaussian   representation, Dual of EKF \\ \hline
		5                & Particle Filter (PF)                          & Non-Parametric (Particles with weights)  & Linear or NL                              & Applicable to non-Gaussian distributions                & Computationally expensive                         & Can handle outliers                              \\ \hline
		6                & Maximum~A~Priori (MAP)                        & Gaussian                                   & Linear or NL                              & Better estimates, Estimates computed for all time steps & Computationally expensive                         & Requires storing past states.                  \\ \hline \hline
	\end{tabular}
\end{table*}

\begin{enumerate}
	\item Speed of propagation: The water temperature in seas and oceans is not a constant function of depth; instead, it varies with it. The warm waters are near the surface, while the cold waters are near the floor. Similarly, a column of water may consist of strata of different salinity at any location and time, thus, different densities. The density of the water is also a function of the depth. These spatiotemporal gradients of temperature, salinity, and depth affect the speed of propagation of acoustic signals \cite{107149}, \cite{1637927}. This is especially severe in communications that involve large distances. While in practice, it is common to assume a constant speed of sound (1500 m/s) underwater, it does not represent the actual speed of sound at the instant of time and space, and thus, will be a source of error in methods that rely on range computations for localization. However, some approaches can estimate the sound speed profile of the acoustic channel during a mission and compensate for any such effects \cite{bo_optimal_2018}, \cite{yan_moving_2015}, \cite{book_Lurton}, \cite{6387620}.
	\item Latency: This is the direct consequence of acoustic signals' speed being much less than RF signals' speed. It leads to time delays between signal transmission and reception. The transmitter or receiver might have moved in that time, which causes an offset between the estimated and the actual positions. This has led to the development of delayed-state-based estimators \cite{doi:10.1177/0278364912446166}, \cite{7828800}, \cite{5779652}.
	\item Propagation Path: The density changes also cause the acoustic signals to travel along a curved path instead of straight lines \cite{book_Lurton}. This introduces errors in range measurements, as the actual traveled distance is greater than the exact Euclidean distance between any two points. Unlike the sound speed profile, this effect is difficult to characterize and compensate for large distances. In most cases, the path is assumed to be a straight line.
	\item Multipath effects: These effects are predominantly encountered in shallow waters, wherein the acoustic signals bounce off the seafloor or surface boundary and arrive at the receiver with a delay. These can also be experienced in deep water missions near the seafloor. Multipath signals give rise to measurement outliers or inaccurate range measurements and cause significant errors in state estimators' output. Outlier mitigation is one of the critical considerations in the performance evaluation of localization algorithms. Some approaches are given in \cite{506191}.
	\item Bandwidth: The acoustic channel is inherently narrow-band since it operates in the audible/ultrasonic frequency bands. This limits the number of bits one can transmit per second. Although recent advancements have achieved up to 64 kbps of throughput \cite{EvoLogics_1} over short distances of 300 m, it is a fraction of what is achievable in terrestrial networks. This calls for CL techniques that are robust against the limited channel capacity.
	\item Measurement noise: For mathematical and computational convenience, it is often the practice to assume any noise source in acoustic communications and uncertainty in measurement as being Gaussian distributed. However, as evident from many practical experiments, the distributions are more often than not heavy-tailed, especially in scenarios where multipath is evident \cite{4089076}. This leads to apparent errors in the location estimates generated by the estimation algorithms.
	\item Lost transmissions/Packet Loss: Due to the harsh underwater environment, underwater acoustic channels are far less reliable than terrestrial RF channels and suffer from intermittent lost transmissions/ packets up to 20-50\% of the total \cite{6942559}. This severely affects the convergence rate of estimation algorithms and can even render them unstable.
\end{enumerate}
As can be seen from above, the underwater acoustic channel has many challenges in terms of communication accuracy and reliability and is an active area of research. A comprehensive survey of communication challenges, solutions, and open problems can be found in \cite{9356608}.

\subsection{State Estimation Techniques}
As evident from the previous section, it is difficult to get complete location information in the harsh and dynamic underwater environment. It necessitates estimation techniques that can predict the vehicle's current location with a high degree of confidence by incorporating all the noisy and outlier-affected measurements from the internal and external sensors of an AUV. State estimators can be classified into three categories: a) Stochastic, b) SLAM based, and c) Deterministic \cite{kinsey_survey_2006}. Stochastic, Bayes filter-based estimators have found wide use due to their simplicity and computational efficiencies. A brief comparison of the various estimators is in Table ~\ref{table_1}.

While there have been many SLAM-based approaches for state estimation for which the reader is referred to \cite{6678293}, \cite{5c7d9431766940dca820263b1dfa293c}, \cite{10.1145/3366194.3366262}, the domain of underwater cooperative SLAM is still unexplored except for a few works \cite{7139227}, \cite{WALTER2004880}, \cite{rashidi2011simultaneous}. Similar is the case with deterministic state estimators that require exact plant and measurement models, which are difficult to model in an uncertain underwater environment.

\subsection{Measurement inputs for the state estimators}

\begin{figure}
	\centering
	\includegraphics[width=\linewidth]{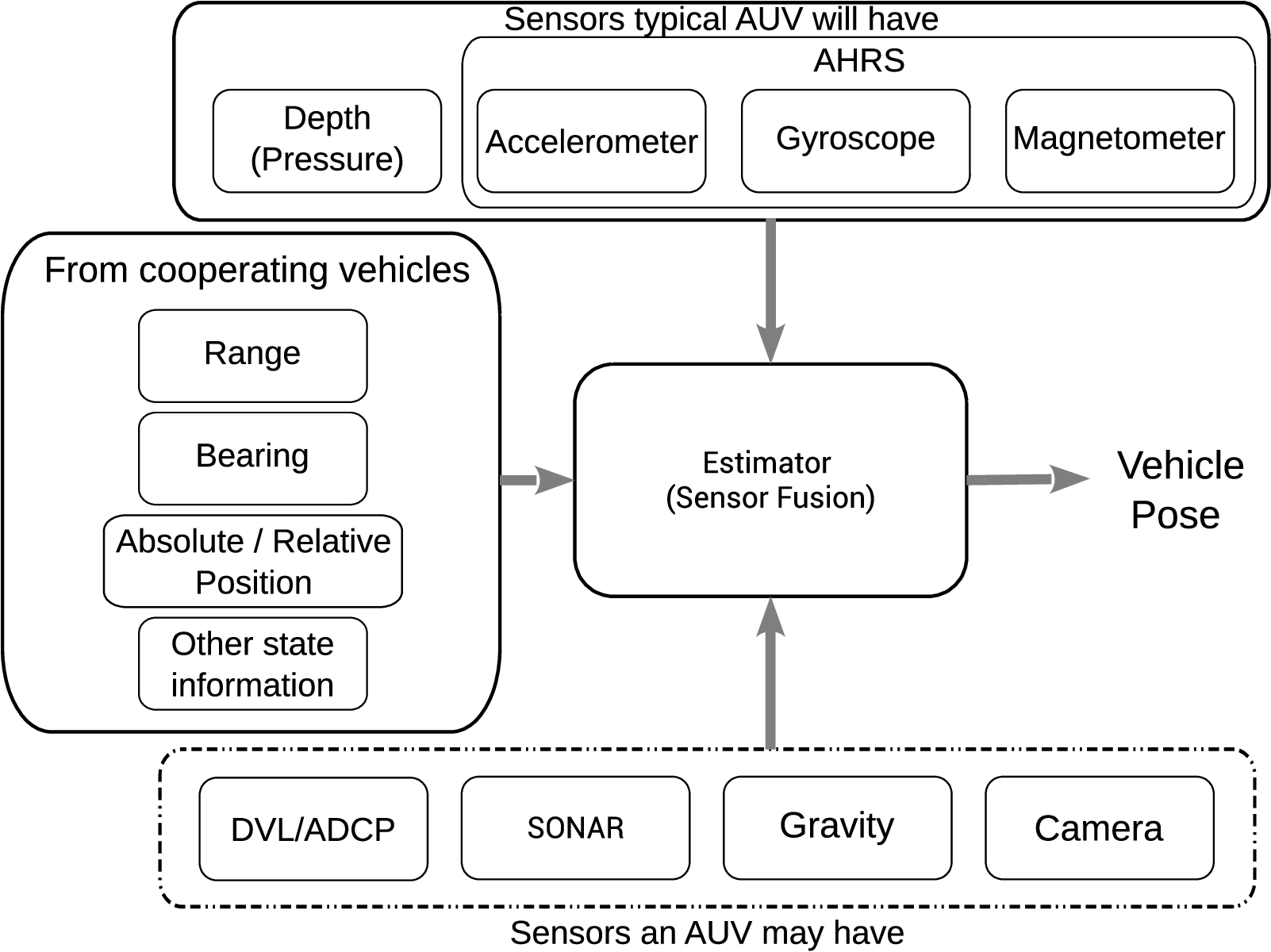}
	\caption{General system architecture for CL based pose estimation}
	\label{fig:fig1}
\end{figure} 

The common state estimation algorithms in Table \ref{table_1} have two stages, predict and update. The prediction stage uses the past state and beliefs to predict the next state. The measurement stage corrects this prediction using information from internal and external sensors. The AUVs general sensor fusion architecture to estimate the position of a vehicle in a cooperative scenario is shown in Fig. \ref{fig:fig1}. An AUV will have a bare minimum sensor suite consisting of the Attitude Heading Reference system (AHRS) and pressure (depth) sensor. The AHRS consists of a gyroscope and compass/magnetometer and provides the state estimators with angular accelerations, velocity, orientation, and heading inputs. Additionally, an Inertial Measurement Unit (IMU) combines an AHRS with accelerometers, providing additional 3D acceleration information which can be integrated for linear velocity and position estimates. The Inertial Navigation System (INS) uses data from AHRS, accelerometers, depth sensors, and DVL/ADCP (Acoustic Doppler current profiler) (if present) to estimate the vehicle's pose, also known as the dead reckoned estimate. The more expensive sensors, such as DVL, SONAR (Side-scan/Multibeam), gravity (Gravity map-based localization), etc., may or may not be present. The availability of cheap and accurate pressure-based depth sensors has rendered the three-dimensional underwater localization problem to two dimensions. This simplifies analysis and subsequent computations. However, over time the position estimate becomes more and more uncertain due to inherent drift in the sensors. 

Fusing INS information with other sensor data, such as from GPS, range, bearing, etc., either eliminates the uncertainty in position or bounds the error within a desired range. While GPS is available only on the sea surface, bearing-based methods either rely on visual information or USBL. Visual information is subject to the water's turbidity and is inherently limited to short distances \cite{7583426}. As mentioned previously, these methods are unsuitable for long-distance cooperative localization. This leaves the range-based methods, which explains its wide popularity in underwater localization. Range information can be acquired through different means. If all the vehicles are synchronized in time, time of flight (ToF) can be used to measure the distance between any two vehicles. This method, also known as one-way travel time (OWTT) ranging, has the benefit that it requires only one acoustic transmission per range measurement and is scalable with the number of vehicles. However, it requires high accuracy and stable clocks that are temperature, bias, and drift compensated, such as chip-scale atomic clocks. If synchronization is not possible, two-way travel time (TWTT) or Time difference of arrival (TDOA) can be used \cite{bo_review_2019}, \cite{su_review_2020}. In TWTT, the first vehicle sends a request ping, to which the second vehicle replies with a finite delay. If the delay is fixed and known, the distance between the two vehicles is a function of the total time from transmission to reception at the first vehicle. Since this method requires two acoustic transmissions for each range measurement, it is not scalable to large teams. In the TDOA method, the transmission from one vehicle is received by two or more vehicles. By knowing the arrival times at each of the receiving vehicles and their respective locations, the location or range of the transmitting vehicle can be estimated by exchanging data between the receiving vehicles, as it is a function of the difference in the arrival times at the receiving vehicles. However, this method requires more acoustic transmissions and is thus also not scalable. For the above reasons, OWTT has emerged as the preferred method for range measurements in underwater environments.


\begin{figure*}[!h]
	\centering
	\begin{tabular}{cccc}
		\includegraphics[width=0.3\linewidth]{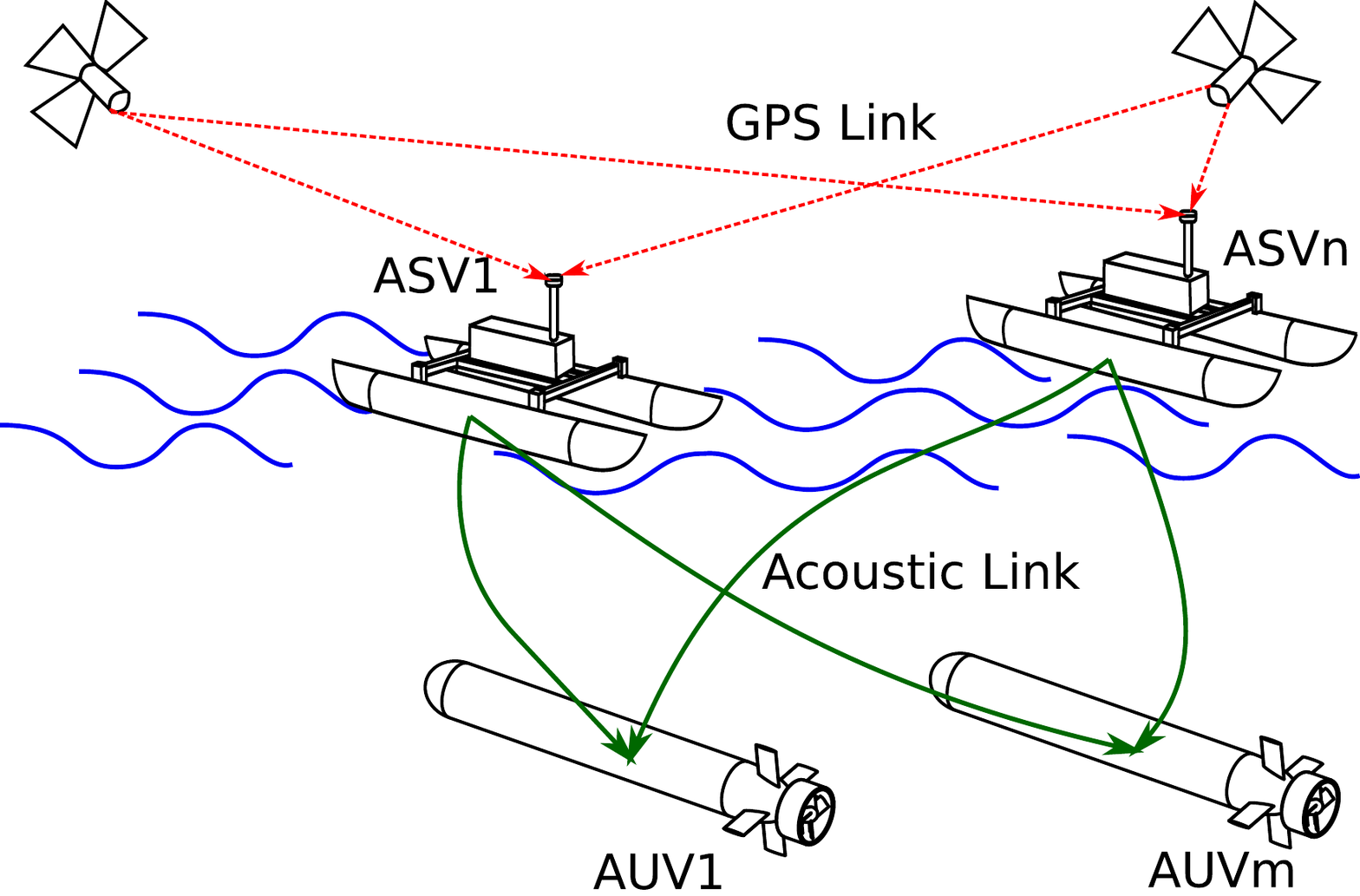} & & & \includegraphics[width=0.3\textwidth]{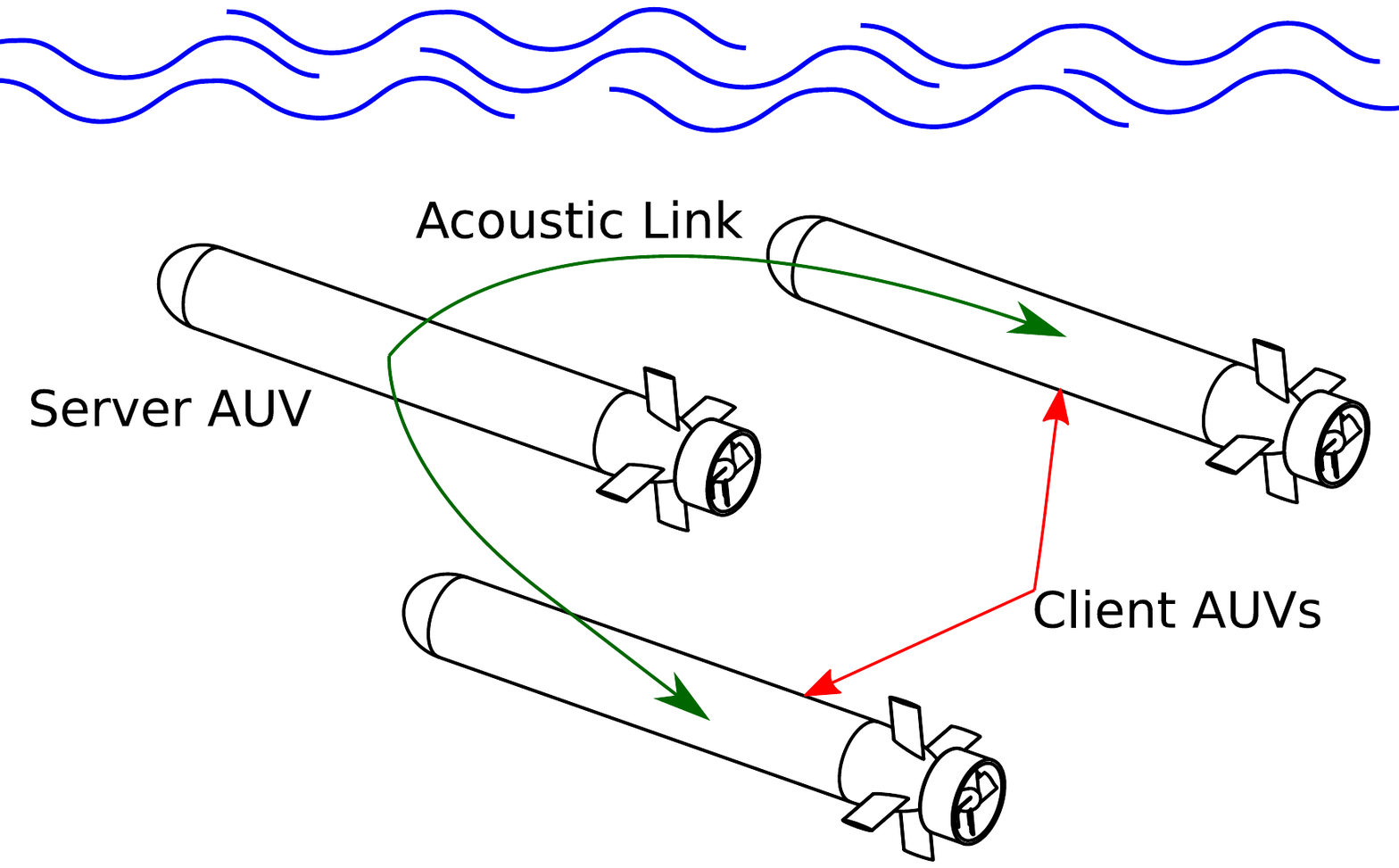}\\
		(a) Surface Vehicle-AUV &&& (b) Server-Client \\[6pt]
		\includegraphics[width=0.3\linewidth]{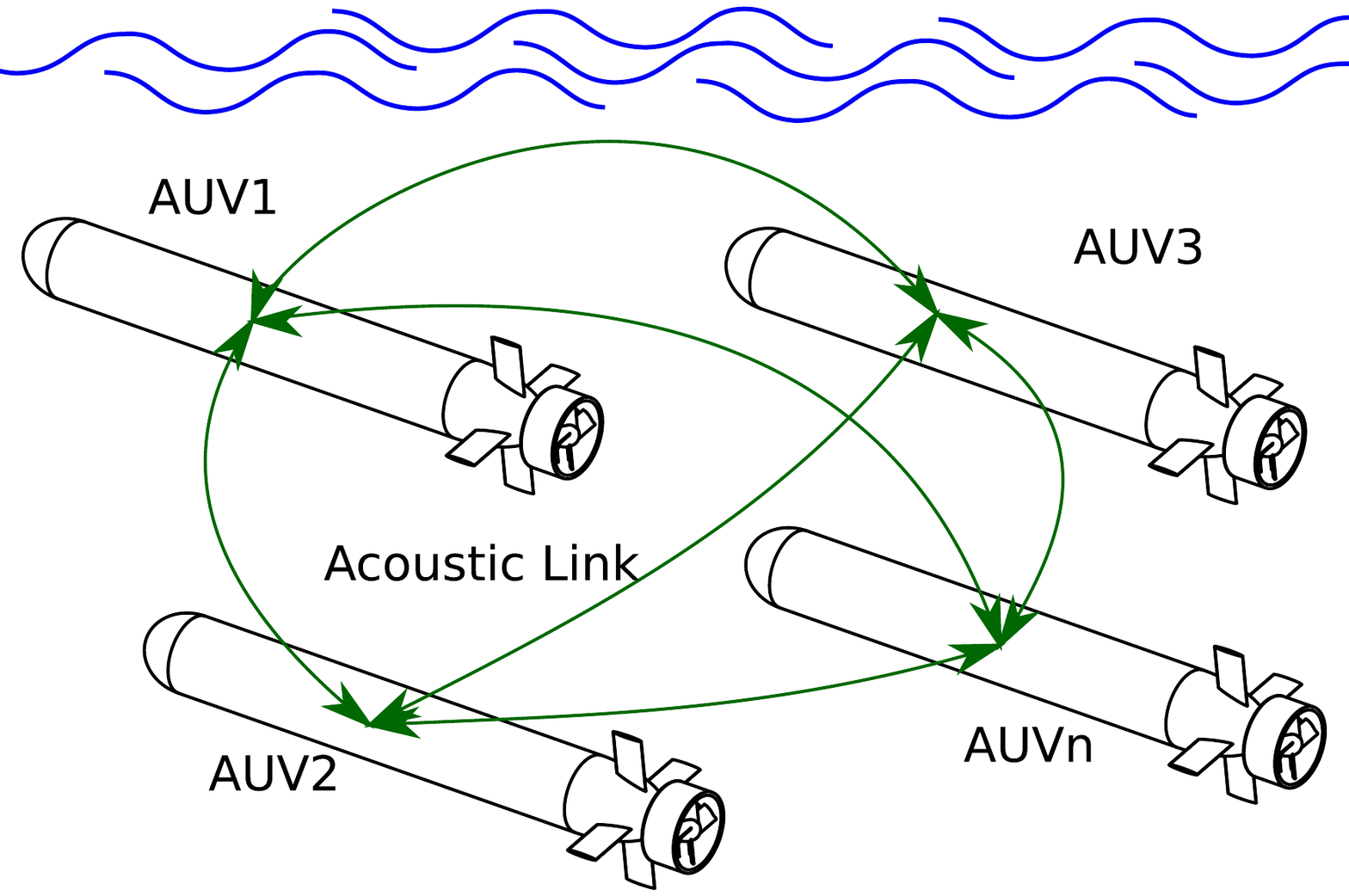}  &&&   \includegraphics[width=0.3\linewidth]{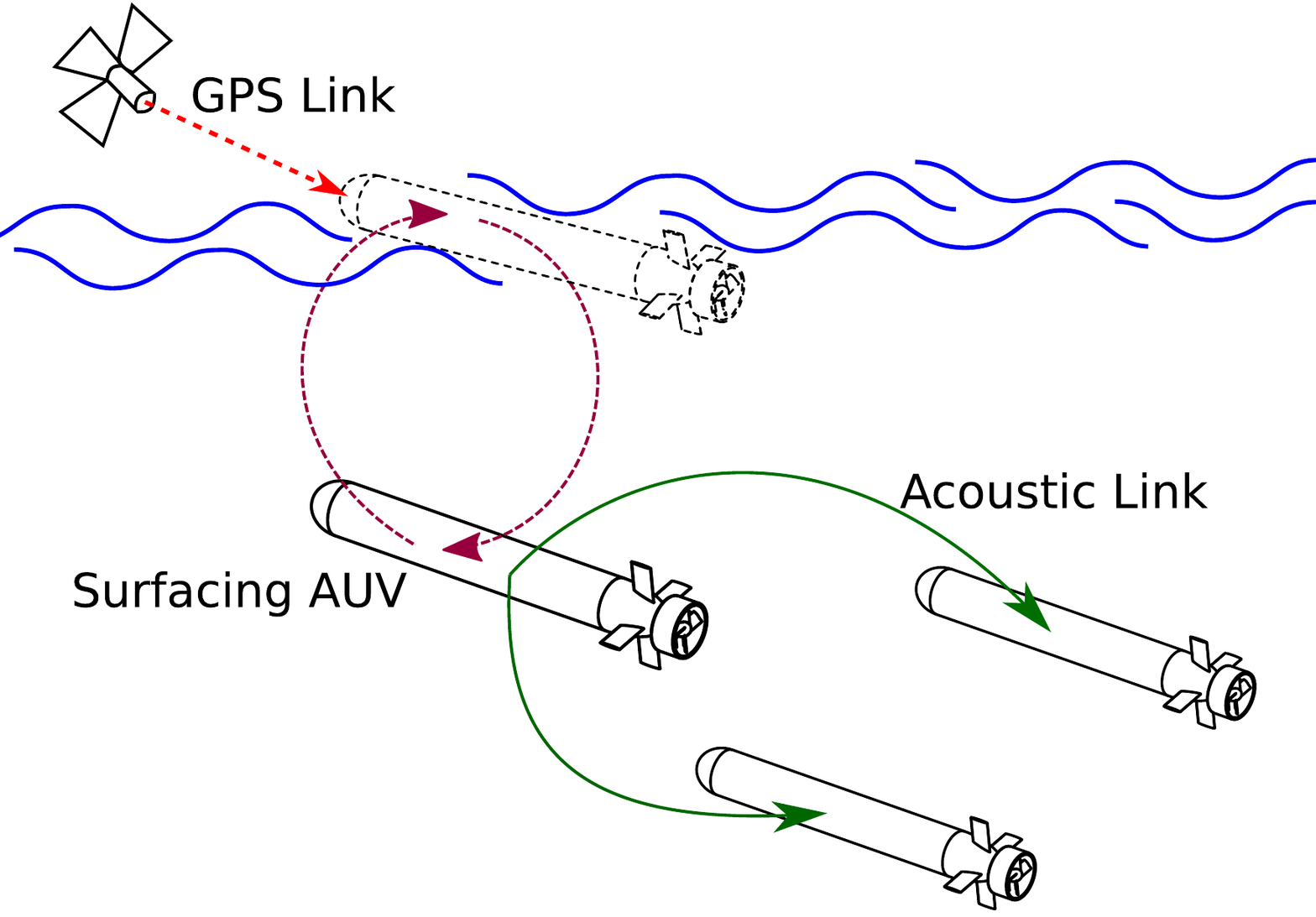} \\
		(c) Mesh &&& (d) Surfacing \\[6pt]
	\end{tabular}
	\caption{Overview of most common underwater cooperative localization approaches}
	\label{fig:copnav}
\end{figure*}

\subsection{Scalability}
The CL approach's scalability is inherently tied to how ranging is performed and how the data is exchanged between the vehicles involved due to the narrow bandwidth of the acoustic channel. Consequently, multiple simultaneous communications cannot exist. Thus, two of the most popular schemes to share data from one to many are a) broadcasting and b) Time-division multiple access (TDMA). The first approach is scalable to any number of receiving vehicles, whereas the latter is not. In a large team, the total time for updating the location information for one vehicle increases linearly, thus leading to considerable delays and can affect the convergence of the estimators. Other approaches, such as data exchange with neighbors-only, have also been reported \cite{de2015multi}, which need fewer transmissions but need a scheduling algorithm. There are also recent approaches of using orthogonal frequency division multiple access (OFDMA) for simultaneous communications between vehicles, as reported in \cite{8227642}. Another aspect that affects scalability is the size of the communication packets. The smaller the packets, the more reliable the communication with shorter communication intervals; thus, more vehicles can communicate in any given duration. But, this contravenes the requirement that for CL, the vehicles must exchange their states with each other, which is a lot of information. Thus, there have been attempts to efficiently manage the bandwidth in CL scenarios, as will be seen later.

\subsection{Other Considerations}
Ocean currents can have a detrimental effect on cooperative localization and, in general, localization of any AUV. Ocean currents tend to exacerbate the drift in the position estimate of the vehicles. They are dynamic, thus, cannot be accurately accounted for apriori and need to be estimated in situ for accurate localization results. To some extent, ocean general circulation models (OGCM) can be preloaded in the AUV prior to a mission, if available \cite{8232328} and can be used to compensate for ocean currents for tasks or missions involving large areas.

\section{UNDERWATER COOPERATIVE LOCALIZATION}
The term cooperative implies that the vehicles involved in localization share some information about their respective locations/state with each other \cite{1067998}. The location information shared could be absolute or could be relative. While absolute location information present with any one of the vehicles in the team can essentially drive down the error to a minimal value, even relative position exchange between teams can prevent the error from unbounded growth \cite{1668252}. \\

Some of the common AUV cooperative localization approaches are shown in Fig. \ref{fig:copnav}. In (a), surface vehicles are employed to aid the underwater vehicle by transmitting its absolute position information through acoustic channels. The surface vehicles could be single or multiple, autonomous or manned, and localized with the help of GPS signals. In (b), a "server/leader" underwater vehicle, which has very high accuracy and expensive sensors for its own localization, aids in the localization of several other "client/follower" AUVs. The client AUVs generally have low-accuracy inertial sensors or an incomplete sensor suite, along with other mission-specific payloads. Approach (c) does away with the aid vehicle altogether and instead relies on inter-vehicle communications to bound their localization error growth. In this type of approach, only the error growth rate can be lowered. This issue can be resolved in (d) type, wherein a team member can surface for GPS fix and then dive back to share the positional information with other team members.

\begin{figure}[!t]
	\centering
	\includegraphics[width=\linewidth]{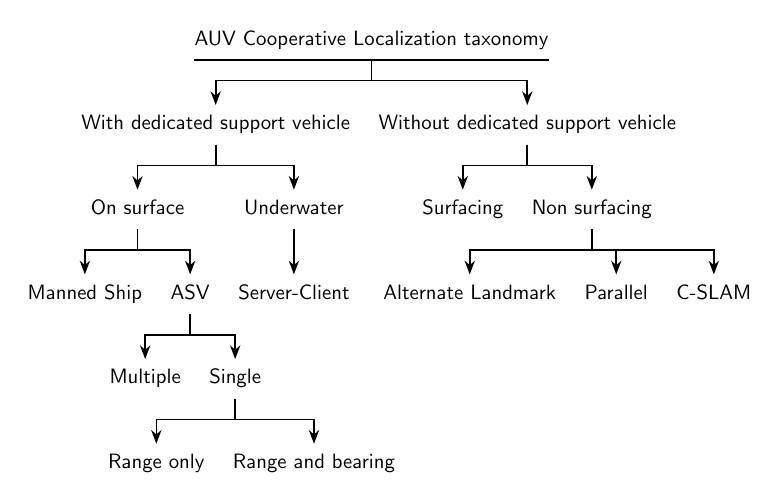}
	\caption{Classification of underwater cooperative localization approaches}
	\label{fig:classification}
\end{figure}

The taxonomy of underwater CL methods is shown in Fig. \ref{fig:classification}, and a brief comparison between the various categories is given in Table \ref{table_2}. In the following sections, we describe each of these approaches in detail and put forth the research contributions in those areas.\\
\textit{Remark 1}: Although cooperation for localization could also be with static sensors on the ocean floor or surface, as in the case with GiB, LBL, or UWSN, we restrict ourselves to the review of cases where cooperation is between moving vehicles underwater with or without help from those at the surface. This is for the reasons mentioned in the prior sections and because the dynamics of the moving beacons pose interesting and challenging problems. For UWSN based localization techniques the reader is referred to \cite{TAN20111663}, \cite{bo_review_2019}, \cite{su_review_2020}.\\

\subsection{With a dedicated support vehicle}
In this approach, dedicated support vehicles are used as navigation aids (NA). These support vehicles can also act as communication gateways between the AUVs and ground or ship-based mission control stations. This configuration of the support vehicles is often termed a communication and navigation aid (CNA). CNAs have the advantage of being able to facilitate mission parameter changes and telemetry relays on the fly. Other benefits of using a dedicated NA include longer mission durations and possibly a facility for docking and recharging. The NA can either be on the surface of the water body or near the other AUVs underwater.\\

While having a NA for localization results in excellent accuracy, they also pose certain challenges. The primary one being the path planning of the aiding vehicles. This has led to investigations into the observability of the localization problem with NA and optimal path planning strategies. This is more prominent in surface-based navigation aids since aid underwater generally has the same mission plan as the AUVs. In the following sections, we discuss these approaches and the related research.\\

\subsubsection{Navigational Aid on the surface}
Having the CNA on the surface has the advantages of being in constant reception of absolute GPS location data and communication network. However, this approach does have its challenges, such as more complex mission planning, longer distances for acoustic communication with AUVs resulting in acoustic channel issues such as packet loss, higher noise, latency, etc., and mitigation of other surface traffic. Furthermore, their performance is dependent on the surface sea states. Some of the earliest results in CL using surface vehicles were with crewed ships and boats, which were non-autonomous. The conducive results from these experiments were then extended to autonomous surface vehicles.\\

\begin{table*}[]
	\renewcommand{\arraystretch}{2}
	\caption{A BRIEF COMPARISON OF UW CL CATEGORIES}
	\label{table_2}
	\centering
	\begin{tabular}{p{0.1\textwidth}p{0.08\textwidth}p{0.3\textwidth}p{0.3\textwidth}}
		\hline
				\hline
		\textbf{Class}                                       & \multicolumn{1}{c}{\textbf{Subclass}} & \multicolumn{1}{c}{\textbf{Positive Aspects}}                                              & \multicolumn{1}{c}{\textbf{Negative Aspects}}                                                          \\ \hline 		\hline
		\multirow{2}{0.1\textwidth}{With a dedicated support vehicle}    & On surface                             & Lower localization error,   No time/energy wasted in surfacing                                          & Requires additional   vehicle at the surface, Ship operations cost, Mission Planning                    \\ \cline{2-4} 
		& Underwater                             & Stealth, Does not require   surface traffic mitigation, Relatively simpler mission planning & Requires additional   vehicles, Localization error reduction not as good as with vehicle on the surface \\ \hline
		\multirow{2}{0.1\textwidth}{Without a dedicated support vehicle} & Surfacing                              & Does not require   additional support vehicle, simpler mission planning                     & Time/energy wasted in resurfacing,   relatively lower localization error reduction                      \\ \cline{2-4} 
		& Non-surfacing                          & No time energy wasted in   surfacing and no additional vehicle required                     & without absolute location information, error can grow unbounded.                                      \\ \hline 		\hline
	\end{tabular}
	
\end{table*}

\paragraph{Crewed Ship as localization aid}
In this section, we review cooperative localization using non-autonomous surface vehicles as a navigational aid. These are mainly crewed ships and boats used for the deployment and retrieval of the AUVs. In one of the earliest and simplest approaches, {Matos and Cruz} \cite{matos_auv_2005} used two boats as moving beacons to localize a single AUV in a river. The AUV runs an EKF estimator, fusing distance information from the two beacons on the boats and its dead reckoning (DR) data. Both the boats move along a path perpendicular to the AUV path, which is a bank-to-bank, back-and-forth motion along the river. In \cite{hartsfield_single_2005}, a single ship was used instead, utilizing a least-squares (LS) based approach. The position is estimated from range information and the known trajectory through a least-squares estimator, and the resulting data is fused with dead reckoning data in a Kalman Filter. The algorithm’s accuracy suffered in cases of poor trilateration geometry. It also required prior knowledge of sound speed profile (SSP), refraction index, and depth errors, the former two being challenging to acquire. McPhail and Pebody \cite{5290050} tackled the problem of position drift during the dive of a deep diving AUV by utilizing range information from a surface ship. The AUV, post-diving, is made to move in a circular orbit about the ship while its true position is estimated using a least mean square (LMS) algorithm. The paper also discusses mitigation of the problems mentioned in \cite{hartsfield_single_2005}, such as the measurement of SSP, depth error, refraction, sensor errors, etc., and others, such as the effects of tides and atmospheric pressure changes. For the same problem, in \cite{8274781}, a strong tracking algorithm in which the prediction is carried out using unscented transform is proposed. The observability with successive measurements was proved using Lie algebra. The proposed algorithm showed marginal improvements over generic UKF. {Folk \textit{et al.}} \cite{5664462} evaluated relative and absolute localization of AUV to a naval ship to measure the latter’s magnetic signatures. In the absolute case, the AUV used EKF for estimating its and the ship’s states, while in the relative case, only its own states were estimated. {Eustice \textit{et al.}} used a ship as a navigation aid for single or multiple AUVs in \cite{doi:10.1002/rob.20365}. The AUVs localized using OWTT of the acoustic signal from the ship and its sensors through a decentralized Least Squares (LS)-maximum likelihood estimator (MLE). The proposed approach was shown to perform comparably to an LBL system. The ship, however, was free-drifting without any specific or optimal path. In \cite{doi:10.1177/0278364912446166}, OWTT information from a ship was used to localize a deepwater AUV but in post-processing. The authors used a delayed state centralized EKF (DS-CEKF) to process the ship, AUV sensor, and range data. The delayed states compensated for the movement of AUV between acoustic transmission by ship and its reception by AUV. Centralized post-processing led to the incorporation of the cross-correlations between the ship and AUV states resulting in the lowest estimation errors compared to distributed estimation. Hence CEKF is often referred to as the gold standard of estimation with other estimators compared against it for performance evaluation. The ship’s motion was confined to a diamond-shaped path to improve observability. The initial position uncertainty was tackled by initializing the EKF with an MLE estimate, as improper EKF initialization leads to its instability. The same authors in \cite{6504537} proposed a decentralized extended information filter (DEIF) for localization of multiple AUVs using a single moving beacon, which is not only scalable but also suited for the low bandwidth, low capacity acoustic channel. The beacon and AUVs maintain separate filters. The beacon broadcasts only the changes in its state and uncertainty, since the last broadcast, to the AUVs asynchronously. The AUV filter reconstructs the beacon state using the information in the acoustic messages. The beacon/AUV, process, and observation models are considered to be linear. The proposed approach is evaluated against CEKF, Egocentric EKF, Interleaved Update (IU) \cite{bahr2009cooperative}, and DR for two cooperating scenarios: a) Ship as a beacon b) AUV (resurfacing for GPS) as a beacon. Results show that DEIF performs similarly to CEKF, although its performance is subject to packet loss. A pre-planned path is chosen for the beacon since the mission is known. {Allota \textit{et al.}} \cite{10.1016/j.robot.2014.03.004} presented a strategy based on geometrically calculating individual locations through inter-AUV and AUV-Ship range information, which is then utilized in the Kalman filter measurement step. The proposed scheme is evaluated using 3 AUVs and one ship, although it can be scaled while ensuring at least one AUV has a DVL sensor. The AUV, which has a DVL sensor, is denoted as the server, and the tetrahedral geometry-based algorithm is run only on it in a centralized manner with state inputs from other vehicles. The calculated locations are then communicated to other AUVs. The server AUV uses a non-linear complimentary filter, while all other AUVs use KF for position and velocity estimates. Intervehicle communication, excluding server, has no information exchange and is only used to calculate the range. The scheme has no limitations on the relative distances or paths, although it requires more than four vehicles to work. Harris and Whitcomb \cite{7487420} proposed range and range rate estimation based on cooperative localization of AUVs, without DVL or ADCP, with the help of surface ship. A  delayed state centralized EKF was used for estimation. It was shown through simulations that including the range rate information didn’t improve the localization error. Only in the case of poor range measurements marginal improvement was observed with range rate. The above approach was extended in \cite{8460970} to use a dynamic model of the vehicle instead of a kinematic model. Compared to AUVs estimating with a kinematic model without DVL, the proposed method gives results comparable to AUVs with DVL and kinematic model. The surface ship was navigating along a circular trajectory about the work area of the AUV. This approach, however, is heavily dependent on accurate modeling of the AUV.
In \cite{costanzi2018estimation}, an AUV was localized using a surface ship and a static beacon. The approach relies on utilizing information exchange within the existing ad-hoc network among AUVs, surface vehicles, and beacon nodes for localization. AUVs are assumed to be operating in deep water without the bottom lock and rely on IMU, relative velocity, and range/bearing information for positioning. The ship localizes the AUV using a high-precision acoustic positioning (HiPAP) system. Each AUV runs two EKF algorithms, one for its localization and the other to estimate neighboring nodes’ positions. Outliers are rejected using the Mahalanobis distance metric, while delays are taken care of using a back-and-forth approach wherein measurements are used in previous estimates and then propagated forward. The approach requires that the AUVs be equipped with very high-accuracy INS and a USBL modem. This not only increases cost, but USBL also limits the size of the team.  

An overview of all the above approaches using crewed vessels as CNA is given in Table \ref{table_3}. While the results using a ship or boat as a CNA show performance that is almost on par with traditional localization approaches such as LBL, operating a ship or a boat is prohibitively expensive. This is true, especially for long-duration missions, due to the crew’s expenses, operations, and maintenance of the vessel \cite{german_2012}, \cite{Kalwa:2016:0025-3324:26}. Furthermore, they are non-autonomous and thus need the mission path of the AUV known a priori and can only move in paths that are simple shapes made up of straight lines or circles.  \\


\begin{table*}[]
	\centering
	\begin{threeparttable}
		\caption{Overview of Crewed ship (non-autonomous) underwater CL}
		\label{table_3}
		\small
		\begin{tabular}{m{0.4cm}m{0.2cm}m{0.2cm}m{0.2cm}m{0.2cm}m{0.2cm}m{1cm}m{0.2cm}m{0.2cm}m{0.6cm}m{0.2cm}m{0.6cm}m{0.6cm}m{0.6cm}m{0.5cm}m{0.15cm}m{4.6cm}} 
			\hline
						\hline
			\multicolumn{1}{c}{\begin{sideways}\textbf{Reference}\end{sideways}} & \multicolumn{1}{p{0.2cm}}{\begin{sideways}\textbf{Propagation Path~} \end{sideways}} & 
			\multicolumn{1}{p{0.2cm}}{\begin{sideways}\textbf{Packet Loss}\end{sideways}} & \multicolumn{1}{p{0.2cm}}{\begin{sideways}\textbf{Packet Latency}\end{sideways}} & \multicolumn{1}{p{0.2cm}}{\begin{sideways}\textbf{Measurement Outliers~~}\end{sideways}} & 
			\multicolumn{1}{p{0.2cm}}{\begin{sideways}\textbf{Bandwidth}\end{sideways}} & \multicolumn{1}{c}{\begin{sideways}\textbf{Estimator}\end{sideways}} & \multicolumn{1}{p{0.2cm}}{\begin{sideways}\textbf{Scalable?}\end{sideways}} & \multicolumn{1}{p{0.2cm}}{\begin{sideways}\textbf{Ocean Currents}\end{sideways}} & \multicolumn{1}{c}{\begin{sideways}\textbf{Results (Sim/Exp)}\end{sideways}} & \multicolumn{1}{c}{\begin{sideways}\textbf{Ranging Method}\end{sideways}} & \multicolumn{1}{c}{\begin{sideways}\textbf{Velocity Sensing}\end{sideways}} & \multicolumn{1}{c}{\begin{sideways}\textbf{Uncertainty\footnote[1]{}} \end{sideways}} & \multicolumn{1}{c}{\begin{sideways}\textbf{Channel Sharing}\end{sideways}} & \multicolumn{1}{c}{\begin{sideways}\textbf{No. of aid vehicles} \end{sideways}} & \multicolumn{1}{p{0.2cm}}{\begin{sideways}\textbf{No. of aided vehicles}\end{sideways}} & \multicolumn{1}{c}{\begin{sideways}\textbf{Remarks}\end{sideways}}                       \\ 
			\hline
			\hline
			\cite{matos_auv_2005} & \tikzred & \tikzred & \tikzred & \tikzred & \tikzred & EKF & N & \tikzred & S & T & TE & \tikzred  & RR/F & 2 & 1 & \vspace{1mm} Two boats moves along predefined path. \\ 
			\hline			
			\cite{hartsfield_single_2005} & \tikzred & \tikzred & \tikzred & \tikzred & \tikzred & NL-LS/ KF & N & \tikzred & EPP & T & D/A & Using LBL & RR & 1 & 1 & \vspace{1mm} Only straight motion of AUV, validation in post processing                 \\ 
			\hline			
			\cite{5290050} & \tikzgreen & \tikzred & \tikzred & \tikzred & \tikzred & LMS & N & \tikzred & EPP & T & A & \tikzred & RR/F & 1 & 1 & \vspace{1mm} Deep diving AUV, Error contributions from changes in Sound Speed profile and sensor errors considered      \\ 
			\hline	
			\cite{8274781} & \tikzred & \tikzred & \tikzred & \tikzred & \tikzred & ST-UKF & N & \tikzred & S & - & N & \tikzred & - & 1 & 1 & \vspace{1mm}  Strong tracking UKF. Localization of deep diving AUV while diving. Rician noise distribution.                 \\ 

			\hline			
			\cite{5664462} & \tikzred & \tikzred & \tikzred & \tikzred & \tikzred & EKF/ CEKF & N & \tikzred & EPP & O & - & \tikzred & B & 1 & 1 & \vspace{1mm}  Relative and Global localization with respect to ship, Result on post processed data  \\ 
			\hline			
			\cite{doi:10.1002/rob.20365} & \tikzred & \tikzred & \tikzred & \tikzred & \tikzred        & LS-MLE  & Y & \tikzred & E & O & D & \tikzred  & B & 1  & 1 & \vspace{1mm}  Ship was either anchored or freely drifting. Decentralized estimator.                            \\ 
			\hline			
			\cite{doi:10.1177/0278364912446166} & \tikzgreen & \tikzred & \tikzgreen & \tikzgreen & \tikzred & DS-CEKF   & Y & \tikzred & EPP & O & D & MLE & T & 1 & 1 &  \vspace{1mm} Ship moved in a Diamond shaped pattern, Centralized Delayed state EKF considers correlation terms, but only works in post processing                  \\ 
			\hline			
			\cite{6504537} & \tikzred & \tikzred & \tikzgreen & \tikzgreen & \tikzred & DEIF & Y & \tikzred & E & O & D & \tikzred & T & 1 & 1 & \vspace{1mm}  Non optimal path for ASV, DEIF sensitive to packet loss.                 \\ 
			\hline			
			\cite{10.1016/j.robot.2014.03.004} & \tikzred & \tikzred & \tikzred & \tikzred & \tikzred & KF & N & \tikzred & S & O & D & \tikzred & T & 1 + 1 AUV  & 3 &  \vspace{1mm} Geometry based approach (Trilateration), Scaling is difficult since all vehicles transmit, Ship is assumed stationary  \\
			
			\hline			
		
			\cite{8460970} & \tikzred & \tikzred & \tikzgreen & \tikzred & \tikzred & DS-CEKF & Y & \tikzred & E & O & - & \tikzred & T & 1 & 1 &  \vspace{1mm} Used dynamic AUV model, Delayed state centralised EKF.  \\
			%
			\hline			
			\cite{costanzi2018estimation} & \tikzred & \tikzred & \tikzred & \tikzgreen & \tikzgreen & EKF & Y & \tikzgreen & E & T & D & \tikzred & T & 2* & 1 &  \vspace{1mm} *Includes 1 static beacon. All assets require USBL. 2 EKF's run on AUVs.  \\
			\hline
						\hline
		\end{tabular}
		\begin{tablenotes}
			\item \tikzgreen \hspace{0.2cm}Considered\hspace{0.5cm} \tikzred \hspace{0.2cm}Not considered. \textbf{Scalability}: Y-Yes, N-No. \textbf{Results}: E - Experimental, S - Simulation, EPP - Post processed experimental data. \textbf{Ranging Method}: O - OWTT, T - TWTT. \textbf{Velocity Sensing}: D - DVL, A - ADCP, TE - Thrust estimation from motor speed, R - Required but not specified, N - No. \textbf{Channel Sharing}: T - TDMA, B - Broadcast, RR - Request/Reply, F - FDMA
			\item[1] Uncertainty in the initial location of the AUV.
		\end{tablenotes}
	\end{threeparttable}
\end{table*}

\paragraph{ASV as localization aid}

The costs associated with crewed ships naturally led to research on surface vehicles that are uncrewed, autonomous, and can reliably perform for long durations at sea. The reader can find a review of uncrewed surface vehicles in \cite{LIU201671}. While the ASV did away with some of the costs associated with the workforce, maintenance, etc., of a large boat or ship, it introduces new challenges in the form of its control, coordination, and mission planning. This has led to new research directions, such as optimal path planning, formation control, and observability analysis in the context of CL. In the following subsections, we look at the current state of the art in CL of AUVs with the help of autonomous surface vehicles.\\

%

\textit{Single ASV and Range information only:} In this approach, the AUVs localize using range information calculated from the ASV's acoustic transmissions and the data therein. In the simplest of these cases, a single ASV is used as a localization aid for a single AUV. To find its X and Y coordinates with respect to a pre-defined frame of reference (since depth is known), the AUV requires at least two distance measurements from two different ASV locations if its current location is known with some uncertainty. For a mobile AUV carrying out its mission, this imposes constraints on the ASV motions and trajectories. {Fallon \textit{et al.}}\cite{fallon2010cooperative} used the current and past range measurements, current and past locations of the ASV, and distance traveled by ASV in-between measurements for estimating position with an EKF. For range measurement updates, the position and covariance of ASV are incorporated, but the cross-correlation is neglected, which can lead to overconfidence in estimates. As EKF fails to converge when the initial uncertainty in the location is large, in \cite{SCHERBATYUK20121}, methods for estimating the initial location of AUV and the ocean currents are discussed. Post diving, when the uncertainty is the largest, the AUV uses a vision system and stored image database to calculate its location offset. While it is stationary on the floor, an ASV or ship with towed beacon is used to estimate the initial position and sound speed profile. In \cite{8206508}, an ASV with USBL was used to localize an AUV without DVL or IMU. The AUV was assumed to have only attitude information and an acoustic model. The range information was fused with speed estimation using thruster motor current measurements in an EKF with state augmented to include unknown current velocities. 

Since EKF linearizes the system, its estimate is less accurate. To mitigate this, {Gao \textit{et al.}} \cite{gao2014robust} combined an iterative divided difference filter (I-DDF) algorithm with Huber M-estimator for localizing an underwater vehicle. The advantage of the DDF filter is that it does not linearize the system, and compared to UKF, its covariance matrix estimate is more accurate. The slow convergence of DDF in systems with weak observability and large initial error is mitigated through iteration. The Huber-based M estimator is employed to take care of outliers in range measurements. The proposed Huber-M-based DDF (HIDDF) algorithm is shown to perform better than EKF, DDF, and IDDF alone, albeit with a higher computational burden. In \cite{wu2019cooperative}, a factor graph (FG) based approach is proposed to estimate AUVs location and current velocity in the absence of bottom lock/DVL using range from a surface vehicle and neighboring AUVs. A factor graph is a graphical representation of the joint probability density functions of all the unknown positions of vehicles given the measurements from sensors. To solve the nonlinear estimation problem, a Maximum-A-Priori (MAP) algorithm is used, and to maintain the observability of the whole system, a formation-switching strategy is employed. The effects of packet loss and clock drift were also evaluated in field trials. The factor graphs, however, have high memory requirements and can get complex with increasing team size. In \cite{7004622}, an EKF and MAP based Moving horizon estimation algorithm (MHE) is proposed wherein the EKF is used to generate high-frequency estimates using depth and IMU data, while MHE is used to fuse the low-frequency range information along with its history to generate consistent estimates that do not suffer from linearization errors. To keep the computational costs in check, MHE is implemented as a moving window version of MAP.

The particle filter is another popular filter for state estimation. In \cite{7003048}, \cite{dubrovin_studying_2016}, the performance of PF against EKF is compared. Simulations show similar performance of PF and EKF due to the assumption that the measurement errors are Gaussian distributed. In \cite{fallon2010cooperative2} the performance of EKF against PF and nonlinear least squares (NLS) estimators is compared. The NLS estimator outperformed both EKF and PF, especially in the case of post-processed data. In \cite{claus2018closed}, \cite{8390710}, a comparison between DR, distributed EKF, and loosely coupled PF for estimating vehicle states using OWTT ranges and dynamic vehicle model is presented. The distributed EKF on every vehicle is augmented with other vehicles' states. The sum of covariance is used in the augmented covariance matrix to reduce errors due to overconfidence. In the loosely coupled PF case, the PF was only used for the measurement update, while the prediction stage used the output of EKF. A KF-based velocity (due to ocean currents) and synchronization bias estimator were used to correct errors due to ocean currents and clock offset. Experimental results show that the PF performed marginally better than EKF. Including bias and ranges from multiple sources (AUVs other than on the surface) improved the estimates even further. While newer techniques, such as MHE, PF, IDDF, etc., are reported, as seen from the above discussion, EKF is widely popular due to its simplicity and effectiveness in most cases.

Another interesting challenge with ASV-based cooperative localization, especially single ASV, is the path planning of ASV to minimize the positional error of the aided AUVs. In \cite{fallon2010cooperative}, the ASV is made to follow a
simple pre-planned zig-zag path which can be easily parameterized and implemented, but it is not optimal and unsuitable for more than one AUV. Extension of this work was presented in \cite{fallon2010cooperative2}, where it was also shown
that using nonlinear estimation, the system can be observable under less stringent conditions, unlike in the case of a linearized version of the system, which causes EKF to diverge. The ASV paths were generated using AUV position estimates
and uncertainty to maintain the observability of the system. Two such paths, zig-zag and circular orbit about AUV, were
evaluated. In \cite{german_2012}, the ASV, from its knowledge of the AUVs mission, uses a simple heuristic algorithm based on the minimization of the integral of squared inter-vehicle distance to plan its positions for acoustic transmissions. The approach, however, is not scalable, as each iteration of the algorithm requires three transmissions. In \cite{SCHERBATYUK20121,7003048, dubrovin_studying_2016}, to find the optimal waypoint for the next acoustic transmission, the ASV calculates the error uncertainty ellipse using the state information transmitted by AUV. The waypoint is then selected such that the error ellipse is minimized, which is along the direction of the major axis. The above results were extended in \cite{Sergeenko_2013} for ASV aiding multiple underwater AUVs by incorporating inter-vehicle range measurements. Two cases were considered. One, where the ASV helped minimize the positional error of the AUV with the worst error, and second, where the ASV helped minimize the positional error of the whole group. For a similar scenario, Chitre \cite{5547044} proposed dynamic programming (DP) based approach to generate ASV paths. It minimizes the localization error ellipse for the AUV's along the major axis, which is orthogonal to the range measurement vector. The Bellman equation for the optimal solution of DP is solved using an approximation of value function, i.e., planning over a finite horizon. The approach provides a globally optimal trajectory for ASV from the knowledge of AUV paths. This is similar to the work in \cite{1302422}, \cite{1545230}, \cite{batista2011single} for localization of AUVs using a single static beacon. In \cite{6107044}, authors tackle the path planning problem using the Markov decision process (MDP) framework. An MDP policy maps a state to action. In this case, the probability of choosing the bearing angle of ASV given its current state, AUV path, and relative range and angle. The ASV is further made to adaptively "learn" to position itself through Cross-Entropy (C-E) method over a segment of the AUVs path. A smoothing filter is applied to prevent the C-E method from converging to a local minimum. The ASV path was computed using three different strategies: a) proximity to the AUV with a larger error, b) along the centroid of the AUVs formation, and c) based on the sum of errors squared of all AUVs. The third approach was concluded to be simpler and produced better results. Computationally, this approach is only efficient once the learning is done offline and has a linear increase in complexity with the number of AUVs. A comparison of DP and MDP methods is given in \cite{6727582}. In \cite{teck2014direct}, a genetic algorithm (GA) based policy search approach that is computationally more efficient than the MDP-CE approach is proposed. The state space is divided using Voronoi tessellations to reduce the number of representative states compared to MDP. A variable-length GA is used to select the appropriate state-action pair. {Seto \textit{et al.}} \cite{seto_three-dimensional_2011, 6942873} proposed an optimal path planner in 3D for ASV aiding multiple AUVs performing mine sweeping tasks using an approach similar to \cite{SCHERBATYUK20121}. The path planner involves a look ahead strategy that also incorporates distance penalty, which helps to bound the error and reduces the computational effort. In \cite{6942873}, two cases are considered, a) the ASV knows AUV paths a priori, and b) it estimates the AUV paths in real-time using the next three waypoints communicated by the AUV. In the latter case, the cost of heading decisions at L future times steps is calculated, and the path is chosen such that the AUV position errors are minimized. The latter case is useful in tasks where the AUV may have to change its path in between the mission. In \cite{6859032},  the condition number of observability gramian and empirical observability gramian of the linearized discrete system are used to optimize the trajectory of ASV. In the former, the condition number is minimized by minimizing the difference between the trajectory inertia matrix's eigenvalues. For single ASV localizing multiple AUVs, the condition number-based approach is shown to be better than the empirical gramian based while being less computationally taxing. Walls and Eustice \cite{7003099} proposed an information maximization-based approach, similar to the maximization of the determinant of Fisher information matrix (FIM), to compute optimal trajectories for an ASV localizing multiple AUVs. Only those ASV trajectories are selected, which can be parameterized by diamond and/or zig-zag patterns instead of searching over the full space of trajectories. This is in contrast to \cite{6727582}, where the approach was based on a future segment of AUV paths at any time 't'. The performance was evaluated against the approaches in \cite{6727582} and \cite{german_2012}. It was shown that the proposed approach could attain higher information gain than others. A problem with such easily parameterizable trajectories is that if the pattern is large, the measurements appear to be from a straight line segment, while if the pattern is small, the difference between successive measurements may not be much for a vehicle that is far away. {Sousa \textit{et al.}} \cite{Sousa_2018} presented two FIM-based approaches for finding the optimal path, wherein the ASV calculates its next location with or without using the estimated AUV position. The calculation involves selecting the one point having the maximum determinant of FIM within an estimated set of all reachable points from the current location until the next communication. The effect of AUV depth and the horizontal range is evaluated on localization performance, with the localization error increasing with depth and decreasing with radius. In \cite{mandic2015range}, an extremum seeking (ES) based approach is proposed. In ES, the optimal input for an unknown input-output map is found using online gradient estimation. The proposed approach is vehicle model agnostic, subsumes constant disturbances such as gravity, currents, etc., requires minimal acoustic data transmission, and does not require AUV trajectory apriori. The cost function is formulated in terms of the estimation filter's covariance matrix, thus ensuring low computation complexity. Its optimal value minimizes the maximum eigenvalue of the covariance matrix. This, in turn, maximizes the minimum eigenvalue of the observability gramian. However, the system model considered requires the ASV to move arbitrarily in the horizontal plane, which may not always be the case. Also, the approach requires to and fro communications between AUV and ASV, which will introduce delays and will not scale well in the case of multiple AUVs. An extension to the case of underactuated beacon vehicles moving in a 2D plane was presented in \cite{mandic2016mobile}. In \cite{8755392}, an algorithm that combines priority-based expansion of a search tree with random sampling-based exploration to adaptively plan multiple future waypoints is presented. Sampling-based exploration is chosen to reduce the number of states that need to be evaluated. At the same time, the expansion of the search tree is done such that the angle between the distance vector and the major axis of AUV uncertainty is minimized. The optimal locations are such that they also correspond to the optimal time for transmission and are calculated by considering the limitations of the ASV dynamics. The optimal time for transmission is chosen from a set of TDMA time slots in which the ASV is allowed to transmit. In the case of multiple AUVs, the sum of total uncertainties is minimized since optimal locations for all AUVs will not be the same. However, priorities, if required, could be assigned. {Rua \textit{et al.}} \cite{rua2019cooperative} proposed a novel solution to the single beacon, range-based, cooperative localization of an AUV  wherein the beacon is mounted on a rotating arm, which in turn is mounted on the hull of an ASV or at a static location. However, the AUV motion is considered to be in trimming trajectories, i.e., only constant linear and angular velocities. It is shown that the system is observable in the case of AUV moving with a) constant linear and angular velocity or b) constant linear velocity under initial condition constraints, while the system is unobservable when the AUV is not moving. Further optimality analysis to find the optimal motions of only the beacon, beacon, and vehicle, optimal fixed rotation rate of beacon, and optimal energy and rotation rate are carried out in \cite{rua2020enhanced}. 

Another aspect of cooperative localization between vehicles is the observability analysis of the states of AUVs. Authors have reported analysis using nonlinear weak observability \cite{5650250,7004622} and nonlinear to linear time-varying transformations (NL-LTV) \cite{6224634,viegas2014position}. {Viegas \textit{et al.}} \cite{6224634} presented observability analysis using NL-LTV transformations using state augmentation. The authors considered two cases under the influence of unknown ocean currents: a) ASV transmits velocity and position information to AUV, and b) ASV transmits only position information, and AUV estimates ASV velocity through some sensors. A Kalman filter was then proposed for the LTV system, which guarantees global asymptotic stability of error dynamics. In \cite{7004622}, nonlinear observability analysis in the discrete domain was presented. It was shown that the AUV with only IMU, depth sensor, and range information is weakly observable in nonlinear case but is unobservable in linearized case. 

There also have been attempts at efficient and optimal use of the acoustic channel to share information between cooperating vehicles. In \cite{meira_cooperative_2011}, {Meira \textit{et al.}} coupled CL algorithm from \cite{bahr2009cooperative} with a logic-based communication approach that transmits location information from ASV to AUV depending on a threshold instead of a pre-determined periodic transmission. This threshold is based on the difference between ASV's position estimate and GPS data. While the authors analytically proved the position error's boundedness under certain assumptions on formation, velocity, and currents, the experimental implementation had no such assumptions. It was demonstrated that the approach gives only marginally worse performance than periodic transmission but with almost 62.5\% fewer transmissions. It was assumed that the AUV runs an ASV model parallel to its own to estimate better filter parameters. In \cite{8867537}, the effects of adaptive time-of-launch (TOL) of localization packets within the TDMA time slot were studied. By choosing the TOL based on a criterion, it is shown that the localization error can be reduced compared to a static TOL. EKF and NLS trilateration-based estimators were compared in the case of static single/three beacons, a follower ASV and a lawn mowing ASV, wherein the EKF performed better than the NLS-based method.

Table \ref{table_4} gives an overview of all the above approaches using a single ASV and only range information for localization. \\

\begin{table*}
	\begin{threeparttable}
		\caption{Overview of Single ASV range only CL}
		\label{table_4}
		\centering
		\begin{tabular}{m{0.4cm}m{0.1cm}m{0.1cm}m{0.1cm}m{0.1cm}m{0.1cm}m{1cm}m{0.1cm}m{0.1cm}m{0.1cm}m{0.3cm}m{0.3cm}m{0.4cm}m{0.6cm}m{0.1cm}m{0.3cm}m{0.3cm}m{0.1cm}m{0.1cm}m{5cm}} 
			\hline
						\hline
			\multicolumn{1}{c}{\begin{sideways}\textbf{Reference}\end{sideways}} & \multicolumn{1}{p{0.1cm}}{\begin{sideways}\textbf{Propagation Path~~~}\end{sideways}} & 
			\multicolumn{1}{p{0.1cm}}{\begin{sideways}\textbf{Packet Loss}\end{sideways}} & \multicolumn{1}{p{0.1cm}}{\begin{sideways}\textbf{Packet Latency}\end{sideways}} & \multicolumn{1}{p{0.1cm}}{\begin{sideways}\textbf{Measurement Outliers}\end{sideways}} & \multicolumn{1}{p{0.1cm}}{\begin{sideways}\textbf{Bandwidth}\end{sideways}} & \multicolumn{1}{c}{\begin{sideways}\textbf{Estimator}\end{sideways}} & \multicolumn{1}{p{0.1cm}}{\begin{sideways}\textbf{Scalable?}\end{sideways}} & \multicolumn{1}{p{0.1cm}}{\begin{sideways}\textbf{Ocean Currents}\end{sideways}} & \multicolumn{1}{p{0.1cm}}{\begin{sideways}\textbf{Results (Sim/Exp)}\end{sideways}} & 
			\multicolumn{1}{c}{\begin{sideways}\textbf{ASV knows Mission Plan?~~}\end{sideways}} & 
			\multicolumn{1}{c}{\begin{sideways}\textbf{Ranging Method}\end{sideways}} & \multicolumn{1}{c}{\begin{sideways}\textbf{Velocity Sensing}\end{sideways}} & \multicolumn{1}{c}{\begin{sideways}\textbf{Obs. Analysis \textbackslash{} Metric}\end{sideways}} & \multicolumn{1}{p{0.1cm}}{\begin{sideways}\textbf{Model}\end{sideways}} & \multicolumn{1}{p{0.3cm}}{\begin{sideways}\textbf{Uncertainty}\end{sideways}} & \multicolumn{1}{c}{\begin{sideways}\textbf{Channel Sharing}\end{sideways}} & \multicolumn{1}{p{0.1cm}}{\begin{sideways}\textbf{No. of aid vehicles}\end{sideways}} & \multicolumn{1}{p{0.1cm}}{\begin{sideways}\textbf{No. of aided vehicles}\end{sideways}} & \multicolumn{1}{c}{\begin{sideways}\textbf{Remarks}\end{sideways}}                                                  \\ 
			\hline
						\hline
			\cite{german_2012} & \tikzred & \tikzred & \tikzred & \tikzred & \tikzred & EKF & N & \tikzred & E & Y & T & D & \tikzred & - & - & RR & 1 & 1 & \vspace{1mm}Communication intensive approach \\ 
			\hline
			\cite{fallon2010cooperative} & \tikzred & \tikzgreen & \tikzgreen & \tikzred & \tikzred & EKF & N & \tikzred & E & N & O & \tikzgreen & NL & C & \tikzred & RR & 1 & 1 & \vspace{1mm}Predefined Zig-Zag path for ASV, Long communication times \\ 
			\hline
			\cite{SCHERBATYUK20121} & \tikzred & \tikzred & \tikzgreen & \tikzred & \tikzred & EKF & N & \tikzgreen & S & N & O & D & \tikzred & D & \tikzgreen$^{\#}$ & B & 1 & 1 & \vspace{0.5mm}$^{\#}$Initial position estimation using vision and towed beacon. Optimal path for ASV. \\ 
			\hline
			\cite{8206508} &\tikzred & \tikzred & \tikzred & \tikzred & \tikzred & EKF & N & \tikzgreen & E & N & - & TE & \tikzred & C & \tikzred & - & 1 & 2 & \vspace{1mm}ASV has USBL. No DVL/IMU on AUV.\\
			\hline	
			\cite{gao2014robust} &\tikzred & \tikzred & \tikzred & \tikzgreen & \tikzred & HIDDF & Y & \tikzred & E & N & O/T & D & \tikzred & D & \tikzred & - & 1 & 1 &\vspace{0.5mm} Iterative DDF estimator for Nonlinear model. Huber M estimator for outliers. Requires high computations. \\
			\hline			
			\cite{wu2019cooperative} &\tikzred & \tikzgreen & \tikzred & \tikzred & \tikzred & FG /MAP & Y & \tikzgreen & E & N & O & A & NL & D & \tikzred & T & 1 & 4 & \vspace{0.5mm} Mid water column with only relative velocity estimate. High memory requirements. \\
			\hline
			\cite{7004622} &\tikzred & \tikzred & \tikzred & \tikzred & \tikzred & MHE-EKF & N & \tikzred & S & N & - & N & NL & D & \tikzred & - & 1 & 1 & \vspace{0.5mm}Nonlinear observability analysis. Static beacon used in simulations.\\
			\hline	
			\cite{7003048}, \cite{dubrovin_studying_2016} & \tikzred & \tikzred & \tikzgreen & \tikzred & \tikzred & \vspace{1mm}EKF/ PF & N & \tikzred & S  & N & O & D & \tikzred & D & \tikzred & B & 1 & 1 & \vspace{0.5mm}Single ASV-AUV, PF derived trajectory is not smooth. Experimental results in \cite{dubrovin_studying_2016} \\ 
			\hline
			\cite{fallon2010cooperative2} & \tikzred & \tikzred & \tikzred & \tikzred & \tikzred & \vspace{1mm}NLS/ EKF/PF & N & \tikzred  & E  & N & O/T & \tikzgreen & NL & C & \tikzred & RR & 1 & 2 & Zig-Zag and Circular path for ASV, Long communication times \\ 

			\hline	
			\cite{claus2018closed}, \cite{8390710} &\tikzred & \tikzred & \tikzred & \tikzred & \tikzred & EKF /PF & Y & \tikzgreen & E & N & O & TE & \tikzred & D & \tikzred & T & 1 & 2 & \vspace{1mm}Distributed EKF. Considered clock offset effects. \\
			\hline

			\cite{Sergeenko_2013} &\tikzred & \tikzred & \tikzred & \tikzred & \tikzred & EKF & Y & \tikzred & S & N & O & D & \tikzred & D & \tikzgreen$^{\#}$ & - & 1 & 3 &\vspace{1mm} AUV share inter vehicle ranges. Cross correlations in the position estimates ignored.\\
			\hline
			\cite{5547044} & \tikzred & \tikzred & \tikzred & \tikzred & \tikzred & - & N & \tikzred & S & Y & O & - & \tikzred & D & - & B & 1  & 2  & \vspace{0.5mm}ASV path planning using Dynamic Programming approach. \\ 
			\hline

			\cite{6107044} & \tikzred & \tikzred & \tikzred & \tikzred & \tikzred & -  & N & \tikzred & S & Y & O & - & \tikzred & D & - & B & 1 & 2 & \vspace{1mm}Heuristic MDP-CE based method, sub-optimal paths, learning can take time. \\ 
			\hline
			\cite{6727582} & \tikzred & \tikzgreen & \tikzred & \tikzred & \tikzred & EKF & N  & \tikzgreen & S  & Y & O & D/ TE & \tikzred & D & - & B & 1 & 1 & \vspace{0.5mm}Comparison of DP \& MDP-CE. ASV estimates AUV position.                   \\ 
			\hline
			\cite{teck2014direct} &\tikzred & \tikzred & \tikzred & \tikzred & \tikzred & - & Y & \tikzred & S & Y & - & - & \tikzred & D & - & B & 1 & 4 &\vspace{0.5mm} Genetic algorithm based policy search approach.\\
			\hline
			\cite{seto_three-dimensional_2011}, \cite{6942873} & \tikzred  & \tikzred & \tikzred & \tikzred & \tikzred & EKF & Y  & \tikzred & E & Y/N & O & - & \tikzred & D & \tikzred & T & 1 & 2 & \vspace{0.5mm}Limited Scalability. AUV depth is considered variable. Doesn't need mission plan. \\ 
			\hline
			\cite{6859032} & \tikzred & \tikzred & \tikzred & \tikzred & \tikzred & - & - & \tikzred & S & N & - & - & COG/ CEG & C & \tikzred & - & 1 & 3 & \vspace{0.5mm}ASV Optimal trajectory using path inertia and Empirical Observability Gramian. \\ 
			\hline
			\cite{7003099} & \tikzred & \tikzred & \tikzred & \tikzred & \tikzred & Delayed State IF & Y & \tikzred & S & Y & O & - & M & C & - & T & 1 & 2 & \vspace{0.5mm}ASV Optimal path in terms of zigzag or diamond shapes. No localization error performance details are provided.             \\ 
			\hline
			\cite{Sousa_2018} & \tikzred & \tikzred & \tikzred & \tikzred & \tikzred & EKF & N & \tikzred & S & N & - & - & D & D & \tikzred & B & 1 & 1 & \vspace{0.5mm}AUV is remotely operated. ASV is assumed holonomic.                     \\ 
			\hline
			\cite{mandic2015range}, \cite{mandic2016mobile} & \tikzred & \tikzred & \tikzred & \tikzred & \tikzgreen & KF & N & \tikzgreen & E & N & - & D & COG & C & \tikzred & - & 1 & 1 & \vspace{0.5mm}Extremum Seeking based path planning for ASV, Model agnostic approach.                                               \\ 
			\hline

			\cite{8755392} & \tikzred & \tikzred & \tikzred & \tikzred  & \tikzred & EKF & N & \tikzred & S & N & O & - & \tikzred & - & \tikzred & T & 1 & 2 & \vspace{0.5mm}Adaptive ASV path Planning, Computationally expensive for multiple AUVs. \\ 
			\hline
			\cite{rua2019cooperative}  & \tikzred & \tikzred & \tikzred & \tikzred & \tikzred & - & Y & \tikzred & S & N & - & - & NLTV & C & \tikzred & - & 1 & 1 & \vspace{0.5mm}Beacon mounted on rotating arm in ASV. Trimming trajectories only.  \\ 
			\hline	
			
			\cite{viegas2014position} & \tikzred & \tikzred & \tikzred & \tikzred & \tikzred & KF & - & \tikzgreen & S & N & - & D & NLTV & C  & \tikzred & - & 1 & 1 & \vspace{1mm}NLTV Observability analysis and Observer design presented for single ASV-AUV pair.          \\ 
			\hline
			\cite{meira_cooperative_2011} & \tikzred & \tikzred & \tikzred & \tikzred & \tikzgreen & \vspace{1mm}IU/ Hybrid KF & Y & \tikzgreen & E & N & O  & \tikzgreen & \tikzred & C & \tikzred & LB$^{*}$ & 1 & 1 & \vspace{0.5mm}$^{*}$Logic based communication. AUV runs a ASV model in parallel. \\ 
			\hline
			\cite{8867537} &\tikzred & \tikzred & \tikzred & \tikzred & \tikzred & \vspace{0.5mm}EKF/ NLS & N & \tikzred & S & N & O & - & - & - & \tikzred & T & 1 & 1 & Adaptive time of launch within TDMA slot. No optimal path.\\
			\hline
			\hline
		\end{tabular}
		\begin{tablenotes}
			\item \tikzgreen \hspace{0.2cm}Considered\hspace{0.5cm} \tikzred \hspace{0.2cm}Not considered. \textbf{Results}: E - Experimental, S - Simulation, EPP - Post processed experimental data. \textbf{Ranging Method}: O - OWTT, T - TWTT. \textbf{Velocity Sensing}: D - DVL, A - ADCP, TE - Thrust estimation from motor speed, N - No.\textbf{Obs. Analysis \textbackslash{} Metric}: NL - Non linear weak observability, NLTV - Non linear to linear time varying transformation, COG - Condition number of observability gramian, CEG - Condition number of empirical observability gramian, M - Mutual information, D - Determinant of FIM.  \textbf{Model}: C - Continuous, D - Discrete. \textbf{Channel Sharing}: T - TDMA, B - Broadcast, RR - Request/Reply, F - FDMA. 
		\end{tablenotes}
	\end{threeparttable}
\end{table*}

\textit{Single ASV with Range and bearing information:}%

In this approach, localization between vehicles is carried out with sensors that provide both range and bearing information, such as USBL, SBL, etc. When range and bearing information is combined with depth sensor data, localization of any vehicle in 3D ideally requires only one acoustic transmission if the clocks are synchronized. In practice, considering a single ASV-AUV case, the ASV initially sends an interrogation signal to which the AUV replies. Using the two-way travel time and the calculated bearing, the ASV can localize the AUV. The estimated position is then communicated back to the AUV. In \cite{Glotz2015}, a hierarchical cooperative localization scheme between one ASV and two AUVs is presented. One of the AUVs acted as a guide/server for the other AUVs and was stationed between the latter and the ASV in the water column. It localized relative to the ASV and the other AUV using USBL along with absolute position information from ASV and velocity/depth data from acoustic packets. It used a linearized system model with KF for state estimation of the ASV and the other AUV. The other AUV only had an AHRS and estimated its states using speed estimation and data communicated by the middle AUV. In \cite{7401992},  USBL in inverted mode (iUSBL) with OWTT is used to localize an AUV using a surface craft. In iUSBL, the USBL modem is onboard the AUV instead of the ASV. The AUV interrogates the ASV, which replies with its position data. This data, along with the calculated range and bearing, is then used for localization. This provides position information for the AUV independent of INS. Using OWTT further alleviates the need for back-and-forth communication; thus, multiple vehicles can be localized simultaneously. In \cite{glotzbach2016acoustic}, a UKF-based state estimator is proposed for AUV having access to the range, bearing, and elevation information relative to ASV, along with its velocity estimate and velocity of the ASV. Packet latency issues are resolved by back-calculating the estimates using current measurements. {Salavasidis \textit{et al.}} \cite{salavasidis2016co} proposed an algorithm that uses EKF for state estimation of AUV, but is run partially on AUV and ASV, such that the measurement update part is carried out on the ASV using USBL, to reduce the computational burden on the AUV. The computed location estimate is then communicated to the AUV. The approach, however, has a high communication overhead. In \cite{phillips2018autonomous}, the ASV computes the AUV location through USBL and communicates back only the error measured with respect to the GPS position instead of the absolute position. It also tracks the AUV using a virtual target approach while maintaining a given offset. The communicated error is used in a KF onboard the AUV along with DR sensors to compute its position. A maximum-a-posteriori estimation-based approach is presented in \cite{franchi2021maximum}. The range and bearing information are acquired using a USBL on the surface vehicle stationed at a fixed point. In \cite{nad2016cooperative}, an artificial potential field (APF) based controller for ASV is presented to support the localization of multiple underwater agents, which could be AUVs or human divers. The agents use iUSBL and exchange location, velocity, and course data with the ASV. As iUSBL works only when communicating nodes are vertically aligned, the APF is created such that it has an attraction basin between the agents and repulsive fields directly above them. In \cite{zhang2020cooperative}, MDP is combined with a reinforcement-based Q-learning approach. However, Q learning complexity increases exponentially with the increasing number of AUVs.

The back-and-forth communication involved in this approach limits its scalability to a few vehicles. The bearing calculations are further dependent on the roll and pitch moments of the ASV when the USBL is mounted on it. The limited range of the USBL also imposes constraints on the operating area of the team. Since USBL systems are very costly, utilizing them on each team member as iUSBL increases the cost drastically. These are the primary reasons why range-only localization is more popular, and there are very few works in CL utilizing USBL.

Table \ref{table_5} gives an overview of approaches using a single ASV with range and bearing-based CL. \\

\begin{table*}
	\begin{threeparttable}
		\caption{Overview of single ASV (range and bearing) and multiple ASV CL}
		\label{table_5}
		\centering
		\renewcommand{\arraystretch}{1.5}
		\begin{tabular}{m{0.2cm}m{0.4cm}m{0.1cm}m{0.1cm}m{0.1cm}m{0.1cm}m{0.1cm}m{0.5cm}m{0.2cm}m{0.1cm}m{0.1cm}m{0.3cm}m{0.3cm}m{0.2cm}m{0.1cm}m{0.2cm}m{0.2cm}m{0.3cm}m{0.1cm}m{0.2cm}m{5cm}} 
			\hline
			\hline
			\multicolumn{1}{p{0.1cm}}{\begin{sideways}\textbf{Category}\end{sideways}} & \multicolumn{1}{c}{\begin{sideways}\textbf{Reference}\end{sideways}} & \multicolumn{1}{p{0.1cm}}{\begin{sideways}\textbf{Propagation Path~~~}\end{sideways}} & \multicolumn{1}{p{0.1cm}}{\begin{sideways}\textbf{Packet Loss}\end{sideways}} & \multicolumn{1}{p{0.1cm}}{\begin{sideways}\textbf{Packet Latency}\end{sideways}} & \multicolumn{1}{p{0.1cm}}{\begin{sideways}\textbf{Measurement Outliers}\end{sideways}} & \multicolumn{1}{p{0.1cm}}{\begin{sideways}\textbf{Bandwidth}\end{sideways}} & \multicolumn{1}{c}{\begin{sideways}\textbf{Estimator}\end{sideways}} & \multicolumn{1}{p{0.2cm}}{\begin{sideways}\textbf{Scalable?}\end{sideways}} & \multicolumn{1}{p{0.1cm}}{\begin{sideways}\textbf{Ocean Currents}\end{sideways}} & \multicolumn{1}{p{0.1cm}}{\begin{sideways}\textbf{Results (Sim/Exp)}\end{sideways}} & \multicolumn{1}{c}{\begin{sideways}\textbf{ASV knows Mission Plan?~~}\end{sideways}} & \multicolumn{1}{c}{\begin{sideways}\textbf{Ranging Method}\end{sideways}} & \multicolumn{1}{p{0.2cm}}{\begin{sideways}\textbf{Velocity Sensing}\end{sideways}} & \multicolumn{1}{p{0.1cm}}{\begin{sideways}\textbf{Obs. Analysis \textbackslash{} Metric}\end{sideways}} & 
			\multicolumn{1}{p{0.1cm}}{\begin{sideways}\textbf{Model}\end{sideways}} & \multicolumn{1}{p{0.2cm}}{\begin{sideways}\textbf{Uncertainty}\end{sideways}} & \multicolumn{1}{c}{\begin{sideways}\textbf{Channel Sharing}\end{sideways}} & \multicolumn{1}{p{0.1cm}}{\begin{sideways}\textbf{No. of aid vehicles}\end{sideways}} & \multicolumn{1}{p{0.2cm}}{\begin{sideways}\textbf{No. of aidded vehicles}\end{sideways}} & \multicolumn{1}{c}{\begin{sideways}\textbf{Remarks}\end{sideways}}                                                  \\ 
			\hline
			\hline
			\multirow[c]{5}{*}[-0.5cm]{\begin{sideways}{Single ASV – Range and Bearing}\end{sideways}}             & 
			\cite{Glotz2015} &\tikzred & \tikzgreen & \tikzred & \tikzred & \tikzred & KF & N & \tikzgreen & S & N & U & TE & \tikzgreen & D  & \tikzred & T & 1 & 1-1 & Hierarchical cooperative localization scheme. Uses USBL in middle vehicles.\\
			\cline{2-21}
			&\cite{7401992} &\tikzred & \tikzred & \tikzred & \tikzred & \tikzgreen & - & Y & \tikzred & E & - & U & - & - & -  & - & - & 1 & 1 & Inverted USBL concept. Scalable but costly in terms of hardware.\\
			\cline{2-21}
			&	\cite{glotzbach2016acoustic} & \tikzred & \tikzgreen & \tikzgreen & \tikzred & \tikzred & UKF & N & \tikzgreen & S & N & U & TE & \tikzred & D & \tikzred & B & 1 & 1 &  UKF for estimation and considers packet latency. Rest same as \cite{Glotz2015}.  \\ 
			\cline{2-21}
			& \cite{salavasidis2016co} & \tikzred & \tikzred & \tikzred & \tikzred & \tikzred & EKF & N & \tikzred  & S & N & U & A & \tikzred & D & \tikzred & RR & 1 & 1 & EKF partially run on AUV and ASV. High communication overhead. \\ 
			\cline{2-21}
			& \cite{phillips2018autonomous} & \tikzred & \tikzred & \tikzgreen & \tikzgreen & \tikzred & KF & N & \tikzred & E & N & U  & A & \tikzred & D & \tikzred & B  & 1 & 1 & Considers packet latency, Thresholding for outliers.  \\ 
			\cline{2-21}
			& \cite{franchi2021maximum} & \tikzred & \tikzred & \tikzred & \tikzred & \tikzred & MAP & N & \tikzred & E & Y & U  & TE & \tikzred & D & \tikzred & RR  & 1 & 1 & MAP based estimation. ASV is stationary.  \\ 
			\cline{2-21}
			& \cite{nad2016cooperative} & \tikzred & \tikzred & \tikzgreen & \tikzred & \tikzred  & -  & N & \tikzred & E & N & U & - & \tikzred & - & \tikzred & RR & 1 & 1 & ASV path planning using Artificial Potential Field. Large acoustic period.  \\ 
			\cline{2-21}
			&\cite{zhang2020cooperative} & \tikzred & \tikzred & \tikzred & \tikzred & \tikzred & - & N  & \tikzred & S  & Y & U & - & \tikzred & D & \tikzred & - & 1 & 2 & ASV path planning using MDP with Q learning based approach.\\ 
			\hline
			\hline
			\multirow[b]{5}{*}[-3cm]{\begin{sideways}{Multiple ASV}\end{sideways}}                               
			& \cite{curcio2005experiments} & \tikzred & \tikzred & \tikzred & \tikzred & \tikzred  & - & Y & \tikzred & E & - & O/T & D & - & - & - & T & 2 & 1 & No AUV was used in experiments, all 3 Vehicles were kayaks. \\ 
			\cline{2-21}
			& \cite{bahr2009cooperative} & \tikzred & \tikzred & \tikzgreen & \tikzgreen & \tikzgreen & IU & Y${^\#}$ & \tikzred & E & N & O & - & \tikzred &- & \tikzred & RR & 2 & 1 & ${^\#}$Limited scalability of less than 4 vehicles.                           \\ 
			\cline{2-21}
			& \cite{glotzbach2012underwater} &\tikzred & \tikzgreen & \tikzgreen & \tikzred & \tikzred & EKF & Y & \tikzred & E & N & T & - & - & D & \tikzred & B & 3 & 1 & Tracking of diver. Delay between ping reception and reply is ignored.\\
			\cline{2-21}
			& \cite{chen_moving_2016} &\tikzred & \tikzred & \tikzred & \tikzred & \tikzred & MHE & Y & \tikzred  & S & - & O & - & - & D & \tikzred & T & 4 & 1 & MLBL based approach.Moving horizon estimator. Uses linear model.\\
			\cline{2-21}
			&\cite{8084581} &\tikzred & \tikzred & \tikzred & \tikzgreen & \tikzred & UKF & N & \tikzred & E & - & T & - & - & D & \tikzred & -  & 4 & 1 & Acoustic channel modeling for sound speed profile. Consensus current estimation. Shore based centralized approach. High communication overhead.\\
			\cline{2-21}
			& \cite{crasta2017range}, \cite{crasta2018multiple} & \tikzred & \tikzred & \tikzred & \tikzred & \tikzred & - & Y & \tikzred & E & Y/N & - & - & D & C & \tikzred & - & 2  & 1 & Observability analysis. Cooperative Tracking of target with Companion that has its own mission which is known or unknown. Continuous Kinematic Model, Discrete Measurement Model.                      \\ 
			\cline{2-21}
			& \cite{bo_optimal_2018} & \tikzgreen & \tikzred & \tikzred & \tikzred & \tikzred & - & Y & \tikzred & S & N & D & - & D  & C & \tikzgreen & - & 6* & 3 & 3D model, Distance Dependent Noise, Multiple ASV tracking Single/Multiple Targets, no. of ASV$>$AUV. *1 static sensor.  \\ 
			\cline{2-21}
			& \cite{quraishi2019flexible} & \tikzred & \tikzgreen & \tikzred & \tikzred & \tikzred & EKF & Y & \tikzred & E & Y  & O & - & \tikzred & - & - & B & 2 & 1 & Static beacon used as one of the ASV \\
			\cline{2-21}
			& \cite{bai2020novel} & \tikzred & \tikzred & \tikzred & \tikzgreen & \tikzred & HTM-EKF & Y & \tikzred & E & Y  & T & D & \tikzred & D & - & RR & 2 & 1 & Outlier mititgation using heavy tailed Gaussian mixture model \\
			\hline
			\hline
		\end{tabular}
		\begin{tablenotes}
			\item \tikzgreen \hspace{0.2cm}Considered\hspace{0.5cm} \tikzred \hspace{0.2cm}Not considered. \textbf{Results}: E - Experimental, S - Simulation, EPP - Post processed experimental data. \textbf{Ranging Method}: O - OWTT, T - TWTT, U - USBL, D -TDOA. \textbf{Velocity Sensing}: D - DVL, A - ADCP, TE - Thrust estimation from motor speed. \textbf{Obs. Analysis \textbackslash{} Metric}: NL - Non linear weak observability, NLTV - Non linear to linear time varying transformation, COG - Condition number of observability gramian, CEG - Condition number of empirical observability gramian, M - Mutual information, D - Determinant of FIM.  \textbf{Model}: C - Continuous, D - Discrete. \textbf{Channel Sharing}: T - TDMA, B - Broadcast, RR - Request/Reply, F - FDMA. 
		\end{tablenotes}
	\end{threeparttable}
\end{table*}

{\textit{Multiple ASV’s as localization aids:}}

For a single ASV to aid even a single AUV for localization requires that the ASV moves to optimal waypoints, as seen previously. The primary advantage of having multiple ASVs to aid in localization is thus the relaxed requirements on the motions and trajectories that an ASV has to execute to keep the localization error of AUVs within bounds \cite{6859032, mandic2016mobile, Sousa_2018}. It is also beneficial when the AUVs' team size is large and spread over a larger area, as it would be impossible for a single AUV to satisfactorily aid all the AUVs at the same time and will necessitate a priority-based approach. It also increases the probability of acoustic data reception from AUVs, minimizing packet loss errors. This, however, has additional costs in terms of hardware, mission planning complexities, and computations required. Although the performance benefits outweigh the costs, only a few works have addressed this problem. 

One of the first works to propose multiple autonomous surface craft as a CNA is \cite{curcio2005experiments}. This is an extension of moving long baseline (MLBL) work done in \cite{vaganay2004experimental} by using ASVs instead of boats. MLBL is similar in concept to LBL, but the beacons are considered to be mobile. {Bahr \textit{et al.}} \cite{bahr2009cooperative} used two ASVs for cooperative localization of an actual AUV and proposed an Interleaved Update (IU) algorithm that works in the presence of measurement outliers and can scale to large team size. At each instant probable position, candidates are evaluated, and by minimizing a cost function based on Kullback-Leibler divergence (KLD), the most appropriate one is selected. In \cite{glotzbach2012underwater}, a diver/AUV position estimation technique using multiple ASVs was proposed. The maximum velocity, acceleration, and turning rate of the diver target are assumed to be known. The position is deduced using transmit/reply TWTT scheme through a CEKF algorithm that uses the back-and-forth technique to take care of delays and vehicles' motion during measurements. {Chen \textit{et al.}} \cite{chen_moving_2016}, extended the work in \cite{7004622}, which uses single ASV, to use MHE for localizing a single AUV with multiple surface vehicles in an MLBL approach similar to \cite{yan2015moving, yan2015optimal}. The proposed approach is compared against a KF, showing marginal improvements. The same author in \cite{chen2016optimal} investigated the optimal number of ASVs, the range between AUV-ASV, and the effect of distance-dependent noise factor using a cost formulation. It was shown that the cost is inversely proportional to the number of ASVs and directly proportional to the distance-dependent noise factor. An optimal value of range to minimize the localization error for AUV is also found. 

In \cite{8084581}, a shore-based centralized approach that uses the knowledge of the ocean current model is presented. The AUVs employ a UKF, with drift, due to ocean currents modeled as a random walk, used as one of the states. The AUVs' estimates are processed using a consensus algorithm at a shore-based centralized server via surface vehicles as communication intermediaries. The consensus current estimate is then communicated back to the AUVs to improve their estimation. In \cite{crasta2017range, crasta2018multiple}, a concept of using a companion vehicle to aid ASV for localization of a target using range information is proposed. The companion vehicle can be solely used as an aid or could be performing an independent mission of its own. Furthermore, the companion vehicle's location may or may not be known to the ASV, which is measured in the latter case. In all cases, the companion vehicle shares the information about its range to the target with the ASV. The ASV optimal trajectories are calculated by maximizing the determinant of FIM. Except when the companion is also cooperatively tracking the target, the target is assumed to be stationary. In \cite{bo_optimal_2018}, an approach for localizing N target AUVs by M cooperating surface AUVs, where M$>$N, using the TDOA technique in 3D, is presented. The effects of distance-dependent noise, uncertainty in the target AUV's initial location, and curved path in acoustic signal transmission are taken into account. The optimal formation/locations for the surface AUVs are arrived at by sequentially evaluating the determinant of FIM at each time step and moving only if it's higher than the current value. The optimal formations are evaluated in simulations against different depths for target AUV, centralized and decentralized sensor pairings, different numbers of surfaces, and target AUVs. However, only the cases with static target AUVs are considered. In \cite{quraishi2019flexible}, a multi-ASV system for localization of multiple AUVs that is scalable in the number of AUVs was proposed. The AUVs are made to be passive listeners, while the ASVs transmit their location information using TDMA. The ASV is assumed to know the AUV path for it to be in its vicinity to aid in localization. The optimality of the ASV location, although, has not been considered. In \cite{bai2020novel} a robust KF is proposed that uses heavy-tailed mixture distribution for outlier mitigation in the case of two ASVs aiding a single AUV. It is shown using experimental data that the proposed approach is better than several contemporary approaches for outlier-affected acoustic communication. 

Table \ref{table_5} gives an overview of approaches using multiple ASVs. \\

\subsubsection{Navigational Aid underwater}

While the CNAs, on the surface, have the advantages of lower localization error and communication, they are unsuitable in certain applications, especially in defense such as espionage, target tracking, etc. Furthermore, surface craft can be affected by sea states and other surface traffic. However, the acoustic channel's challenges, particularly with deep-diving AUVs, can have a far more detrimental effect on CL performance. An alternative approach is to have the navigation aid (NA) vehicles close to the AUV team performing specified tasks. This also benefits from outfitting the survey AUVs with high-accuracy task-specific sensors and medium or low accuracy (thus low cost) navigation sensors. The aiding AUVs meanwhile have high accuracy inertial navigation sensors, which are fewer in number. In case absolute positioning information is required, the NA AUVs can then resurface for a GPS fix. While different terminology has been used for this category in literature, such as leader-follower, mother-daughter, and master-slave, we will refer to it as the server-client approach. Here the server vehicles are the ones that have high-accuracy navigational sensors and provide localization support for the client survey AUVs by sharing their current location information. 

One of the earliest solutions resembling cooperative strategies for localization was presented in \cite{singh1996integrated}, wherein a server-client approach was used. The client AUVs would localize using USBL with respect to a server AUV, which in turn would localize itself using LBL. In \cite{baccou2001cooperative}, a similar approach was proposed, but the client AUVs use only range and location data from the server through an acoustic modem instead of USBL, along with data from their DR sensors. {Vaganay \textit{et al.}} \cite{vaganay2004experimental} is the first work to present the MLBL concept wherein two AUVs perform the role of CNA for other survey AUVs. The time-synchronized survey AUVs calculate their positions by passively listening to location update pings from CNA, which flank the former on both sides. Since the survey AUVs are passive listeners, this approach is highly scalable. In \cite{5357852}, a centralized delayed state EKF (DSEKF) running on a server was proposed for fusing information in one server, multiple client configuration, which takes into account the delay in propagation of acoustic signals. With one filter instance for every client vehicle, the computational complexity increases with the number of vehicles. 
In \cite{song2013cooperative}, a distributed approach combining Dynamic SLAM and cooperative localization of client AUVs is proposed. The server AUVs are assumed to be localized with very low error and are used as dynamic landmarks for client AUVs' SLAM algorithm. In the absence of server AUVs in proximity, CL with other client AUVs is used to bound the error. Consistency is preserved by using the client AUV with the least covariance as the beacon. Both DSLAM and CL are formulated as independent Bayes filters and solved using EKF. The consistency issue when using server AUVs is resolved in \cite{Song_2013b} by formulating a distributed modified EKF (MEKF). Here, MEKF takes care of the cross-correlations by employing Jacobian multipliers that the client AUVs use to track the server AUV's location changes during each DSLAM correction phase. But since the AUV model is non-linear, the error performance of MEKF is limited, which is why PF is then proposed in \cite{6859344}. It is shown that the PF performs marginally better with similar or lower computational complexity than MEKF. In \cite{walls2014origin}, an algorithm based on the origin state method is proposed to estimate the states of multiple client vehicles using acoustic range measurements and pose-graph information from multiple server vehicles. It is robust against packet loss and is bandwidth-efficient. The server communicates incremental pose-graph information relative to a server state known by the client, termed the origin state. The server vehicles are assumed to have their absolute position information, for example, through a surface vehicle. It was shown that the performance using a DEIF estimator on client vehicles produced consistent results similar to centralized estimation schemes without any overconfident estimates. The above results were extended in \cite{7139030} by constructing the local state in the form of factor chains using a factor graph framework with odometry and/or GPS as factors. These factor chains are then approximated through composition, unlike approximations relative to the origin pose as in the previous paper, and are shared with the other vehicles. This approach does not require shifting the origin pose and allows vehicles without odometry to join the network. While it is similar to \cite{6942559}, it has lower communication overhead and better performance. { Ben \textit{et al.}} \cite{ben2021novel} used factor graphs for the case when both range and bearing information are used for client position. Since cycles may exist within the graph, a clustering method is used to obtain cycle-free graphs. {Zhang \textit{et al.}} \cite{zhang2016optimal} presented a triangulation-based approach to localize a single client AUV using three server AUVs. The server paths were independent of each other and the client's mission. This approach fails if the server tasks are such that it takes them beyond the AUV's communication range. In \cite{8003160}, a parallel projection algorithm (PPA) based approach is proposed wherein the global pose of the vehicle to be localized (rotation matrix and translation) is estimated through MLE formulation that is convexified and solved using PPA. The approach is compared against a semi-definite programming-based approach for the coordinate alignment-based formulation. The proposed approach is shown to have similar or better performance while having a faster convergence rate. {Zhang \textit{et al.}} \cite{zhang2019cooperative} extended the MDP-CE approach proposed in \cite{6107044} to the server-client configuration, using two server AUVs to aid multiple client AUVs. However, the approach requires training the CE algorithm every time the trajectory of the clients is changed. In \cite{kim2020cooperative}, the effects of unknown constant current are investigated. Only the server is equipped with DVL and uses USBL to measure the clients' range and orientation. Locations of the server and clients are estimated by the server using a hybrid UKF-KF estimator, wherein the prediction step uses UKF while the measurement step uses linear KF. The authors also consider the case where the clients also communicate among themselves but don't consider the cross-correlation, thus giving overconfident results that are worse than no intercommunication. {Yan \textit{et al.}} \cite{yan2018polar} proposed a cooperative localization approach for server-client UUVs in polar regions where navigation is difficult as meridians converge faster, thus leading to calculation overflow and error magnification in conventional SINS. For this, a polar grid algorithm-based state formulation is used, and a delayed state Adaptive KF is employed for state estimation of follower UUVs. In \cite{5603992} \cite{6608182}, multiple AUVs are used as localization aids for the main AUV performing a critical task. Centralized processing is carried out on the main AUV, which estimates its state and the state of the aiding AUVs. The aiding AUVs take turns to resurface for GPS fix and use EKF to estimate their states using this data and information received from the main AUV. 

In \cite{7752795}, a KF-based solution to the time delay problem in server-client cooperative localization is presented. The client AUV is localized by server AUV using USBL. The time delays are taken care of by modifying the update step of KF in terms of the delayed measurement updates, i.e., time delays are converted into a bias in the observation equation. Simulations show that the proposed approach reduces the error in estimates by orders of magnitude. In \cite{qi2016cooperative}, an augmented EKF (AEKF) for mitigating time delays in the acoustic channel was proposed. With the knowledge of the delay, the client AUV's state vector was augmented with all the states from the actual transmitted time till the current time, using which the estimate was then propagated to the current time. Results presented indicate AEKF can bound the localization error better than EKF in the presence of time delays at the expense of computational cost.

In \cite{zhang2017cooperative}, cooperative localization of a single client vehicle with multiple servers using a probability hypothesis density filter is presented. The filter runs on all server vehicles and estimates the client location using TWTT. The estimates are communicated back to the client, which uses an information entropy-based approach to fuse and obtain the best estimate of its position. However, the proposed method has high communication overhead and is computationally complex. In \cite{gao2014improved}, an improved TWTT communication scheme for the server-client CL approach is presented. The server interrogates each client AUV using TWTT, but instead of re-transmitting measurement updates to each client AUV, the complete state information is broadcast. Thus requiring $2N+1$ acoustic transmissions instead of $3N$, where $N$ is the number of client AUVs. In \cite{8084717}, localization of a single client AUV with two server AUVs without time synchronization is presented. The range calculated using TWTT from both server AUVs was used in EKF to estimate the clients' position. While the proposed method is not affected by clock drift, it is not scalable. Qu \textit{et al.} \cite{qu2021optimal} investigated the optimal formation for multiple AUVs acting as servers for a single client AUV. The formulation uses FIM and area of information ellipse. This approach requires clock synchronization among the vehicles and bearing information. In \cite{bo2020optimal}, these shortcomings are mitigated by relying on RSSI information instead.

{Fan \textit{et al.}} \cite{fan2018maximum} proposed a maximum correntropy (MCC) based unscented PF to mitigate outliers in range measurements wherein it used KLD-based re-sampling. Compared to PF, CKF, MCC-UKF, UPF, Huber-based UPF, and cubature KF, the proposed algorithm produced the least average error but is computationally intensive. In \cite{8049492}, an adaptive extended Kalman filter that estimates the unknown process and measurement noise covariances using the expectation-maximization method was presented. For a localization problem involving two servers and one client, the proposed method is compared against DR, EKF, Innovation AKF, and Sage-Husa AKF. The proposed algorithm converges the fastest while having better error performance than other methods except for EKF, where it is marginally better with higher computational complexity. In \cite{zhao2016collaborative}, an unscented PF (UPF) based estimator was proposed to take care of the non-Gaussian nature of noise and depletion of particles in PF. In UPF, UKF was used to update the state of each particle in the client AUV. While the proposed algorithm has the lowest error compared to EKF, UKF, and PF, it requires ten times more computational time. In \cite{8327867}, a Student-t based EKF (SEKF) for outlier mitigation is presented. The students-T distribution is used for process and measurement noises instead of the usual Gaussian distribution. Results indicate that the proposed method has almost $30-40\%$ better error reduction than Threshold EKF and $60-50\%$ better than standard EKF, albeit with slightly higher computational requirements. In \cite{xu2019cooperative}, a maximum correntropy criterion (MCC), adaptive neuro-fuzzy inference system (ANFIS), and CKF-based approach for mitigation of outliers and acoustic packet loss is proposed. The packet loss is taken care of by ANFIS, which is a fuzzy system wherein the fuzzy membership function and rules are trained using the neural network from a large amount of data instead of being selected arbitrarily. For the outliers, the MCC is used. State estimation when there is no packet loss is carried out using a cubature Kalman filter based on the MCC data. The same is used for training ANFIS. The trained ANFIS model predicts the location when there is packet loss for more than 3 seconds. The proposed approach improves error performance by over 60\% against only CKF and by over 40\% against ANFIS-CKF. While MCC-CKF computational requirements are almost the same as CKF, ANFIS requires a large amount of training data. This can be an issue if the packet loss is frequent and not enough data is acquired to train ANFIS. The MCC kernel bandwidth is chosen heuristically. To mitigate this, an adaptive version of the algorithm is proposed in \cite{li2020improved}. The same authors in \cite{xu2021novel} combined adaptive cubature KF to track ranging errors and used ANFIS for detecting anomalies in ranging data. In \cite{xu2020novel}, a robust Gaussian approximate smoother based on expectation maximization is presented for outliers mitigation and sensor faults. A faulty DVL is considered, and any bias in acoustic modem data is treated as unknown inputs.

As for the path planning of the aiding vehicle, there are very few papers that investigate the problem. This is because the server's path is generally the same as the client's mission. However, there are few works where this assumption is not held true. In \cite{rui2010cooperative}, the server plans its path using a dynamic programming-based approach from \cite{5547044}. The server AUV is assumed to know the survey plan a priori. In \cite{bahr2012dynamic}, an algorithm for optimal positioning of server AUVs, without prior knowledge of client AUVs paths, is presented. The server AUV calculates the optimal positions to broadcast position information that will minimize the combined location uncertainty for all client AUVs. This is done by dividing the area into grid points that can be reached by server AUV before the next broadcast and estimating the minimization of uncertainty by broadcasting from all those grid points. The point that leads to the minimum error estimates is chosen. While the proposed approach is distributed and robust against the number of beacons/survey AUVs at any instant, correlations among the vehicles are ignored. In \cite{7353681}, a belief space path planner based on a partially observable Markov decision process (POMDP) model was proposed, which uses a probabilistic acoustic channel model that accounts for randomness in measurements such as packet loss. Optimal open-loop control actions or parameterized paths are generated for the server using EKF, the proposed model, and the known client trajectory. 

Table \ref{table_6} gives a summary of the approaches using underwater navigational aid.

\begin{table*}
	\begin{threeparttable}
		\caption{Overview of CL using underwater navigation aid}
		\label{table_6}
		\centering
		\renewcommand{\arraystretch}{1.1}
		\begin{tabular}{m{0.5cm}m{0.1cm}m{0.1cm}m{0.1cm}m{0.1cm}m{0.1cm}m{1cm}m{0.15cm}m{0.1cm}m{0.4cm}m{0.15cm}m{0.2cm}m{0.3cm}m{0.1cm}m{0.3cm}m{0.25cm}m{0.25cm}m{5.8cm}} 
			\hline
			\hline
			\multicolumn{1}{c}{\begin{sideways}\textbf{Reference}\end{sideways}} & 	\multicolumn{1}{p{0.1cm}}{\begin{sideways}\textbf{Propagation Path}\end{sideways}} & 	
			\multicolumn{1}{p{0.1cm}}{\begin{sideways}\textbf{Packet Loss}\end{sideways}} & \multicolumn{1}{p{0.1cm}}{\begin{sideways}\textbf{Packet Latency}\end{sideways}} & \multicolumn{1}{p{0.1cm}}{\begin{sideways}\textbf{Measurement Outliers}\end{sideways}} & \multicolumn{1}{p{0.1cm}}{\begin{sideways}\textbf{Bandwidth}\end{sideways}} & \multicolumn{1}{c}{\begin{sideways}\textbf{Estimator}\end{sideways}} & \multicolumn{1}{p{0.1cm}}{\begin{sideways}\textbf{Scalable?}\end{sideways}} & \multicolumn{1}{p{0.1cm}}{\begin{sideways}\textbf{Ocean Currents}\end{sideways}} & \multicolumn{1}{c}{\begin{sideways}\textbf{Results (Sim/Exp)}\end{sideways}} & \multicolumn{1}{p{0.1cm}}{\begin{sideways}\textbf{LAUV${}^1$ knows mission?~~}\end{sideways}} & 
			\multicolumn{1}{p{0.2cm}}{\begin{sideways}\textbf{Ranging Method}\end{sideways}} & \multicolumn{1}{p{0.3cm}}{\begin{sideways}\textbf{Velocity Sensing}\end{sideways}} & \multicolumn{1}{c}{\begin{sideways}\textbf{Uncertainty}\end{sideways}} & \multicolumn{1}{c}{\begin{sideways}\textbf{Channel Sharing}\end{sideways}} & \multicolumn{1}{c}{\begin{sideways}\textbf{No. of aid vehicles}\end{sideways}} & 
			\multicolumn{1}{c}{\begin{sideways}\textbf{No. of aided vehicles}\end{sideways}} & \multicolumn{1}{c}{\begin{sideways}\textbf{Remarks}\end{sideways}}                                                                                             \\ 
			\hline
			\hline
			\cite{vaganay2004experimental} & \tikzred & \tikzred & \tikzgreen & \tikzgreen & \tikzred & - & Y & \tikzred & EPP & Y & T & D & - & B & 2 & 1 & Boats used as AUV to gather data. Approach calls for CNA-AUV to resurface for GPS fix periodically                                          \\ 
			\hline
			\cite{5357852} &\tikzred & \tikzred & \tikzgreen & \tikzred & \tikzred & DSEKF & Y & \tikzred & S & Y & O & D & - & - & 1 & 2 & Delays in acoustic communication are taken care of using centralised delayed state EKF.\\
			\hline
			\cite{song2013cooperative}, \cite{Song_2013b}, \cite{6859344} &\tikzred & \tikzred & \tikzred & \tikzred & \tikzred & SLAM/ EKF/ PF & Y & \tikzgreen & S & N & T & - & \tikzred & - & 3 & 3 & Leader vehicles are dynamic landmarks for SLAM in background ocean flows. In their absence follower AUVs resort to intra-vehicle state exchange.\\
			\hline
			\cite{walls2014origin} &\tikzred & \tikzgreen & \tikzred & \tikzred & \tikzgreen & DEIF & Y & \tikzred & E & A & O & D & - & T & 1 & 2 & Pose-graph information based. Easily scalable, bandwidth efficient and robust against packet loss. \\
			\hline
			\cite{7139030} &\tikzred & \tikzgreen & \tikzred & \tikzred & \tikzgreen & Factor Graphs & Y & \tikzred & E & A & O & D & - & T & 1 & 2 & Factor graphs based approach. Low communication overhead.\\
			\hline
			\cite{ben2021novel} &\tikzred & \tikzgreen & \tikzred & \tikzred & \tikzgreen & Factor Graphs & N & \tikzred & E & A & O & D & - & B & 2 & 1 & Factor graphs based approach using range and bearing information. SBL type appraoch for bearing.\\
			\hline
			\cite{zhang2016optimal} & \tikzred & \tikzred & \tikzred & \tikzred & \tikzred & EKF & Y & \tikzred & S  & N & O & - & \tikzred & T & 3  & 1  & More number of leaders than followers, Trilateration approach.   \\ 
			\hline
			\cite{8003160} &\tikzred & \tikzred & \tikzred & \tikzred & \tikzred & MLE & N & \tikzred & S & N & - & - & \tikzred & - & 1 & 1 & Uses convexified MLE solved using parallel projection algorithm. Computationally expensive.\\
			\hline
			\cite{zhang2019cooperative} & \tikzred & \tikzred & \tikzred  & \tikzred & \tikzred & EKF & Y & \tikzred & S & Y & O & D & - & B & 2 & 3 & MDP-CE for leader AUVs optimal path planning. Requires training.  \\
			\hline
			\cite{kim2020cooperative} &\tikzred & \tikzred & \tikzred & \tikzred & \tikzred & Hybrid UKF & N & \tikzgreen & S & Y & - &  T & - & T & 1 & 2 & Orientation is measured by leader using USBL. Assumes unknown currents. Does not consider cross correlations.\\
			\hline
			\cite{yan2018polar} &\tikzred & \tikzred & \tikzgreen & \tikzred & \tikzred & DS-AEKF & N & \tikzred & E & Y & U & D & - & - & 1 & 1 & Polar regions. Delayed state adaptive EKF. Requires bearing measurement.\\
			\hline
			\cite{5603992}, \cite{6608182} &\tikzred & \tikzgreen & \tikzred & \tikzred & \tikzgreen & CEKF/ CSLAM EKF & N & \tikzred & EPP & Y & - & N & - & T & 3 & 1 & Cross correlations ignored. Aiding AUVs need higher endurance.\\
			\hline
			\cite{7752795} &\tikzred & \tikzred & \tikzgreen & \tikzred & \tikzred & KF & Y & \tikzred & S & - & U & D & \tikzred & - & 1 & 1 & Time delay compensation. Both leader and followers have USBL. Leader is assumed to be stationary.\\
			\hline
			\cite{qi2016cooperative} & \tikzred & \tikzred & \tikzgreen  & \tikzred & \tikzred & AEKF & Y & \tikzred & S  & Y & O  & - & - & B & 1 & 1 & Augmented EKF for time delayed measurements. Higher computational load at follower AUV. \\ 
			\hline
			\cite{zhang2017cooperative} &\tikzred & \tikzred & \tikzred & \tikzred & \tikzred & PHD/IE & N & \tikzred & S & N & T & - & \tikzred & - & 2 & 1 & Leaders localize follower vehicle using bearing and range information. Follower AUV strictly only 1.\\
			\hline
			\cite{gao2014improved} & \tikzred & \tikzred & \tikzred & \tikzred & \tikzgreen  & EKF & N & \tikzred & EPP & Y & T  & D  & - & RR & 1  & 4 & Roundtrip ranging. Predefined zigzag path for beacon AUV. \\ 
			\hline	
			\cite{8084717} &\tikzred & \tikzred & \tikzred & \tikzred & \tikzred & EKF & N & \tikzred & S & N & T & - & \tikzred & T & 2 & 1 & Localization in absence of clock synchronization. Needs 2 leaders.\\
			\hline	
			\cite{qu2021optimal} &\tikzred & \tikzred & \tikzred & \tikzred & \tikzred & EKF & Y & \tikzred & S & Y & - & - & \tikzred & T & 5 & 1 & Different formations of multiple leaders. Requires clock synchronization and bearing measuremnts.\\
			\hline	
			\cite{bo2020optimal} &\tikzgreen & \tikzred & \tikzred & \tikzred & \tikzred & EKF & Y & \tikzred & S & Y & R & - & \tikzgreen & T & 5 & 1 & Different formations of multiple leaders. Ranging using RSSI.\\
			\hline		
			\cite{fan2018maximum} &\tikzred & \tikzred & \tikzred & \tikzgreen & \tikzred & MCC-UPF & N & \tikzred & EPP & Y & T & D & - & B & 2 & 1 & Maximum correntropy based outlier mitigation. Computationally intensive. Non Gaussian noise.\\
			\hline		
			\cite{8049492} &\tikzred & \tikzred & \tikzred & \tikzred & \tikzred & AEKF & N & \tikzred & EPP & Y & T & D & - & T & 2 & 1 & Expectation maximization based adaptive EKF.\\
			\hline				
			\cite{zhao2016collaborative} &\tikzred & \tikzred & \tikzred & \tikzgreen & \tikzred & UPF & Y & \tikzred & S & N & O & - & \tikzred & T & 2 & 1 & 2 leader AUVs. Non Gaussian noise assumption.\\
			\hline
			\cite{8327867} &\tikzred & \tikzred & \tikzred & \tikzgreen & \tikzred & Students T-EKF & N & \tikzred & EPP & Y & - & D & - & B & 1 & 1 & Students-T based EKF for outlier mitigation. Leader has a zigzag path. High computational requirements.\\
			\hline				
			\cite{xu2019cooperative,xu2021novel} & \tikzred & \tikzgreen & \tikzred & \tikzgreen & \tikzred & Cubature KF  & Y  & \tikzred & E & Y & O & D & -  & B & 2  & 1 & Adaptive neuro-fuzzy inference based packet loss \& Maximum correntropy based outlier mitigation.  \\ 
			\hline
			\cite{rui2010cooperative} & \tikzred & \tikzred & \tikzred & \tikzred & \tikzred & EKF & Y & \tikzred & E & A & O & D, TE  & - & - & 1 & 1 & Aid AUV knows survey plan a priori and has better sensor suite. Optimal path planning for Aid AUV. \\ 
			\hline
			\cite{bahr2012dynamic} & \tikzred & \tikzred & \tikzred & \tikzred & \tikzgreen & EKF & Y & \tikzred & S & N & - & - & \tikzred  & T & 1/2 & 1/2 & Optimal path planning. Inter vehicle ranging used in estimation. Ignores correlations in EKF.  \\ 
			\hline
			\cite{7353681} &\tikzred & \tikzgreen & \tikzred & \tikzred & \tikzred & EKF & Y & \tikzred & EPP & Y & O & - & - & - & 1 & 1 &  Parameterized paths for the server. Probabilistic channel model. \\
			\hline	
			\cite{5658626} &\tikzred & \tikzred & \tikzred & \tikzred & \tikzred & EKF & - & \tikzred & S & - & - & D & \tikzred & - & 1 & 2 & Simple EKF implementation.\\
			\hline	
			\hline
		\end{tabular}
		\begin{tablenotes}
			\item \tikzgreen \hspace{0.2cm}Considered\hspace{0.5cm} \tikzred \hspace{0.2cm}Not considered. \textbf{Results}: E - Experimental, S - Simulation, EPP - Post processed experimental data. \textbf{Ranging Method}: O - OWTT, T - TWTT, U - USBL, D -TDOA, R-RSSI. \textbf{Velocity Input}: D - DVL, A - ADCP, T - Thrust estimation from motor speed. \textbf{Channel Sharing}: T - TDMA, B - Broadcast, RR - Request/Reply, F - FDMA. \textbf{Mission plan}: Y- Yes, same as follower, A - Aware, not same as Follower, N- Not aware. {${}^1$}Leader AUV.
		\end{tablenotes}
	\end{threeparttable}
\end{table*}

\subsection{Without a dedicated support vehicle}
Although having a dedicated aid vehicle has its merits in better localization and communications, there are several demerits. In addition to their path planning, aiding vehicles may not carry any mission-specific sensors and thus don't contribute other than in localization. Also, if the aiding vehicle fails, the whole mission can get compromised. In this section, we look at the approaches that do away with support vehicles altogether. This can be done either with one of the vehicles surfacing for GPS or through other sensors, such as vision, SONAR, gravity, etc., that aid in localization.\\

\subsubsection{Surfacing approach}
In this approach, one of the AUVs in the team resurfaces to get absolute GPS position information. With this information, the AUV dives back and shares its absolute position with other team members resulting in a reduction in their localization error. In all the following works, one of the vehicles surfaces for a GPS fix. The earliest results in this category are by {Maczka \textit{et al.}} \cite{4449404}, wherein they demonstrate cooperative navigation by sharing inter-vehicle ranges over acoustic communication channels to complement DR estimates. To mitigate the inability to transmit the full estimation error covariance matrix due to insufficient bandwidth, only a scalar function of the main diagonal elements is shared instead. Acoustic latency is taken care of by recalculating past estimates using the range measurement and propagating it to the current time. In \cite{6942559}, a MAP-based scheme is proposed that computes consistent estimates of the full multi-robot trajectory with a communication strategy involving constant packet size, adaptive performance with respect to acoustic channel, and which scales linearly with the number of AUVs. Every AUV maintains two-factor graphs, multi-AUV, and its own DR. Instead of all the raw sensor data, only change in position factor and associated covariances from the DR factor graph, depth, range data, acknowledgment bits, and GPS fix (if available) are communicated. To maintain consistency, bookkeeping-based tracking is employed. Backlogging due to communication channel issues is taken care of by combining multiple data. {Liu \textit{et al.}} \cite{6847925} describes the 'SUAVE' algorithm for localization among a swarm of AUVs. In it, the AUVs with average tracking variance above a certain threshold resurface and remain stationary to act as beacons for other AUVs. The authors propose an iterative multiple model (IMM) based estimator utilizing a fusion of KF and EKF for linear and angular motion, respectively. In \cite{8084652}, a decentralized, opportunistic communication-based CL within a team of AUVs, wherein the members can join or leave at any time, is proposed. There is no dedicated timeslot for vehicles to communicate. The localization is performed through trilateration among the team members through the non-linear least-squares approach using OWTT ranging and data. Time delay in data packets is compensated by taking into account the vehicle's own motion during the time difference between the received timestamp and its clock. In \cite{allison2020resilient}, an approach that uses a measure of each AUV's confidence of location (LC) estimate to fuse relative pose information through KF for reducing localization error is presented. When LC is below a limit for any of the swarm members, they return to the surface for a GPS fix. The effect of a rouge AUV with high LC in propagating wrong information is also considered. Swarm subsections that have low LC are aided by specially deployed AUVs with high LC to improve their localization, but this was not validated by the authors in the simulation. 

Although this approach does not need any support vehicles, as mentioned before, resurfacing for absolute position information wastes time and energy. Also, once the surfacing AUV dives back down, depending upon the depth it has to dive, its position estimate would have drifted by a significant amount. Hence the total useful contribution in error reduction from resurfacing will not be as good as with a surface vehicle. 

An overview of all the papers using this approach is given in Table \ref{table_7}.\\

\subsubsection{Non Surfacing}
In this approach, there are no aiding vehicles nor resurfacing for GPS. The team either relies only on their inter-vehicle range and exchanged data or SLAM to keep the localization error in check. However, it may be possible that some of the team members have high accuracy sensors than others, but unlike the server-client approach, they have their separate mission. In SLAM, the AUVs rely on other geophysical information such as gravity, magnetic field, and bathymetry map (using vision, side-scan SONAR, multibeam SONAR, etc.) to bound their localization error. The advantage of this approach is that it does not require a support vehicle or surfacing. However, the localization error growth will be the worst among all the strategies due to the absence of absolute position information.\\

\paragraph{Alternating Landmark}

{Matsuda \textit{et al.}} \cite{matsuda2012performance, 6964386} proposed a novel cooperative strategy in which a group of AUVs alternatively performs the role of static landmarks (beacons) and survey vehicles. A particle filter (PF) is used for estimating the horizontal position and yaw of the vehicles from onboard sensors, including DVL, together with relative range and angle measurements. The vehicles are assumed to be hover-capable and equipped with accurate but expensive fiber optic-based gyro sensors. While this approach is able to keep the error growth in check, the performance is not guaranteed in the presence of ocean currents, as the group designated to act as a landmark will drift in such a case. In \cite{6405084}, a communication scheme to reduce the communication overhead in the previous work is proposed wherein the particles are clustered using K-means clustering. Only the averages and standard deviations are then shared across other vehicles leading to lower data transmission. In \cite{Matsuda_2018}, the requirement of hover-capable AUVs is mitigated by having the landmark AUVs remain stationary by landing on the seafloor. The vehicles are divided into two separate groups of landmark and survey vehicles, with only the landmark vehicles alternatively remaining stationary. In \cite{rashidi2011simultaneous}, two methodologies are considered. In 1st, one of the AUVs acts as a static landmark while the other two moves; later, the other two are static and estimate the position of 1st AUV. In the second method, a server-client approach is used, in which one of the AUVs acts as a moving beacon, and the other two localize with respect to the server AUV. However, the EKF estimation is carried out in a centralized manner on one of the robots, including estimating the landmark robot state, which is then shared with it. This, however, leads to overconfident estimates. 

An overview of all the works in this approach is given in Table \ref{table_7}.\\

\begin{table*}
	\centering
	\begin{threeparttable}
		\caption{Overview of CL using surfacing and Alternating landmark based methods}
		\label{table_7}
		\renewcommand{\arraystretch}{1.3}
		\begin{tabular}{m{0.1cm}m{0.5cm}m{0.1cm}m{0.1cm}m{0.1cm}m{0.1cm}m{0.1cm}m{1.1cm}m{0.1cm}m{0.1cm}m{0.4cm}m{0.1cm}m{0.2cm}m{0.3cm}m{0.1cm}m{0.6cm}m{0.2cm}m{0.3cm}m{5cm}} 
			\hline
						\hline
			\multicolumn{1}{p{0.1cm}}{\begin{sideways}\textbf{Category}\end{sideways}}& \multicolumn{1}{c}{\begin{sideways}\textbf{Reference}\end{sideways}} & \multicolumn{1}{p{0.1cm}}{\begin{sideways}\textbf{Propagation Path~ }\end{sideways}} & 
			\multicolumn{1}{p{0.1cm}}{\begin{sideways}\textbf{Packet Loss}\end{sideways}} & \multicolumn{1}{p{0.1cm}}{\begin{sideways}\textbf{Packet Latency}\end{sideways}} & \multicolumn{1}{p{0.1cm}}{\begin{sideways}\textbf{Measurement Outliers~~~}\end{sideways}} & 
			\multicolumn{1}{p{0.1cm}}{\begin{sideways}\textbf{Bandwidth}\end{sideways}} & \multicolumn{1}{c}{\begin{sideways}\textbf{Estimator}\end{sideways}} & \multicolumn{1}{p{0.1cm}}{\begin{sideways}\textbf{Scalable?}\end{sideways}} & \multicolumn{1}{p{0.1cm}}{\begin{sideways}\textbf{Ocean Currents}\end{sideways}} & \multicolumn{1}{c}{\begin{sideways}\textbf{Results (Sim/Exp)}\end{sideways}} & \multicolumn{1}{p{0.1cm}}{\begin{sideways}\textbf{Ranging Method}\end{sideways}} & \multicolumn{1}{c}{\begin{sideways}\textbf{Velocity Sensing}\end{sideways}} & \multicolumn{1}{c}{\begin{sideways}\textbf{Obs. Analysis\textbackslash{} Metric~}\end{sideways}} & \multicolumn{1}{p{0.1cm}}{\begin{sideways}\textbf{Model}\end{sideways}} & \multicolumn{1}{p{0.1cm}}{\begin{sideways}\textbf{Uncertainty}\end{sideways}} & \multicolumn{1}{c}{\begin{sideways}\textbf{Channel Sharing}\end{sideways}} & \multicolumn{1}{p{0.1cm}}{\begin{sideways}\textbf{Team Size}\end{sideways}} & \multicolumn{1}{c}{\begin{sideways}\textbf{Remarks}\end{sideways}}                                                                                                                         \\ 
			\hline
						\hline
			\multirow[c]{3}{*}[-2cm]{\begin{sideways}{Surfacing}\end{sideways}}                      & \cite{4449404} & \tikzred & \tikzred & \tikzgreen & \tikzred & \tikzgreen & Distributed Egocentric EKF & Y & \tikzred & E & O & R & - & D & GPS & TB & 2 &\vspace{1mm} Overconfident estimates as correlations not shared, One of the vehicles resurfaced for GPS.   \\ 
			\cline{2-19}
			& \cite{6942559} & \tikzred & \tikzgreen & \tikzgreen & \tikzred & \tikzgreen & MAP & Y & \tikzred & EPP & O & D  & - & D & GPS & T & 2 & \vspace{1mm}Decentralized approach. One or some AUV's surface intermittently for GPS Fix. Consistent Estimates. \\ 
			\cline{2-19}
			&\cite{6847925} &\tikzred & \tikzred & \tikzred & \tikzred & \tikzred & IMM(KF) & Y & \tikzgreen & S & O & TE & - & D & GPS & S & 10 & Assumes AUV can receive 3 messages simultaneously which is difficult in practice.\\
			\cline{2-19}
			& \cite{8084652} & \tikzred & \tikzred & \tikzgreen & \tikzred & \tikzred & NLS & Y & \tikzgreen & S & O & - & - & -  & - & - & 5 & \vspace{1mm}Trilateration based centralized approach. AUV may surface or the team can take support from ASV. Opportunistic Communication. \\ 
			\cline{2-19}
			& \cite{allison2020resilient} &\tikzred & \tikzred & \tikzred & \tikzred & \tikzred & KF & Y & \tikzred & S & O & R & - & - & GPS & T & 150 & Large swarms. Effect of wrong information being shared also considered.\\
			
			\hline
			\hline
			\multirow[c]{1}{*}{\begin{sideways}{Alt. LM}\end{sideways}}  
			& \cite{matsuda2012performance}, \cite{6964386}, \cite{Matsuda_2018} & \tikzred & \tikzred & \tikzred & \tikzgreen & \tikzgreen & PF & Y & \tikzred & E & U & D & - & D & LM & B & 3 & \vspace{1mm}Alternating Landmark-Moving AUV architecture, Requires relative range and angle measurement. Vehicles requires hovering capability or land on ocean floor. \\ 
			\cline{2-19}
			& \cite{rashidi2011simultaneous} &\tikzred & \tikzred & \tikzred & \tikzgreen & \tikzgreen & EKF & N & \tikzred & S & - & N & - & D & LM & B & 3 & Heterogeneous team, Leader-follower and Static LM approaches. Centralised Estimation.\\
						\hline
			\hline
		\end{tabular}
		\begin{tablenotes}
			\item \tikzgreen \hspace{0.2cm}Considered\hspace{0.5cm} \tikzred \hspace{0.2cm}Not considered. \textbf{Results}: E - Experimental, S - Simulation, EPP - Post processed experimental data. \textbf{Ranging Method}: O - OWTT, T - TWTT, U - USBL, D -TDOA. \textbf{Velocity Sensing}: D - DVL, A - ADCP, TE - Thrust estimation from motor speed, R - Required but not specified. \textbf{Obs. Analysis \textbackslash{} Metric}: NL - Non linear weak observability.  \textbf{Model}: C - Continuous, D - Discrete. \textbf{Channel Sharing}: T - TDMA, B - Broadcast, RR - Request/Reply, TB - TDMA Broadcast, S - Simultaneous. 
		\end{tablenotes}
	\end{threeparttable}
\end{table*}

\paragraph{Parallel}

In this approach, the team members share range and position information with their immediate neighbors or all the team members through broadcast. In the prior case, we have directed graph-based topology, while in the latter, we have mesh topology. When absolute position information is not available with any of the vehicles, only the error growth rate can be reduced with this method. Eventually, at least one vehicle will have to acquire absolute positioning information from GPS, ASV, LBL, etc., to bound the error.

In \cite{5152859}, the authors present an interleaved update (IU) algorithm that ensures consistency of estimates free from overconfidence that could be induced due to the reception of multiple instances of the same information from different vehicles. For this, the authors suggest a bookkeeping approach to properly keep track of measurements to be incorporated, including the cross-correlations of position estimates between vehicles. Every vehicle has multiple estimation filters that track the source of each range measurement. Only those estimates are used which are known to be uncorrelated. The team is assumed not to have a structure, and any member can join or leave at any time. The approach can also take care of lost packets. The disadvantage is that the method cannot be scaled beyond three to four vehicles due to the large covariance information that needs to be transmitted. Also, the estimates are quite conservative. In \cite{liu_convex_2010}, a linear programming and convex optimization-based solution for multiple AUVs equipped with low-cost sensors is presented. Each AUV is assumed to have a sensor that can measure range and bearing with respect to other AUVs. The complexity of this algorithm increases rapidly for a team size of more than ten. In \cite{chen2013minimizing}, a probabilistic method to minimize the localization uncertainty for AUVs working under ice sheets is proposed. The acoustic packets exchanged between AUVs are used to estimate the ranges between them, and using Doppler shifts, the velocity of own vehicle is estimated. Using them and the uncertainties in other vehicles' locations transmitted in packets, a probabilistic method minimizes the location uncertainty. An algorithm to optimize the trade-off between communication overhead and localization error is also presented. In \cite{7017127}, an approach using inter-vehicle ranges and range differences that do not need time synchronization is proposed. Each vehicle interrogates other vehicles sequentially and calculates the relative range from the reply using TWTT. Other vehicles not in the current communicating pair eavesdrop on the broadcasts and use the range information to calculate range differences, which are then used along with their ranging information to construct Euclidian Distance Matrices (EDM) for localization. To mitigate noisy and incomplete data, which will lead to ill-defined EDM, three optimization-based techniques are proposed and evaluated against a least-squares-based approach and non-optimized EDM. Results indicated EDM with plain ranges and Lower-Bounded Epigraph performs the best. In \cite{de2015multi},  geometric constraints that may exist within the team are exploited through the projection approach. The paper uses nonlinear to LTV transformation from \cite{parlangeli2012relative}, which lends to the use of simpler linear KF for state estimation. The inclusion of the geometric constraints gives much lower positional errors and covariance than without. It is required that there exists a cyclic connectivity within the connection graph, which may not always be true. In \cite{viegas2015distributed}, decentralized state estimation in formations of vehicles with time-varying topologies is presented. Each AUV relies on a local observer using local measurements and limited communications with neighboring vehicles for its state estimation. Some of the vehicles are assumed to have access to absolute position information from LBL/USBL system. Sufficient conditions for global exponential stability of error dynamics are derived using switching systems theory. The approach is extended to acyclic formations with fixed topologies In \cite{viegas2016decentralized}. The performance of KF is shown to be similar to EKF but with observability and stability guarantees. This is important as EKF is known to diverge rapidly if the choice of initial conditions is poor. The same authors addressed the problem of distributed state estimation in a multi-vehicle fixed formation framework using discrete KF formulation in \cite{viegas2018discrete}. Two algorithms, one-step and finite horizon, are proposed to find the steady-state discrete-time KF gains with sparsity constraints. The first one, which calculates current gain based only on current covariance, is computationally less intensive and simple; its error performance is worse. The latter has better error performance but has a higher computational load. In \cite{7778673}, a DEIF-based decentralized cooperative localization strategy that is well suited for low bandwidth acoustic communication and is robust against packet loss is proposed. Its performance is compared against distributed Naive KF and Single KF for different packet loss scenarios. It is shown that the proposed method provides better estimates compared to the other two and is robust against packet loss. In \cite{7583426}, an approach using relative concurrent range and bearing measurements from vision-based sensors is presented. The localization problem is solved using a variant of convex disk relaxation. A set of position/bearing reference nodes or anchors are assumed to be deployed at fixed locations and accessible to vehicles over time. The approach is most effective when the vehicle trajectories are predominantly linear and suffers from position /attitude ambiguity when anchors are not accessible. In \cite{8232328,8206528}, localization among a team of AUVs in the mid-ocean zone in the presence of ocean background flows is investigated. The large background flows are preloaded from ocean general circulation models (OGCMs) and are used in localization, while the local flows are measured using ACDP/DVL sensor. The vehicles communicate state information when in range of each other. Cross-correlation is taken care of with the covariance intersection method. A marginalized PF/Rao-Blackwellized PF is used for state estimation utilizing an EKF for position and velocity estimation to reduce the number of particles otherwise needed considering the large state vector size. In \cite{8604531}, an approach that uses information entropy-based criteria to evaluate and select information from neighboring vehicles to update its state is proposed. The performance was evaluated based on mutual information, relative distance, and estimated covariance in two cases, leader-follower and parallel architecture. The simulations indicate that selecting the AUVs closest for updates gave the best performance. A fuzzy logic-based localization scheme for large swarms is presented in \cite{sabra2020fuzzy}. The localization is carried out using a trilateration approach using PSO at each AUV. Some of the AUVs are updated from a boat using USBL. These AUVs then communicate and localize others in the swarms.

\begin{table*}
	\centering
	\begin{threeparttable}
		\caption{Overview of CL using parallel method and CSLAM}
		\label{table_8}
		\renewcommand{\arraystretch}{1.3}
		\begin{tabular}{m{0.5cm}m{0.1cm}m{0.1cm}m{0.1cm}m{0.1cm}m{0.1cm}m{1.4cm}m{0.2cm}m{0.1cm}m{0.4cm}m{0.2cm}m{0.4cm}m{0.6cm}m{0.1cm}m{0.3cm}m{0.2cm}m{5.4cm}} 
			\hline
						\hline
			\multicolumn{1}{c}{\begin{sideways}\textbf{Reference}\end{sideways}} & \multicolumn{1}{p{0.1cm}}{\begin{sideways}\textbf{Propagation Path~ }\end{sideways}} & 
			\multicolumn{1}{p{0.1cm}}{\begin{sideways}\textbf{Packet Loss}\end{sideways}} & \multicolumn{1}{p{0.1cm}}{\begin{sideways}\textbf{Packet Latency}\end{sideways}} & \multicolumn{1}{p{0.1cm}}{\begin{sideways}\textbf{Measurement Outliers~~~}\end{sideways}} & 
			\multicolumn{1}{p{0.1cm}}{\begin{sideways}\textbf{Bandwidth}\end{sideways}} & \multicolumn{1}{c}{\begin{sideways}\textbf{Estimator}\end{sideways}} & \multicolumn{1}{p{0.2cm}}{\begin{sideways}\textbf{Scalable?}\end{sideways}} & \multicolumn{1}{p{0.1cm}}{\begin{sideways}\textbf{Ocean Currents}\end{sideways}} & \multicolumn{1}{p{0.4cm}}{\begin{sideways}\textbf{Results (Sim/Exp)}\end{sideways}} & \multicolumn{1}{p{0.2cm}}{\begin{sideways}\textbf{Ranging Method}\end{sideways}} & \multicolumn{1}{c}{\begin{sideways}\textbf{Velocity Sensing}\end{sideways}} & \multicolumn{1}{c}{\begin{sideways}\textbf{Obs. Analysis\textbackslash{} Metric~}\end{sideways}} & \multicolumn{1}{p{0.1cm}}{\begin{sideways}\textbf{Model}\end{sideways}} & 
			\multicolumn{1}{p{0.3cm}}{\begin{sideways}\textbf{Channel Sharing}\end{sideways}} & \multicolumn{1}{p{0.2cm}}{\begin{sideways}\textbf{Team Size}\end{sideways}} & \multicolumn{1}{c}{\begin{sideways}\textbf{Remarks}\end{sideways}}                                                                                                    \\ 
			\hline
			\hline
			\cite{5152859} & \tikzred & \tikzgreen & \tikzred & \tikzred & \tikzred & IU/ EKF & Y & \tikzred & S & O & - & - & D  & TB & 3 & Bookkeeping for tracking measurement origins ensuring estimate consistency.                  \\ 
			\hline
			\cite{liu_convex_2010} &\tikzred & \tikzred & \tikzred & \tikzred & \tikzred & LP/ Convex Opt. & Y & \tikzred & S & - & - & - & -  & - & 13 & Complex algorithm, requires bearing measurements along with range.\\
			\hline
			\cite{chen2013minimizing} &\tikzred & \tikzred & \tikzred & \tikzred & \tikzgreen & UKF & Y & \tikzgreen & S & O & P/A & - & -  & T & 12 & Incorporates Doppler shifts in communication for reducing uncertainty in under ice autonomous navigation.\\
			\hline
			\cite{7017127} & \tikzred & \tikzgreen & \tikzgreen & \tikzred & \tikzred & Euclidean Distance Matrices & N & \tikzred & E & T & - & - & -  & RR & 3 & Ranges and range differences are used to localize. Requires large number of acoustic transmissions. AUV’s eavesdrop on other transmissions. Experiments with towed nodes.  \\ 
			\hline
			\cite{de2015multi} & \tikzred & \tikzred & \tikzred & \tikzred & \tikzred & KF & N & \tikzred & S & - & R & NLTV & C  & - & 4 & RPMG based approach. Requires cyclic connections in graph for geometric constraints (for observability).  \\ 
			\hline
			\cite{viegas2015distributed}, \cite{viegas2016decentralized} &\tikzred & \tikzred & \tikzred & \tikzred & \tikzred & Luenberger (Decentralised) & Y & \tikzred & S & - & - & LTI & C & - & 6 & Decentralized Observer. Time varying topologies \cite{viegas2015distributed}. Acyclic fixed topologies\cite{viegas2016decentralized}\\
			\hline
			\cite{7778673} & \tikzred & \tikzgreen & \tikzred & \tikzred & \tikzgreen & DEIF & Y & \tikzred & EPP & O & - & - & D  & RB & 3 & Low bandwidth acoustic communication that is robust against packet loss.  \\ 
			\hline
			\cite{7583426} & \tikzred & \tikzred & \tikzred & \tikzred & \tikzred & LS & N & \tikzred & S & - & - & - & D  & - & 10* & Hybrid approach. *Requires static beacons. Visual bearing sensors.   \\ 
			\hline
			\cite{8232328} &\tikzred & \tikzred & \tikzred & \tikzred & \tikzred & RBPF & Y & \tikzgreen & S & O & A & - & D & - & 4 & Uses Rao-Backwellized PF. Requires ocean general circulation models. Mid water zone.\\
			\hline
			\cite{8604531} &\tikzred & \tikzred & \tikzred & \tikzred & \tikzred & EKF & N & \tikzred & S & - & D & - & D  & B & 9 & Requires bearing information. Information entropy based neighbor selection.\\
			\hline
			\cite{sabra2020fuzzy} &\tikzgreen & \tikzgreen & \tikzgreen & \tikzred & \tikzgreen & EKF/PSO-Fuzzy & Y & \tikzred & S & O & D & - & D  & T & 50+ & Fuzzy logic based. PSO trilateration at each node.\\
			\hline
			\cite{5278138} & \tikzred & \tikzred & \tikzred & \tikzred & \tikzred & EKF & Y & \tikzred & S & - & N & DJ & D  & - & 2 & AUV’s are assumed to not suddenly turn or accelerate, Only one AUV localizes with respect to other. Uses linearised model.  \\
			\hline 
			\cite{5509573} & \tikzred & \tikzred & \tikzred & \tikzred & \tikzred & EKF & - & \tikzred & EPP & O & R & NL & C  & - & 2 & \vspace{1mm}Observability analysis. 3D Model. Experiment using two surface crafts instead of AUV \\ 
			\hline
			\cite{6094466}, \cite{arrichiello2013observability} & \tikzred & \tikzred & \tikzred & \tikzred & \tikzred & Ext. Luenberger \cite{6094466}/ EKF\cite{arrichiello2013observability} & - & \tikzred & EPP & T & D & NL & C  & - & 2 & \vspace{1mm}Experiment using one AUV and one static beacon. Uses inverse of condition no. for optimality.     \\ 
			\hline
			\cite{parlangeli2012relative}, \cite{parlangeli2015single}  & \tikzred & \tikzred & \tikzred & \tikzred & \tikzred  & - & - & \tikzred  & S & -  & R & NLTV & C  & - & 2 & \vspace{1mm}3D Model. Observability of relative positioning between 2 AUV’s. Cases where at least one AUV has zero linear speed. Point mass model \\ 
			\hline
			\hline
			\cite{WALTER2004880}  & \tikzred & \tikzred & \tikzred & \tikzred & \tikzred & DSEKF & Y & \tikzred & E & - & - & - & D  & - & 3 & \vspace{1mm}Heterogeneous server-client. Doesn't consider bandwidth limits. Experiments on land robots.\\
			\hline
			\cite{7139227}  & \tikzred & \tikzred & \tikzgreen & \tikzred & \tikzgreen & FG & Y & \tikzred & S & O & D & - & D  & T & 2 & \vspace{1mm}Uses side scan SONAR imagery. Low communication overhead. \\
			\hline
			\cite{allotta2016cooperative}  & \tikzred & \tikzred & \tikzred & \tikzred & \tikzred & EKF & N & \tikzred & S & T & D & - & D  & T & 7* & \vspace{1mm}*Uses 5 beacons as landmarks. Centralized processing.           \\
			\hline
			\cite{tan2016cooperative}  & \tikzred & \tikzred & \tikzred & \tikzred & \tikzgreen & PF & Y & \tikzgreen & EPP & O & TE & - & D  & RB & 3 & \vspace{1mm}Bathymetry based, uses Decentralized Marginalized PF, Requires a priori map.           \\
			\hline
			\cite{wiktor2020collaborative}  & \tikzred & \tikzred & \tikzred & \tikzred & \tikzgreen & PF & Y & \tikzred & EPP & O & D & - & D  & RB & 3 & \vspace{1mm}Terrain Relative Navigation based PF with Covariance intersection.           \\
			\hline
			\hline
		\end{tabular}
		\begin{tablenotes}
			\item \tikzgreen \hspace{0.2cm}Considered\hspace{0.5cm} \tikzred \hspace{0.2cm}Not considered. \textbf{Results}: E - Experimental, S - Simulation, EPP - Post processed experimental data. \textbf{Ranging Method}: O - OWTT, T - TWTT, U - USBL, D -TDOA. \textbf{Velocity Sensing}: D - DVL, A - ADCP, TE - Thrust estimation from motor speed, R - Required but not specified, P-Doppler shift in communication packets. \textbf{Obs. Analysis \textbackslash{} Metric}: NL - Non linear weak observability, NLTV - Non linear to linear time varying transformation, DJ - Determinant of Jacobian.  \textbf{Model}: C - Continuous, D - Discrete. \textbf{Channel Sharing}: T - TDMA, B - Broadcast, RR - Request/Reply, TB - TDMA Broadcast, RB - Round Robin. 
		\end{tablenotes}
	\end{threeparttable}
\end{table*}

Quite a few works have also investigated the observability conditions in the parallel approach. In \cite{5278138}, the observability analysis of localization using DR and range measurements between two AUVs is presented under the assumptions of no velocity measurements and zero latency of acoustic signals in a 2D scenario. The local weak observability condition is derived from the determinant of the Jacobian of two consecutive range measurements, which requires that the two vehicles should not move parallel to each other at the same velocity. In other words, the team members should exhibit sufficiently exciting relative manoeuvrers for the network to be locally weakly observable. This is extended to 3D space in \cite{5509573}. The nonlinear model-based weak observability condition is shown to be less stringent than the conditions using a linear model. The linear observability condition requires non-null angular velocity; in contrast, the weak observability condition does not. The authors use EKF-based observer for state estimation using their own and other vehicles' linear and angular velocity, orientation, and depth data. Instead of the binary rank condition for observability presented in \cite{5509573}, \cite{6094466, arrichiello2013observability} propose inverse of the condition number of observability matrix as a better measure of observability. The analysis is carried out in 3D space, and it is shown that the state's vertical component does not affect observability if it can be directly measured. Further, the system is observable as long as the relative position and velocity vectors are not parallel. In \cite{parlangeli2012relative}, observability analysis using the state augmentation technique to transform a nonlinear system into a higher dimensional LTV system is presented for a case of two AUVs. Linear system observability analysis is used to find the indistinguishable states of the nonlinear system. This is done for a limited case of constant angular and linear velocities and a point-mass model to keep the analysis mathematically tractable. It is shown that the approach leads to global observability conditions rather than the weak local notion of observability and that there is a one-to-one correspondence between the trajectories of the augmented LTV system and the original nonlinear system. In \cite{parlangeli2015single}, the above analysis is extended to the case wherein both the AUVs move with constant nonzero linear velocities and the observing vehicle with constant nonzero angular velocity. Also, the relative angular velocity, as seen from observing the vehicle, is assumed constant. The observability analysis of the augmented state vector belonging to $R^{25}$ is carried out using Popov - Belevitch - Hautus (PBH) Lemma. However, it is assumed that the vehicle can turn on the spot, i.e., the angular and linear velocities are independent. This restricts the analysis of a limited set of underwater vehicles. 

An overview of parallel approaches is given in Table \ref{table_8}.\\

\paragraph{Cooperative SLAM}

In simultaneous localization and mapping (SLAM), an extensive review of which can be found in \cite{7747236}, the autonomous vehicle is assumed to be equipped with some sensor with which the vehicle can map its surroundings. The most popular are LIDAR, ultrasonic distance sensors, monocular or stereo vision systems, RADARs, and SONARs. In an underwater environment, SONAR-based sensors such as side scan, multibeam, and echo sounders are much more prevalent than LIDARs and vision. This is due to the reasons mentioned in section II. However, in clear waters and narrow structures such as flooded mines/caves, etc., LIDARs and vision sensors are finding increasing acceptance. With a single vehicle, though, mapping and subsequent localization can take a large amount of time. This has led to an interest in the field of cooperative SLAM. Here, each of the vehicles is outfitted with some mapping sensor, information from which is shared with others for localization. 

{Walter \textit{et al.}} \cite{WALTER2004880} presented an approach using a heterogeneous server-client AUV configuration. The sensor data of environmental observations made by all the vehicles are fused using a SLAM algorithm in a centralized manner on one of the leader vehicles. The output is communicated back for the navigation of other parent and child vehicles. Data association is carried out using a joint compatibility branch and bound (JCBB) test. In \cite{7139227}, a CSLAM algorithm utilizing side-scan sonar images and INS is proposed. The proposed algorithm is robust against packet loss and generates acoustic packets that are small enough to be transmitted in the underwater acoustic channel. The method employs factor graph-based SLAM with data reduction using intermediate (between communications) state marginalization by Schur's complement and further consistent sparsification by convex optimization using KLD as the cost between original and sparsified data. In \cite{allotta2016cooperative}, an acoustic-SLAM approach is proposed in which a client AUV and static beacons are localized using a server AUV having USBL. The static beacons' location, which later acts as landmarks, is assumed to be initially unknown. The estimation is carried out centrally on the server AUV using EKF, which then communicates the estimates to child and beacon nodes. The approach does not scale with the number of client vehicles due to the TDMA scheme employed and two-way communication. Furthermore, the necessity of static beacons requires deployment and retrieval. {Tan \textit{et al.}} \cite{tan2016cooperative} proposed a bathymetry based multi AUV cooperative localization scheme wherein the AUVs utilize only a low-cost sensor set of an altimeter, depth sensor, and an acoustic modem. The collected sensor data is fused in a decentralized Marginalized PF (DMPF), i.e., the marginalized linear dynamical states are estimated using KF while the others with PF reduce the computational and communication load. The other vehicles' beliefs, along with the estimated inter-vehicle range, are used to influence the particle distribution and likelihood computation and are not fused directly in the filter to prevent the effects of large errors in the beliefs/position on estimates. The bathymetric map data is used to update the measurement model of the PF. This approach requires the bathymetry map of the area a priori, and the proposed DMPF algorithm does not take cross-correlation into account. In \cite{wiktor2020collaborative}, terrain relative navigation with inter-robot measurements is proposed for multi-robot localization. A particle filter is used along with a covariance intersection filter. The complexity only grows linearly with the number of vehicles. However, it is assumed that the terrain map is available and clocks are synchronized.

The major impediment to this approach is the large amount of data that needs to be exchanged among the vehicles. Given the bandwidth of the acoustic channel, this is a very difficult challenge and remains an active area of research. 

\subsection{Other works}
Several works have also investigated mixed strategies or combinations of the approaches discussed in the previous sections. In \cite{5509869}, the authors proposed a modular measurement distribution framework that is scalable and allows any cooperating team member to share measurements using TDMA, enabling the whole team to estimate positions consistently and accurately. The framework is amenable to any cooperative approach, such as ASV/CNA-based, server-client, parallel/mesh, or surfacing type, and is independent of any state estimation scheme. It does not need an entire covariance matrix to be transmitted to ensure no overconfidence in position estimates and ensures scalability, although the algorithm is sensitive to packet loss. In \cite{6107061}, the experimental comparison between 3 estimation algorithms, i.e., simple distributed EKF, interleaved update, and distributed extended information filter (DEIF) against a centralized EKF (Post-processed), are reported. In the case of DEIF, the assumptions made render it useful only in two-node unidirectional topologies. Different cooperating strategies considered between one ASV and two AUVs are AUV aiding AUV, ASV aiding two AUV, 3 Node mesh, and ASV aiding AUV, which in turn is a server for other AUVs. Results indicate that the distributed EKF produces overconfident estimates when a strong correlation persists, while IU and DEIF give consistent results. Otherwise, DEKF performs nearly as well as CEKF. While DEKF and DEIF bounds error, IU does not. DEIF requires the largest packet size among the three and is the least robust against packet loss.

There have been other approaches as well that combine cooperating vehicles with one or more static beacons. For example, Mirza \textit{et al.} \cite{mirza2015real} presented a factor graph-based approach using maximum likelihood for CL between multiple underwater vehicles and beacons in a distributed setup. Real-time and non-real-time centralized setups were evaluated against real-time distributed setup for cases wherein a) all states were shared with all neighbors (RTD-A), b) all states were shared with an immediate neighbor only (RTD-B), and c) current states shared with an immediate neighbor only (RTD-C), the latter being the one with the least communication overhead and thus preferred in an underwater scenario. The RTD-C scheme was evaluated for different no of beacons and vehicles, indicating more collaborating vehicles and beacons, the lower the error. When all the vehicles are in intermittent contact with the beacons, it was reported that collaboration did not provide any improvement in error, while minimum error for all vehicles is achieved when the beacons are uniformly distributed. Maximum gain is for those vehicles which are not in contact with the beacon. However, the proposed approach tends to deteriorate collaborative localization performance when any one or more vehicles are consistently not in contact with the beacon in a team of more than two vehicles. {Rego \textit{et al.}} \cite{rego2021cooperative} evaluated the performance of estimating the position and velocity of vehicles under stringent communication bandwidth constraints. It is assumed that there is a fixed beacon at a known location. The vehicles either exchange measurements or state estimates. Adaptive quantization is used to limit the amount of data sent over a communication link under a zero packet loss assumption. In \cite{7271663}, an algorithm for optimal placement of multiple heterogeneous beacon vehicles (including static nodes such as GIB) not capable of rapid motions relative to AUVs is proposed. Optimum locations are found by minimizing the trace of appropriately defined CRLB matrix using range-only information. The locations are always estimated on the perimeter of a circle, and at each step, the dynamic beacons move to the optimal location. In \cite{8546733}, a hierarchical beacon-server-client cooperative localization scheme using range information from time delays is presented. The server AUV localizes with respect to the beacon, while the client AUV localizes with respect to the server AUV and the single beacon. The proposed architecture reduces the number of acoustic transmissions required relative to each AUV localizing with respect to the beacon. However, the server AUV errors are not discussed; the motions are considered to be stop-and-go and require the server AUV speed and heading to be different from the team members. In the coordinate fusion technique, all AUVs have to communicate, which negates the original proposals' advantage of lower communication overhead.

This concludes the review of all the works in the domain of cooperative localization in the underwater scenario. In the next section, we briefly discuss the issues and open problems in this area.

\section{DISCUSSION AND OPEN PROBLEMS}

Before we discuss the open challenges in the cooperative localization of underwater vehicles, we briefly summarize the research presented in the preceding section. 

As noted, the estimation algorithm forms the heart of the localization problem. While EKF \cite{matos_auv_2005,fallon2010cooperative} is widely popular, as can be gauged from the tables, several different estimation techniques have been proposed across all categories, such as least squares \cite{hartsfield_single_2005}, 
least mean squares	\cite{5290050},	UKF \cite{8274781, glotzbach2016acoustic}, Decentralised LS-MLE \cite{doi:10.1002/rob.20365}, Decentralised EIF \cite{6504537}, distributed extended information filter  \cite{6107061}, Delayed State Centralized EKF \cite{doi:10.1177/0278364912446166, 5357852},  centralised EKF \cite{glotzbach2012underwater},	Interleaved Update (IU) \cite{bahr2009cooperative, bahr2009cooperative}, iterative divided difference filter \cite{gao2014robust}, Factor graphs: \cite{wu2019cooperative,ben2021novel,mirza2015real}, Maximum-A-Priori (MAP) \cite{wu2019cooperative, franchi2021maximum}, moving horizon estimation  \cite{7004622, chen_moving_2016}, 	particle filters \cite{7003048, fallon2010cooperative2}, distibuted EKF \cite{claus2018closed}, distributed modified EKF \cite{Song_2013b}, 	origin state method \cite{walls2014origin}, Parallel projection algorithm+MLE \cite{8003160}, hybrid UKF-KF estimator \cite{kim2020cooperative}, delayed state Adaptive KF \cite{yan2018polar}, augmented EKF \cite{qi2016cooperative}, 	probability hypothesis density filter \cite{zhang2017cooperative}, unscented PF \cite{zhao2016collaborative}, Student-t based EKF \cite{8327867}, Linear programming \cite{liu_convex_2010}, fuzzy logic \cite{sabra2020fuzzy} and	SLAM \cite{7139227, allotta2016cooperative, tan2016cooperative}. Authors have also exploited geometric properties of the localization problem, such as in \cite{10.1016/j.robot.2014.03.004,zhang2016optimal,de2015multi}. Often the initial location of the vehicle may not be known; for this, a few methods are suggested in \cite{SCHERBATYUK20121}, which are useful for initializing EKF-based estimators. 

As for the challenges presented by the acoustic channel, several authors have presented solutions for low bandwidth, such as by using estimators \cite{6504537,7778673}, logic-based communication \cite{meira_cooperative_2011}, adaptive time-of-launch  \cite{8867537}, reducing communication overhead \cite{chen2013minimizing}  and adaptive quantization \cite{rego2021cooperative}. For outlier mitigation most approaches use Mahalanobis distance metric \cite{costanzi2018estimation} while others have proposed different noise distributions such as heavy-tailed mixture distribution \cite{bai2020novel}, Student-t based EKF \cite{8327867} or different estimators like Huber based M estimator \cite{gao2014robust}, maximum correntropy-PF \cite{fan2018maximum}, unscented PF \cite{zhao2016collaborative} and MCC-ANFIS \cite{xu2019cooperative}. Managing delays in communication is carried out either through back and forth technique \cite{glotzbach2012underwater,costanzi2018estimation} or delayed state estimators \cite{doi:10.1177/0278364912446166,7752795,qi2016cooperative}. Some techniques to estimate sound speed profile and refraction are mentioned in \cite{5290050}. 

For path planning of ASVs and server vehicles the approaches include simple paths such as diamond \cite{doi:10.1177/0278364912446166}, zig-zag \cite{fallon2010cooperative} or circular \cite{fallon2010cooperative2} or online path planning through heuristics \cite{german_2012}, uncertainty ellipse \cite{SCHERBATYUK20121,7003048, dubrovin_studying_2016}, Dynamic programming \cite{5547044, rui2010cooperative}, Markov Decision process \cite{6107044,zhang2019cooperative}, Genetic algorithm \cite{teck2014direct}, condition number of observability gramian and empirical observability gramian \cite{6859032}, FIM: \cite{7003099, Sousa_2018}, extremum seeking: \cite{mandic2015range, mandic2016mobile}, priority-based expansion of a search tree\cite{8755392}, artificial potential field \cite{nad2016cooperative}, Q-learning \cite{zhang2020cooperative} and partially observable Markov decision process \cite{7353681}. Authors have also evaluated optimal formations of aiding vehicles in \cite{qu2021optimal, bo2020optimal}, optimal number of ASV \cite{chen2016optimal}, and optimal positioning  \cite{bahr2012dynamic,7271663}. Related observability analysis is covered in \cite{8274781,5650250,7004622,6224634,viegas2014position,5278138,5509573,6094466, arrichiello2013observability,parlangeli2012relative,parlangeli2015single}.

Several works have also tackled problems that deal with other aspects such as centralized \cite{5603992}, hierarchical \cite{Glotz2015} or decentralized \cite{viegas2015distributed, 7778673} estimation, the inclusion of known \cite{8084581, 8232328,8206528} and unknown \cite{kim2020cooperative} ocean current models, combining with other localization sources such as static beacons \cite{costanzi2018estimation}, divers \cite{glotzbach2012underwater} and companion vehicles \cite{crasta2017range, crasta2018multiple} and ranging in the absence of time synchronization \cite{8084717,7017127}.

The underwater cooperative localization problem was first investigated with the help of crewed ships and boats as communication and navigation aids. However, due to the associated low costs and advancements in ASV, acoustic communication capabilities, and better, more cost-effective sensors, the research has since been focused on utilizing ASV as a CNA or relying entirely on the underwater team itself for localization and navigation. Both categories, with and without navigational aid, have seen an almost equal amount of interest and have generated substantial research output over the past decade, as summarized above. However, there are still areas that have not been explored much, especially considering the attention given to the problem of target tracking.

\paragraph{Acoustic Channel}

As evident from the tables, very few results incorporate the challenges afforded by the underwater acoustic channel. In most cases, the acoustic channel is assumed lossless, instantaneous, and Gaussian distributed. The presence of outliers in acoustic communication essentially renders the noise PDF heavy-tailed, and the Gaussian assumption is no longer valid. Many works report the usage of EKF state estimation, which is simple and computationally less taxing but unsuitable in the presence of outliers. Furthermore, EKF is prone to instability in the event of wrong initialization. If computational power requirements are not a concern, the particle filter outperforms EKF, especially in non-Gaussian noise. While little can be done about the acoustic channel's bandwidth from the control perspective, estimation algorithms need to be made robust against the acoustic packet latency and losses. While there are a few approaches that mitigate these issues, there is still scope for more improvements in this area. Estimation of sound speed profile, latency effects on the stability of the estimator, bandwidth and packet size optimization for SLAM-based approaches, TDMA optimization, and event-triggered communications are some of the areas for future research. Scalability to large teams is heavily dependent on solutions to problems in this area.

\paragraph{Optimal paths and formations}

There have been approaches that optimize the trajectory of a single ASV to ensure the observability of the cooperative localization problem. However, in large teams and multiple ASV, optimal path planning and formations have not been explored to the degree seen from investigations into optimal formations of static underwater sensor networks. Optimal trajectories for a single ASV in terms of energy and control input have not been explored. For multiple ASVs aiding a large team of AUVs, problems involving their optimal number, formation, and collision avoidance strategies can be explored. Applications of neural network-based learning methods have not been applied to the path planning problem so far. Furthermore, observability analysis for multiple vehicle teams is still in the nascent stage and has quite a lot of scope.

\paragraph{Practical results}

The basic premise of cooperative localization is the ability of a team of vehicles to exchange data and improve their location estimate. This is even more useful in a large team of vehicles. While many papers have proposed algorithms that are, in general, applicable to large robot teams, only a handful show results with more than 3 or 4 team members. This number is even less in the case of experimental results, which could be explained by the high costs of underwater hardware. However, this is expected to improve in the future with the cost reductions in hardware and more cost-friendly vendors in underwater robotics coming into the picture. One specific area is SLAM utilizing SONAR data for cooperative localization as it requires efficient but high computational capability.

\paragraph{Ocean effects}

Another area that needs attention is the ocean effects, such as ocean currents and tides. Modeling ocean currents has received very little attention in both approaches, more so in the category without a dedicated navigation aid. Considering that most ocean surveys span large areas, this is an important consideration that cannot be overlooked. While DVL sensors can help counter the drifts induced by currents, they are not always useful, such as in the mid-water column where a ground lock is not available. In such instances, ACDP can help to some extent. Tidal effects are essential considerations for long-duration missions, especially for those approaches that use a depth sensor to project 3D localization problems onto a 2D plane. The tidal changes will introduce bias in the depth measurements that would affect the computed location's accuracy. Another effect that has been ignored is the effect of sea states on the performance of ASV as a navigational aid. All the works assume that ASV communicates while in constant motion, which is not valid, as the sea state will induce time-varying changes in the range measurements and increase the difficulty of ASV trajectory planning.

\section{Conclusion}
This paper has presented an exhaustive review of the literature in the underwater cooperative localization domain. A brief overview of the challenges for localization in the underwater acoustic channel was followed by a glimpse of popular state estimation algorithms used in the current state of the art approaches. The CL approaches were classified and evaluated on different parameters, and their salient features were highlighted. Finally, we presented a brief discussion on the open problems in the context of underwater cooperative localization.


%





\ifCLASSOPTIONcaptionsoff
  \newpage
\fi



%
%
%

\bibliographystyle{IEEEtran}
\bibliography{IEEEabrv,IEEEexample}

\begin{thebibliography}{100}
\providecommand{\url}[1]{#1}
\csname url@samestyle\endcsname
\providecommand{\newblock}{\relax}
\providecommand{\bibinfo}[2]{#2}
\providecommand{\BIBentrySTDinterwordspacing}{\spaceskip=0pt\relax}
\providecommand{\BIBentryALTinterwordstretchfactor}{4}
\providecommand{\BIBentryALTinterwordspacing}{\spaceskip=\fontdimen2\font plus
\BIBentryALTinterwordstretchfactor\fontdimen3\font minus
  \fontdimen4\font\relax}
\providecommand{\BIBforeignlanguage}[2]{{%
\expandafter\ifx\csname l@#1\endcsname\relax
\typeout{** WARNING: IEEEtran.bst: No hyphenation pattern has been}%
\typeout{** loaded for the language `#1'. Using the pattern for}%
\typeout{** the default language instead.}%
\else
\language=\csname l@#1\endcsname
\fi
#2}}
\providecommand{\BIBdecl}{\relax}
\BIBdecl

\bibitem{6380737}
M.~E. {Furlong}, D.~{Paxton}, P.~{Stevenson}, M.~{Pebody}, S.~D. {McPhail}, and
  J.~{Perrett}, ``Autosub long range: A long range deep diving auv for ocean
  monitoring,'' in \emph{2012 IEEE/OES Autonomous Underwater Vehicles (AUV)},
  2012, pp. 1--7.

\bibitem{775301}
X.~{Yun}, E.~R. {Bachmann}, R.~B. {McGhee}, R.~H. {Whalen}, R.~L. {Roberts},
  R.~G. {Knapp}, A.~J. {Healey}, and M.~J. {Zyda}, ``Testing and evaluation of
  an integrated gps/ins system for small auv navigation,'' \emph{IEEE Journal
  of Oceanic Engineering}, vol.~24, no.~3, pp. 396--404, 1999.

\bibitem{tan_survey_2011}
\BIBentryALTinterwordspacing
H.-P. Tan, R.~Diamant, W.~K.~G. Seah, and M.~Waldmeyer, ``A survey of
  techniques and challenges in underwater localization,'' \emph{Ocean
  Engineering}, vol.~38, no.~14, pp. 1663 -- 1676, 2011. [Online]. Available:
  \url{http://www.sciencedirect.com/science/article/pii/S0029801811001624}
\BIBentrySTDinterwordspacing

\bibitem{4099086}
R.~M. {Eustice}, L.~L. {Whitcomb}, H.~{Singh}, and M.~{Grund}, ``Recent
  advances in synchronous-clock one-way-travel-time acoustic navigation,'' in
  \emph{OCEANS 2006}, 2006, pp. 1--6.

\bibitem{emami_taban_2018}
M.~Emami and M.~R. Taban, ``A low complexity integrated navigation system for
  underwater vehicles,'' \emph{Journal of Navigation}, vol.~71, no.~5, p.
  1161–1177, 2018.

\bibitem{whitcomb1999combined}
L.~L. Whitcomb, D.~R. Yoerger, H.~Singh, and J.~Howland, ``Combined doppler/lbl
  based navigation of underwater vehicles,'' in \emph{Proceedings of the 11th
  international symposium on unmanned untethered submersible technology}, 1999,
  pp. 1--7.

\bibitem{kinsey_survey_2006}
J.~C. Kinsey, R.~M. Eustice, and L.~L. Whitcomb, ``\BIBforeignlanguage{en}{A
  survey of underwater vehicle navigation: {Recent} advances and new
  challenges},'' \emph{\BIBforeignlanguage{en}{Proc. Conf. Manoeuvering Control
  Marine Craft}}, p.~12, 2006.

\bibitem{kebkal_auv_2017}
\BIBentryALTinterwordspacing
K.~G. Kebkal and A.~I. Mashoshin, ``\BIBforeignlanguage{en}{{AUV} acoustic
  positioning methods},'' \emph{\BIBforeignlanguage{en}{Gyroscopy and
  Navigation}}, vol.~8, no.~1, pp. 80--89, Jan. 2017. [Online]. Available:
  \url{http://link.springer.com/10.1134/S2075108717010059}
\BIBentrySTDinterwordspacing

\bibitem{6678293}
L.~{Paull}, S.~{Saeedi}, M.~{Seto}, and H.~{Li}, ``Auv navigation and
  localization: A review,'' \emph{IEEE Journal of Oceanic Engineering},
  vol.~39, no.~1, pp. 131--149, 2014.

\bibitem{gonzalez-garcia_autonomous_2020}
\BIBentryALTinterwordspacing
J.~González-García, A.~Gómez-Espinosa, E.~Cuan-Urquizo, L.~G.
  García-Valdovinos, T.~Salgado-Jiménez, and J.~A.~E. Cabello,
  ``\BIBforeignlanguage{en}{Autonomous {Underwater} {Vehicles}: {Localization},
  {Navigation}, and {Communication} for {Collaborative} {Missions}},''
  \emph{\BIBforeignlanguage{en}{Applied Sciences}}, vol.~10, no.~4, p. 1256,
  Feb. 2020. [Online]. Available:
  \url{https://www.mdpi.com/2076-3417/10/4/1256}
\BIBentrySTDinterwordspacing

\bibitem{8744517}
I.~{Masmitja}, S.~{Gomariz}, J.~{Del-Rio}, B.~{Kieft}, T.~{O’Reilly},
  P.~{Bouvet}, and J.~{Aguzzi}, ``Range-only single-beacon tracking of
  underwater targets from an autonomous vehicle: From theory to practice,''
  \emph{IEEE Access}, vol.~7, pp. 86\,946--86\,963, 2019.

\bibitem{107149}
J.~A. {Catipovic}, ``Performance limitations in underwater acoustic
  telemetry,'' \emph{IEEE Journal of Oceanic Engineering}, vol.~15, no.~3, pp.
  205--216, 1990.

\bibitem{1637927}
{Jun-Hong Cui}, {Jiejun Kong}, M.~{Gerla}, and {Shengli Zhou}, ``The challenges
  of building mobile underwater wireless networks for aquatic applications,''
  \emph{IEEE Network}, vol.~20, no.~3, pp. 12--18, 2006.

\bibitem{bo_optimal_2018}
\BIBentryALTinterwordspacing
X.~Bo, A.~Razzaqi, and X.~Wang, ``\BIBforeignlanguage{en}{Optimal {Sensor}
  {Formation} for {3D} {Cooperative} {Localization} of {AUVs} {Using} {Time}
  {Difference} of {Arrival} ({TDOA}) {Method}},''
  \emph{\BIBforeignlanguage{en}{Sensors}}, vol.~18, no.~12, p. 4442, Dec. 2018.
  [Online]. Available: \url{http://www.mdpi.com/1424-8220/18/12/4442}
\BIBentrySTDinterwordspacing

\bibitem{yan_moving_2015}
\BIBentryALTinterwordspacing
W.~Yan, W.~Chen, and R.~Cui, ``\BIBforeignlanguage{en}{Moving long baseline
  positioning algorithm with uncertain sound speed},''
  \emph{\BIBforeignlanguage{en}{Journal of Mechanical Science and Technology}},
  vol.~29, no.~9, pp. 3995--4002, Sep. 2015. [Online]. Available:
  \url{http://link.springer.com/10.1007/s12206-015-0845-z}
\BIBentrySTDinterwordspacing

\bibitem{book_Lurton}
X.~Lurton, \emph{An Introduction to Underwater Acoustics: Principles and
  Applications}.\hskip 1em plus 0.5em minus 0.4em\relax Berlin Heidelberg:
  Springer-Verlag, 2010.

\bibitem{6387620}
H.~{Ramezani}, H.~{Jamali-Rad}, and G.~{Leus}, ``Target localization and
  tracking for an isogradient sound speed profile,'' \emph{IEEE Transactions on
  Signal Processing}, vol.~61, no.~6, pp. 1434--1446, 2013.

\bibitem{doi:10.1177/0278364912446166}
\BIBentryALTinterwordspacing
S.~E. Webster, R.~M. Eustice, H.~Singh, and L.~L. Whitcomb, ``Advances in
  single-beacon one-way-travel-time acoustic navigation for underwater
  vehicles,'' \emph{The International Journal of Robotics Research}, vol.~31,
  no.~8, pp. 935--950, 2012. [Online]. Available:
  \url{https://doi.org/10.1177/0278364912446166}
\BIBentrySTDinterwordspacing

\bibitem{7828800}
{Yuantao Qi}, {Bo Wang}, {Shunting Wang}, and {Mengyin Fu}, ``Cooperative
  navigation for multiple autonomous underwater vehicles with time delayed
  measurements,'' in \emph{2016 IEEE Chinese Guidance, Navigation and Control
  Conference (CGNCC)}, 2016, pp. 295--299.

\bibitem{5779652}
M.~J. {Stanway}, ``Delayed-state sigma point kalman filters for underwater
  navigation,'' in \emph{2010 IEEE/OES Autonomous Underwater Vehicles}, 2010,
  pp. 1--9.

\bibitem{506191}
J.~{Vaganay}, J.~J. {Leonard}, and J.~G. {Bellingham}, ``Outlier rejection for
  autonomous acoustic navigation,'' in \emph{Proceedings of IEEE International
  Conference on Robotics and Automation}, vol.~3, 1996, pp. 2174--2181 vol.3.

\bibitem{EvoLogics_1}
\BIBentryALTinterwordspacing
H.~D.~. EvoLogics. (15- Jun- 2020) Evologics gmbh, 2020. [online]. [Online].
  Available: \url{https://evologics.de/acoustic-modem/hs}
\BIBentrySTDinterwordspacing

\bibitem{4089076}
E.~{Olson}, J.~J. {Leonard}, and S.~{Teller}, ``Robust range-only beacon
  localization,'' \emph{IEEE Journal of Oceanic Engineering}, vol.~31, no.~4,
  pp. 949--958, 2006.

\bibitem{6942559}
L.~{Paull}, M.~{Seto}, and J.~J. {Leonard}, ``Decentralized cooperative
  trajectory estimation for autonomous underwater vehicles,'' in \emph{2014
  IEEE/RSJ International Conference on Intelligent Robots and Systems}, 2014,
  pp. 184--191.

\bibitem{9356608}
Y.~Yang, Y.~Xiao, and T.~Li, ``A survey of autonomous underwater vehicle
  formation: Performance, formation control, and communication capability,''
  \emph{IEEE Communications Surveys Tutorials}, vol.~23, no.~2, pp. 815--841,
  2021.

\bibitem{5c7d9431766940dca820263b1dfa293c}
F.~Hidalgo and T.~Braunl, ``\BIBforeignlanguage{English}{Review of underwater
  slam techniques},'' in \emph{\BIBforeignlanguage{English}{Proceedings of the
  6th International Conference on Automation, Robotics and Applications}},
  vol.~I.\hskip 1em plus 0.5em minus 0.4em\relax United States: IEEE, Institute
  of Electrical and Electronics Engineers, 2015, pp. 306--311.

\bibitem{10.1145/3366194.3366262}
\BIBentryALTinterwordspacing
W.~Zhao, T.~He, A.~Y.~M. Sani, and T.~Yao, ``Review of slam techniques for
  autonomous underwater vehicles,'' in \emph{Proceedings of the 2019
  International Conference on Robotics, Intelligent Control and Artificial
  Intelligence}, ser. RICAI 2019.\hskip 1em plus 0.5em minus 0.4em\relax New
  York, NY, USA: Association for Computing Machinery, 2019, p. 384–389.
  [Online]. Available: \url{https://doi.org/10.1145/3366194.3366262}
\BIBentrySTDinterwordspacing

\bibitem{7139227}
L.~{Paull}, G.~{Huang}, M.~{Seto}, and J.~J. {Leonard},
  ``Communication-constrained multi-auv cooperative slam,'' in \emph{2015 IEEE
  International Conference on Robotics and Automation (ICRA)}, 2015, pp.
  509--516.

\bibitem{WALTER2004880}
\BIBentryALTinterwordspacing
M.~Walter and J.~Leonard, ``An experimental investigation of cooperative
  slam,'' \emph{IFAC Proceedings Volumes}, vol.~37, no.~8, pp. 880 -- 885,
  2004, iFAC/EURON Symposium on Intelligent Autonomous Vehicles, Lisbon,
  Portugal, 5-7 July 2004. [Online]. Available:
  \url{http://www.sciencedirect.com/science/article/pii/S1474667017320918}
\BIBentrySTDinterwordspacing

\bibitem{rashidi2011simultaneous}
A.~J. Rashidi and S.~Mohammadloo, ``Simultaneous cooperative localization for
  auvs using range-only sensors,'' \emph{International Journal of Information
  Acquisition}, vol.~8, no.~02, pp. 117--132, 2011.

\bibitem{7583426}
B.~Q. {Ferreira}, J.~{Gomes}, C.~{Soares}, and J.~P. {Costeira},
  ``Collaborative localization of vehicle formations based on ranges and
  bearings,'' in \emph{2016 IEEE Third Underwater Communications and Networking
  Conference (UComms)}, 2016, pp. 1--5.

\bibitem{bo_review_2019}
\BIBentryALTinterwordspacing
X.~Bo, A.~A. Razzaqi, and G.~Farid, ``\BIBforeignlanguage{en}{A {Review} on
  {Optimal} {Placement} of {Sensors} for {Cooperative} {Localization} of
  {AUVs}},'' \emph{\BIBforeignlanguage{en}{Journal of Sensors}}, vol. 2019, pp.
  1--12, Jul. 2019. [Online]. Available:
  \url{https://www.hindawi.com/journals/js/2019/4276987/}
\BIBentrySTDinterwordspacing

\bibitem{su_review_2020}
\BIBentryALTinterwordspacing
X.~Su, I.~Ullah, X.~Liu, and D.~Choi, ``\BIBforeignlanguage{en}{A {Review} of
  {Underwater} {Localization} {Techniques}, {Algorithms}, and {Challenges}},''
  \emph{\BIBforeignlanguage{en}{Journal of Sensors}}, vol. 2020, pp. 1--24,
  Jan. 2020. [Online]. Available:
  \url{https://www.hindawi.com/journals/js/2020/6403161/}
\BIBentrySTDinterwordspacing

\bibitem{de2015multi}
D.~De~Palma, G.~Indiveri, and G.~Parlangeli, ``Multi-vehicle relative
  localization based on single range measurements,'' \emph{IFAC-PapersOnLine},
  vol.~48, no.~5, pp. 17--22, 2015.

\bibitem{8227642}
J.~{Aparicio} and T.~{Shimura}, ``Ofdma communication system for cooperative
  localization of underwater vehicles,'' in \emph{2017 IEEE 18th International
  Workshop on Signal Processing Advances in Wireless Communications (SPAWC)},
  2017, pp. 1--5.

\bibitem{8232328}
Z.~{Song} and K.~{Mohseni}, ``Cooperative mid-depth navigation aided by ocean
  current prediction,'' in \emph{OCEANS 2017 - Anchorage}, 2017, pp. 1--8.

\bibitem{1067998}
S.~I. {Roumeliotis} and G.~A. {Bekey}, ``Distributed multirobot localization,''
  \emph{IEEE Transactions on Robotics and Automation}, vol.~18, no.~5, pp.
  781--795, 2002.

\bibitem{1668252}
A.~I. {Mourikis} and S.~I. {Roumeliotis}, ``Performance analysis of multirobot
  cooperative localization,'' \emph{IEEE Transactions on Robotics}, vol.~22,
  no.~4, pp. 666--681, 2006.

\bibitem{TAN20111663}
\BIBentryALTinterwordspacing
H.-P. Tan, R.~Diamant, W.~K. Seah, and M.~Waldmeyer, ``A survey of techniques
  and challenges in underwater localization,'' \emph{Ocean Engineering},
  vol.~38, no.~14, pp. 1663 -- 1676, 2011. [Online]. Available:
  \url{http://www.sciencedirect.com/science/article/pii/S0029801811001624}
\BIBentrySTDinterwordspacing

\bibitem{matos_auv_2005}
\BIBentryALTinterwordspacing
A.~Matos and N.~Cruz, ``\BIBforeignlanguage{en}{{AUV} navigation and guidance
  in a moving acoustic network},'' in \emph{\BIBforeignlanguage{en}{Europe
  {Oceans} 2005}}.\hskip 1em plus 0.5em minus 0.4em\relax Brest, France: IEEE,
  2005, pp. 680--685 Vol. 1. [Online]. Available:
  \url{http://ieeexplore.ieee.org/document/1511796/}
\BIBentrySTDinterwordspacing

\bibitem{hartsfield_single_2005}
\BIBentryALTinterwordspacing
J.~C. Hartsfield, ``\BIBforeignlanguage{en}{Single {Transponder} {Range} {Only}
  {Navigation} {Geometry} ({STRONG}) applied to {REMUS} autonomous under water
  vehicles},'' Ph.D. dissertation, Massachusetts Institute of Technology and
  Woods Hole Oceanographic Institution, Woods Hole, MA, 2005. [Online].
  Available: \url{https://hdl.handle.net/1912/1637}
\BIBentrySTDinterwordspacing

\bibitem{5290050}
S.~D. {McPhail} and M.~{Pebody}, ``Range-only positioning of a deep-diving
  autonomous underwater vehicle from a surface ship,'' \emph{IEEE Journal of
  Oceanic Engineering}, vol.~34, no.~4, pp. 669--677, 2009.

\bibitem{8274781}
Z.~{Qiang} and Z.~{Wen}, ``Range-only navigation algorithm for positioning of
  deep-diving auv,'' in \emph{2017 IEEE International Conference on Cybernetics
  and Intelligent Systems (CIS) and IEEE Conference on Robotics, Automation and
  Mechatronics (RAM)}, 2017, pp. 243--248.

\bibitem{5664462}
A.~{Folk}, B.~{Armstrong}, E.~{Wolbrecht}, H.~F. {Grip}, M.~{Anderson}, and
  D.~{Edwards}, ``Autonomous underwater vehicle navigation using moving
  baseline on a target ship,'' in \emph{OCEANS 2010 MTS/IEEE SEATTLE}, 2010,
  pp. 1--7.

\bibitem{doi:10.1002/rob.20365}
\BIBentryALTinterwordspacing
R.~M. Eustice, H.~Singh, and L.~L. Whitcomb, ``Synchronous-clock,
  one-way-travel-time acoustic navigation for underwater vehicles,''
  \emph{Journal of Field Robotics}, vol.~28, no.~1, pp. 121--136, 2011.
  [Online]. Available:
  \url{https://onlinelibrary.wiley.com/doi/abs/10.1002/rob.20365}
\BIBentrySTDinterwordspacing

\bibitem{6504537}
S.~E. {Webster}, J.~M. {Walls}, L.~L. {Whitcomb}, and R.~M. {Eustice},
  ``Decentralized extended information filter for single-beacon cooperative
  acoustic navigation: Theory and experiments,'' \emph{IEEE Transactions on
  Robotics}, vol.~29, no.~4, pp. 957--974, 2013.

\bibitem{bahr2009cooperative}
A.~Bahr, J.~J. Leonard, and M.~F. Fallon, ``Cooperative localization for
  autonomous underwater vehicles,'' \emph{The International Journal of Robotics
  Research}, vol.~28, no.~6, pp. 714--728, 2009.

\bibitem{10.1016/j.robot.2014.03.004}
\BIBentryALTinterwordspacing
B.~Allotta, R.~Costanzi, E.~Meli, L.~Pugi, A.~Ridolfi, and G.~Vettori,
  ``Cooperative localization of a team of auvs by a tetrahedral
  configuration,'' \emph{Robot. Auton. Syst.}, vol.~62, no.~8, p. 1228–1237,
  Aug. 2014. [Online]. Available:
  \url{https://doi.org/10.1016/j.robot.2014.03.004}
\BIBentrySTDinterwordspacing

\bibitem{7487420}
Z.~J. {Harris} and L.~L. {Whitcomb}, ``Preliminary study of cooperative
  navigation of underwater vehicles without a dvl utilizing range and
  range-rate observations,'' in \emph{2016 IEEE International Conference on
  Robotics and Automation (ICRA)}, 2016, pp. 2618--2624.

\bibitem{8460970}
------, ``Preliminary evaluation of cooperative navigation of underwater
  vehicles without a dvl utilizing a dynamic process model,'' in \emph{2018
  IEEE International Conference on Robotics and Automation (ICRA)}, 2018, pp.
  4897--4904.

\bibitem{costanzi2018estimation}
R.~Costanzi, D.~Fenucci, A.~Caiti, M.~Micheli, A.~Vermeij, A.~Tesei, and
  A.~Munaf{\`o}, ``Estimation filtering for deep water navigation,''
  \emph{IFAC-PapersOnLine}, vol.~51, no.~29, pp. 299--304, 2018.

\bibitem{german_2012}
C.~R. {German}, M.~V. {Jakuba}, J.~C. {Kinsey}, J.~{Partan}, S.~{Suman},
  A.~{Belani}, and D.~R. {Yoerger}, ``A long term vision for long-range
  ship-free deep ocean operations: Persistent presence through coordination of
  autonomous surface vehicles and autonomous underwater vehicles,'' in
  \emph{2012 IEEE/OES Autonomous Underwater Vehicles (AUV)}, 2012, pp. 1--7.

\bibitem{Kalwa:2016:0025-3324:26}
J.~Kalwa, D.~Tietjen, M.~Carreiro-Silva, J.~Fontes, L.~Brignone, N.~Gracias,
  P.~Ridao, M.~Pfingsthorn, A.~Birk, T.~Glotzbach, S.~Eckstein, M.~Caccia,
  J.~Alves, T.~Furfaro, J.~Ribeiro, and A.~Pascoal, ``The european project
  morph: Distributed uuv systems for multimodal, 3d underwater surveys,''
  \emph{Marine Technology Society Journal}, vol.~50, no.~4, pp. 26--41, 2016.

\bibitem{LIU201671}
\BIBentryALTinterwordspacing
Z.~Liu, Y.~Zhang, X.~Yu, and C.~Yuan, ``Unmanned surface vehicles: An overview
  of developments and challenges,'' \emph{Annual Reviews in Control}, vol.~41,
  pp. 71 -- 93, 2016. [Online]. Available:
  \url{http://www.sciencedirect.com/science/article/pii/S1367578816300219}
\BIBentrySTDinterwordspacing

\bibitem{fallon2010cooperative}
M.~F. Fallon, G.~Papadopoulos, and J.~J. Leonard, ``Cooperative auv navigation
  using a single surface craft,'' in \emph{Field and service robotics}.\hskip
  1em plus 0.5em minus 0.4em\relax Springer, 2010, pp. 331--340.

\bibitem{SCHERBATYUK20121}
\BIBentryALTinterwordspacing
A.~P. Scherbatyuk and F.~S. Dubrovin, ``Some algorithms of auv positioning
  based on one moving beacon,'' \emph{IFAC Proceedings Volumes}, vol.~45,
  no.~5, pp. 1 -- 6, 2012, 3rd IFAC Workshop on Navigation, Guidance and
  Control of Underwater Vehicles. [Online]. Available:
  \url{http://www.sciencedirect.com/science/article/pii/S1474667016305717}
\BIBentrySTDinterwordspacing

\bibitem{8206508}
M.~{Zhou}, R.~{Bachmayer}, and B.~{de Young}, ``Underwater acoustic-based
  navigation towards multi-vehicle operation and adaptive oceanographic
  sampling,'' in \emph{2017 IEEE/RSJ International Conference on Intelligent
  Robots and Systems (IROS)}, 2017, pp. 6091--6097.

\bibitem{gao2014robust}
W.~Gao, Y.~Liu, and B.~Xu, ``Robust huber-based iterated divided difference
  filtering with application to cooperative localization of autonomous
  underwater vehicles,'' \emph{Sensors}, vol.~14, no.~12, pp. 24\,523--24\,542,
  2014.

\bibitem{wu2019cooperative}
D.~Wu, Z.~Yan, and T.~Chen, ``Cooperative current estimation based multi-auvs
  localization for deep ocean applications,'' \emph{Ocean Engineering}, vol.
  188, p. 106148, 2019.

\bibitem{7004622}
S.~{Wang}, L.~{Chen}, D.~{Gu}, and H.~{Hu}, ``Cooperative localization of auvs
  using moving horizon estimation,'' \emph{IEEE/CAA Journal of Automatica
  Sinica}, vol.~1, no.~1, pp. 68--76, 2014.

\bibitem{7003048}
A.~{Gatsenko}, F.~{Dubrovin}, and A.~{Scherbatyuk}, ``Comparing some algorithms
  for auv single beacon mobile navigation,'' in \emph{2014 Oceans - St.
  John's}, 2014, pp. 1--5.

\bibitem{dubrovin_studying_2016}
\BIBentryALTinterwordspacing
F.~S. Dubrovin and A.~F. Scherbatyuk, ``\BIBforeignlanguage{en}{Studying some
  algorithms for {AUV} navigation using a single beacon: {The} results of
  simulation and sea trials},'' \emph{\BIBforeignlanguage{en}{Gyroscopy and
  Navigation}}, vol.~7, no.~2, pp. 189--196, Apr. 2016. [Online]. Available:
  \url{http://link.springer.com/10.1134/S2075108716020024}
\BIBentrySTDinterwordspacing

\bibitem{fallon2010cooperative2}
M.~F. Fallon, G.~Papadopoulos, J.~J. Leonard, and N.~M. Patrikalakis,
  ``Cooperative auv navigation using a single maneuvering surface craft,''
  \emph{The International Journal of Robotics Research}, vol.~29, no.~12, pp.
  1461--1474, 2010.

\bibitem{claus2018closed}
B.~Claus, J.~H. Kepper~IV, S.~Suman, and J.~C. Kinsey, ``Closed-loop
  one-way-travel-time navigation using low-grade odometry for autonomous
  underwater vehicles,'' \emph{Journal of Field Robotics}, vol.~35, no.~4, pp.
  421--434, 2018.

\bibitem{8390710}
J.~H. {Kepper}, B.~C. {Claus}, and J.~C. {Kinsey}, ``A navigation solution
  using a mems imu, model-based dead-reckoning, and one-way-travel-time
  acoustic range measurements for autonomous underwater vehicles,'' \emph{IEEE
  Journal of Oceanic Engineering}, vol.~44, no.~3, pp. 664--682, 2019.

\bibitem{Sergeenko_2013}
N.~{Sergeenko}, A.~{Scherbatyuk}, and F.~{Dubrovin}, ``Some algorithms of
  cooperative auv navigation with mobile surface beacon,'' in \emph{2013 OCEANS
  - San Diego}, 2013, pp. 1--6.

\bibitem{5547044}
M.~{Chitre}, ``Path planning for cooperative underwater range-only navigation
  using a single beacon,'' in \emph{2010 International Conference on Autonomous
  and Intelligent Systems, AIS 2010}, 2010, pp. 1--6.

\bibitem{1302422}
A.~S. {Gadre} and D.~J. {Stilwell}, ``Toward underwater navigation based on
  range measurements from a single location,'' in \emph{IEEE International
  Conference on Robotics and Automation, 2004. Proceedings. ICRA '04. 2004},
  vol.~5, 2004, pp. 4472--4477 Vol.5.

\bibitem{1545230}
------, ``A complete solution to underwater navigation in the presence of
  unknown currents based on range measurements from a single location,'' in
  \emph{2005 IEEE/RSJ International Conference on Intelligent Robots and
  Systems}, 2005, pp. 1420--1425.

\bibitem{batista2011single}
P.~Batista, C.~Silvestre, and P.~Oliveira, ``Single range aided navigation and
  source localization: Observability and filter design,'' \emph{Systems \&
  Control Letters}, vol.~60, no.~8, pp. 665--673, 2011.

\bibitem{6107044}
T.~Y. {Teck} and M.~{Chitre}, ``Single beacon cooperative path planning using
  cross-entropy method,'' in \emph{OCEANS'11 MTS/IEEE KONA}, 2011, pp. 1--6.

\bibitem{6727582}
Y.~T. {Tan}, R.~{Gao}, and M.~{Chitre}, ``Cooperative path planning for
  range-only localization using a single moving beacon,'' \emph{IEEE Journal of
  Oceanic Engineering}, vol.~39, no.~2, pp. 371--385, 2014.

\bibitem{teck2014direct}
T.~Y. Teck and M.~Chitre, ``Direct policy search with variable-length genetic
  algorithm for single beacon cooperative path planning,'' in \emph{Distributed
  Autonomous Robotic Systems}.\hskip 1em plus 0.5em minus 0.4em\relax Springer,
  2014, pp. 321--336.

\bibitem{seto_three-dimensional_2011}
M.~L. Seto, J.~A. Hudson, and Y.~Pan, ``Three-{Dimensional} {Path}-{Planning}
  for a {Communications} and {Navigation} {Aid} {Working} {Cooperatively} with
  {Autonomous} {Underwater} {Vehicles},'' in \emph{Autonomous and {Intelligent}
  {Systems}}, M.~Kamel, F.~Karray, W.~Gueaieb, and A.~Khamis, Eds.\hskip 1em
  plus 0.5em minus 0.4em\relax Berlin, Heidelberg: Springer Berlin Heidelberg,
  2011, pp. 51--62.

\bibitem{6942873}
J.~{Hudson} and M.~L. {Seto}, ``Underway path-planning for an unmanned surface
  vehicle performing cooperative navigation for uuvs at varying depths,'' in
  \emph{2014 IEEE/RSJ International Conference on Intelligent Robots and
  Systems}, 2014, pp. 2298--2305.

\bibitem{6859032}
J.~D. {Quenzer} and K.~A. {Morgansen}, ``Observability based control in
  range-only underwater vehicle localization,'' in \emph{2014 American Control
  Conference}, 2014, pp. 4702--4707.

\bibitem{7003099}
J.~M. {Walls} and R.~M. {Eustice}, ``Toward informative planning for
  cooperative underwater localization,'' in \emph{2014 Oceans - St. John's},
  2014, pp. 1--7.

\bibitem{Sousa_2018}
J.~P. {Sousa}, B.~M. {Ferreira}, and N.~A. {Cruz}, ``Guidance of an autonomous
  surface vehicle for underwater navigation aid,'' in \emph{2018 IEEE/OES
  Autonomous Underwater Vehicle Workshop (AUV)}, 2018, pp. 1--6.

\bibitem{mandic2015range}
F.~Mandi{\'c}, N.~Mi{\v{s}}kovi{\'c}, and Z.~Vuki{\'c}, ``Range--only
  navigation--maximizing system observability by using extremum seeking,''
  \emph{IFAC-PapersOnLine}, vol.~48, no.~16, pp. 101--106, 2015.

\bibitem{mandic2016mobile}
F.~Mandi{\'c}, N.~Mi{\v{s}}kovi{\'c}, N.~Palomeras, M.~Carreras, and
  G.~Vallicrosa, ``Mobile beacon control algorithm that ensures observability
  in single range navigation,'' \emph{IFAC-PapersOnLine}, vol.~49, no.~23, pp.
  48--53, 2016.

\bibitem{8755392}
J.~S. {Willners}, L.~{Toohey}, and Y.~{Petillot}, ``Sampling-based path
  planning for cooperative autonomous maritime vehicles to reduce uncertainty
  in range-only localization,'' \emph{IEEE Robotics and Automation Letters},
  vol.~4, no.~4, pp. 3987--3994, 2019.

\bibitem{rua2019cooperative}
S.~R{\'u}a, N.~Crasta, R.~E. V{\'a}squez, M.~J. Betancur, and A.~M. Pascoal,
  ``Cooperative range-based navigation using a beacon with circular motion
  installed on board the support platform,'' \emph{IFAC-PapersOnLine}, vol.~52,
  no.~21, pp. 390--395, 2019.

\bibitem{rua2020enhanced}
S.~R{\'u}a, N.~Crasta, R.~E. V{\'a}squez, and A.~M. Pascoal, ``Enhanced
  cooperative single-range underwater navigation based on optimal
  trajectories,'' \emph{IFAC-PapersOnLine}, vol.~53, no.~2, pp.
  14\,668--14\,673, 2020.

\bibitem{5650250}
G.~{Papadopoulos}, M.~F. {Fallon}, J.~J. {Leonard}, and N.~M. {Patrikalakis},
  ``Cooperative localization of marine vehicles using nonlinear state
  estimation,'' in \emph{2010 IEEE/RSJ International Conference on Intelligent
  Robots and Systems}, 2010, pp. 4874--4879.

\bibitem{6224634}
D.~{Viegas}, P.~{Batista}, P.~{Oliveira}, and C.~{Silvestre}, ``Position and
  velocity filters for intervention auvs based on single range and depth
  measurements,'' in \emph{2012 IEEE International Conference on Robotics and
  Automation}, 2012, pp. 4878--4883.

\bibitem{viegas2014position}
D.~Viegas, P.~Batista, P.~Oliveira, and C.~Silvestre, ``Position and velocity
  filters for asc/i-auv tandems based on single range measurements,''
  \emph{Journal of Intelligent \& Robotic Systems}, vol.~74, no. 3-4, pp.
  745--768, 2014.

\bibitem{meira_cooperative_2011}
A.~Meira, A.~P. Aguiar, and A.~Pascoal, ``Cooperative {Navigation} of
  {Multiple} {Autonomous} {Underwater} {Vehicles} with {Logic} {Based}
  {Communication},'' Ph.D. dissertation, Instituto Superior Técnico, 2011.,
  Portugal, 2011.

\bibitem{8867537}
J.~S. {Willners}, L.~{Toohey}, and Y.~{Petillot}, ``Improving acoustic
  range-only localisation by selection of transmission time,'' in \emph{OCEANS
  2019 - Marseille}, 2019, pp. 1--6.

\bibitem{Glotz2015}
\BIBentryALTinterwordspacing
T.~Glotzbach, S.~Eckstein, and C.~Ament, ``Cooperative navigation in a team of
  marine robots for a specific ocean mapping mission,'' \emph{at -
  Automatisierungstechnik}, vol.~63, no.~5, pp. 344 -- 354, 28 May. 2015.
  [Online]. Available:
  \url{https://www.degruyter.com/view/journals/auto/63/5/article-p344.xml}
\BIBentrySTDinterwordspacing

\bibitem{7401992}
M.~V. {Jakuba}, J.~C. {Kinsey}, J.~W. {Partan}, and S.~E. {Webster},
  ``Feasibility of low-power one-way travel-time inverted ultra-short baseline
  navigation,'' in \emph{OCEANS 2015 - MTS/IEEE Washington}, 2015, pp. 1--10.

\bibitem{glotzbach2016acoustic}
T.~Glotzbach, A.-M. Grebner, and C.~Ament, ``Acoustic based navigation of
  cooperative marine robots with advanced filter techniques,''
  \emph{IFAC-PapersOnLine}, vol.~49, no.~23, pp. 335--340, 2016.

\bibitem{salavasidis2016co}
G.~Salavasidis, C.~A. Harris, E.~Rogers, and A.~B. Phillips, ``Co-operative use
  of marine autonomous systems to enhance navigational accuracy of autonomous
  underwater vehicles,'' in \emph{Annual Conference Towards Autonomous Robotic
  Systems}.\hskip 1em plus 0.5em minus 0.4em\relax Springer, 2016, pp.
  275--281.

\bibitem{phillips2018autonomous}
A.~B. Phillips, G.~Salavasidis, M.~Kingsland, C.~Harris, M.~Pebody, D.~R.~R.
  Templeton, S.~McPhail, T.~Prampart, T.~Wood, R.~Taylor \emph{et~al.},
  ``Autonomous surface/subsurface survey system field trials,'' in \emph{2018
  IEEE/OES Autonomous Underwater Vehicle Workshop (AUV)}.\hskip 1em plus 0.5em
  minus 0.4em\relax IEEE, 2018, pp. 1--6.

\bibitem{franchi2021maximum}
M.~Franchi, A.~Bucci, L.~Zacchini, A.~Ridolfi, M.~Bresciani, G.~Peralta, and
  R.~Costanzi, ``Maximum a posteriori estimation for auv localization with usbl
  measurements,'' \emph{IFAC-PapersOnLine}, vol.~54, no.~16, pp. 307--313,
  2021.

\bibitem{nad2016cooperative}
D.~Nad, M.~Ribeiro, H.~Silva, J.~Ribeiro, P.~Abreu, N.~Miskovic, and
  A.~Pascoal, ``Cooperative surface/underwater navigation for auv path
  following missions,'' \emph{IFAC-PapersOnLine}, vol.~49, no.~23, pp.
  355--360, 2016.

\bibitem{zhang2020cooperative}
L.~Zhang, D.~Wu, R.~Ren, and R.~Xing, ``Cooperative path planning for single
  leader using q-learning method,'' in \emph{Global Oceans 2020: Singapore--US
  Gulf Coast}.\hskip 1em plus 0.5em minus 0.4em\relax IEEE, 2020, pp. 1--6.

\bibitem{curcio2005experiments}
J.~Curcio, J.~Leonard, J.~Vaganay, A.~Patrikalakis, A.~Bahr, D.~Battle,
  H.~Schmidt, and M.~Grund, ``Experiments in moving baseline navigation using
  autonomous surface craft,'' in \emph{Proceedings of OCEANS 2005
  MTS/IEEE}.\hskip 1em plus 0.5em minus 0.4em\relax IEEE, 2005, pp. 730--735.

\bibitem{glotzbach2012underwater}
T.~Glotzbach, M.~Bayat, A.~P. Aguiar, and A.~Pascoal, ``An underwater acoustic
  localisation system for assisted human diving operations,'' \emph{IFAC
  Proceedings Volumes}, vol.~45, no.~27, pp. 206--211, 2012.

\bibitem{chen_moving_2016}
W.~Chen, W.~Yan, and R.~Cui, ``\BIBforeignlanguage{en}{Moving horizon
  estimation for moving long baseline based on linear positioning model},''
  \emph{\BIBforeignlanguage{en}{IFAC-PapersOnLine}}, vol.~49, no.~5, pp.
  115--119, 2016.

\bibitem{8084581}
R.~P. {Vio}, R.~{Cristi}, and K.~B. {Smith}, ``Uuv localization using acoustic
  communications, networking, and a priori knowledge of the ocean current,'' in
  \emph{OCEANS 2017 - Aberdeen}, 2017, pp. 1--7.

\bibitem{crasta2017range}
N.~Crasta, D.~Moreno-Salinas, A.~Pascoal, and J.~Aranda, ``Range-based
  cooperative underwater target localization,'' \emph{IFAC-PapersOnLine},
  vol.~50, no.~1, pp. 12\,366--12\,373, 2017.

\bibitem{crasta2018multiple}
N.~Crasta, D.~Moreno-Salinas, A.~M. Pascoal, and J.~Aranda, ``Multiple
  autonomous surface vehicle motion planning for cooperative range-based
  underwater target localization,'' \emph{Annual Reviews in Control}, vol.~46,
  pp. 326--342, 2018.

\bibitem{quraishi2019flexible}
A.~Quraishi, A.~Bahr, F.~Schill, and A.~Martinoli, ``A flexible navigation
  support system for a team of underwater robots,'' in \emph{2019 International
  Symposium on Multi-Robot and Multi-Agent Systems (MRS)}.\hskip 1em plus 0.5em
  minus 0.4em\relax IEEE, 2019, pp. 70--75.

\bibitem{bai2020novel}
M.~Bai, Y.~Huang, Y.~Zhang, and F.~Chen, ``A novel heavy-tailed mixture
  distribution based robust kalman filter for cooperative localization,''
  \emph{IEEE Transactions on Industrial Informatics}, vol.~17, no.~5, pp.
  3671--3681, 2020.

\bibitem{vaganay2004experimental}
J.~Vaganay, J.~J. Leonard, J.~A. Curcio, and J.~S. Willcox, ``Experimental
  validation of the moving long base-line navigation concept,'' in \emph{2004
  IEEE/OES Autonomous Underwater Vehicles (IEEE Cat. No. 04CH37578)}.\hskip 1em
  plus 0.5em minus 0.4em\relax IEEE, 2004, pp. 59--65.

\bibitem{yan2015moving}
W.~Yan, W.~Chen, and R.~Cui, ``Moving long baseline positioning algorithm with
  uncertain sound speed,'' \emph{Journal of Mechanical Science and Technology},
  vol.~29, no.~9, pp. 3995--4002, 2015.

\bibitem{yan2015optimal}
W.~Yan, W.~Chen, R.~Cui, and H.~Li, ``Optimal distance between mobile buoy and
  target for moving long baseline positioning system,'' \emph{The Journal of
  Navigation}, vol.~68, no.~4, pp. 809--826, 2015.

\bibitem{chen2016optimal}
W.~Chen, W.~Yan, R.~Cui, and H.~Cui, ``Optimal configuration of usvs for moving
  long baseline positioning system,'' in \emph{2016 International Conference on
  Advanced Robotics and Mechatronics (ICARM)}.\hskip 1em plus 0.5em minus
  0.4em\relax IEEE, 2016, pp. 394--398.

\bibitem{singh1996integrated}
H.~Singh, J.~Catipovic, R.~Eastwood, L.~Freitag, H.~Henriksen, F.~Hover,
  D.~Yoerger, J.~G. Bellingham, and B.~A. Moran, ``An integrated approach to
  multiple auv communications, navigation and docking,'' in \emph{OCEANS 96
  MTS/IEEE Conference Proceedings. The Coastal Ocean-Prospects for the 21st
  Century}, vol.~1.\hskip 1em plus 0.5em minus 0.4em\relax IEEE, 1996, pp.
  59--64.

\bibitem{baccou2001cooperative}
P.~Baccou, B.~Jouvencel, V.~Creuze, and C.~Rabaud, ``Cooperative positioning
  and navigation for multiple auv operations,'' in \emph{MTS/IEEE Oceans 2001.
  An Ocean Odyssey. Conference Proceedings (IEEE Cat. No. 01CH37295)},
  vol.~3.\hskip 1em plus 0.5em minus 0.4em\relax IEEE, 2001, pp. 1816--1821.

\bibitem{5357852}
{Yao Yao}, {Demin Xu}, and {Weisheng Yan}, ``Cooperative localization with
  communication delays for mauvs,'' in \emph{2009 IEEE International Conference
  on Intelligent Computing and Intelligent Systems}, vol.~1, 2009, pp.
  244--249.

\bibitem{song2013cooperative}
Z.~Song and K.~Mohseni, ``Cooperative underwater localization in ocean
  currents,'' in \emph{AIAA Guidance, Navigation, and Control (GNC)
  Conference}, 2013, p. 5111.

\bibitem{Song_2013b}
Z.~{Song} and K.~{Mohseni}, ``Hierarchical underwater localization in
  dominating background flow fields,'' in \emph{2013 IEEE/RSJ International
  Conference on Intelligent Robots and Systems}, 2013, pp. 3356--3361.

\bibitem{6859344}
------, ``A distributed localization hierarchy for an auv swarm,'' in
  \emph{2014 American Control Conference}, 2014, pp. 4721--4726.

\bibitem{walls2014origin}
J.~M. Walls and R.~M. Eustice, ``An origin state method for communication
  constrained cooperative localization with robustness to packet loss,''
  \emph{The International Journal of Robotics Research}, vol.~33, no.~9, pp.
  1191--1208, 2014.

\bibitem{7139030}
J.~M. {Walls}, A.~G. {Cunningham}, and R.~M. {Eustice}, ``Cooperative
  localization by factor composition over a faulty low-bandwidth communication
  channel,'' in \emph{2015 IEEE International Conference on Robotics and
  Automation (ICRA)}, 2015, pp. 401--408.

\bibitem{ben2021novel}
Y.~Ben, Y.~Sun, Q.~Li, and X.~Zang, ``A novel cooperative navigation algorithm
  based on factor graph with cycles for auvs,'' \emph{Ocean Engineering}, vol.
  241, p. 110024, 2021.

\bibitem{zhang2016optimal}
L.-c. Zhang, J.~Wang, W.~Tonghao, M.~Liu, and J.~Gao, ``Optimal formation of
  multiple auvs cooperative localization based on virtual structure,'' in
  \emph{OCEANS 2016 MTS/IEEE Monterey}.\hskip 1em plus 0.5em minus 0.4em\relax
  IEEE, 2016, pp. 1--6.

\bibitem{8003160}
Q.~{Chen}, K.~{You}, and S.~{Song}, ``Cooperative localization for autonomous
  underwater vehicles using parallel projection,'' in \emph{2017 13th IEEE
  International Conference on Control Automation (ICCA)}, 2017, pp. 788--793.

\bibitem{zhang2019cooperative}
L.~Zhang, Y.~Li, L.~Liu, and X.~Tao, ``Cooperative navigation based on cross
  entropy: Dual leaders,'' \emph{IEEE Access}, vol.~7, pp. 151\,378--151\,388,
  2019.

\bibitem{kim2020cooperative}
J.~Kim, ``Cooperative localization and unknown currents estimation using
  multiple autonomous underwater vehicles,'' \emph{IEEE Robotics and Automation
  Letters}, vol.~5, no.~2, pp. 2365--2371, 2020.

\bibitem{yan2018polar}
Z.~Yan, L.~Wang, T.~Wang, Z.~Yang, T.~Chen, and J.~Xu, ``Polar cooperative
  navigation algorithm for multi-unmanned underwater vehicles considering
  communication delays,'' \emph{Sensors}, vol.~18, no.~4, p. 1044, 2018.

\bibitem{5603992}
M.~{Borges Nogueira}, J.~B. {Sousa}, and F.~L. {Pereira}, ``Cooperative
  autonomous underwater vehicle localization,'' in \emph{OCEANS'10 IEEE
  SYDNEY}, 2010, pp. 1--9.

\bibitem{6608182}
M.~{Nogueira}, J.~{Souza}, and F.~{Pereira}, ``An underwater cooperative
  navigation scheme,'' in \emph{2013 MTS/IEEE OCEANS - Bergen}, 2013, pp. 1--7.

\bibitem{7752795}
G.~{Xiao}, B.~{Wang}, Z.~{Deng}, M.~{Fu}, and Y.~{Ling}, ``An acoustic
  communication time delays compensation approach for master–slave auv
  cooperative navigation,'' \emph{IEEE Sensors Journal}, vol.~17, no.~2, pp.
  504--513, 2017.

\bibitem{qi2016cooperative}
Y.~Qi, B.~Wang, S.~Wang, and M.~Fu, ``Cooperative navigation for multiple
  autonomous underwater vehicles with time delayed measurements,'' in
  \emph{2016 IEEE Chinese Guidance, Navigation and Control Conference
  (CGNCC)}.\hskip 1em plus 0.5em minus 0.4em\relax IEEE, 2016, pp. 295--299.

\bibitem{zhang2017cooperative}
L.~Zhang, T.~Wang, F.~Zhang, and D.~Xu, ``Cooperative localization for
  multi-auvs based on gm-phd filters and information entropy theory,''
  \emph{Sensors}, vol.~17, no.~10, p. 2286, 2017.

\bibitem{gao2014improved}
W.~Gao, Y.~Liu, B.~Xu, and Y.~Che, ``An improved cooperative localization
  method for multiple autonomous underwater vehicles based on acoustic
  round-trip ranging,'' in \emph{2014 IEEE/ION Position, Location and
  Navigation Symposium-PLANS 2014}.\hskip 1em plus 0.5em minus 0.4em\relax
  IEEE, 2014, pp. 1420--1423.

\bibitem{8084717}
Z.~{Lichuan}, F.~{Jingxiang}, W.~{Tonghao}, G.~{Jian}, and Z.~{Ru}, ``A new
  algorithm for collaborative navigation without time synchronization of
  multi-uuvs,'' in \emph{OCEANS 2017 - Aberdeen}, 2017, pp. 1--6.

\bibitem{qu2021optimal}
J.~Qu, X.~Li, and G.~Sun, ``Optimal formation configuration analysis for
  cooperative localization system of multi-auv,'' \emph{IEEE Access}, vol.~9,
  pp. 90\,702--90\,714, 2021.

\bibitem{bo2020optimal}
X.~Bo, A.~A. Razzaqi, X.~Wang, and G.~Farid, ``Optimal geometric configuration
  of sensors for received signal strength based cooperative localization of
  submerged auvs,'' \emph{Ocean Engineering}, vol. 214, p. 107785, 2020.

\bibitem{fan2018maximum}
Y.~Fan, Y.~Zhang, G.~Wang, X.~Wang, and N.~Li, ``Maximum correntropy based
  unscented particle filter for cooperative navigation with heavy-tailed
  measurement noises,'' \emph{Sensors}, vol.~18, no.~10, p. 3183, 2018.

\bibitem{8049492}
Y.~{Huang}, Y.~{Zhang}, B.~{Xu}, Z.~{Wu}, and J.~A. {Chambers}, ``A new
  adaptive extended kalman filter for cooperative localization,'' \emph{IEEE
  Transactions on Aerospace and Electronic Systems}, vol.~54, no.~1, pp.
  353--368, 2018.

\bibitem{zhao2016collaborative}
Y.~Zhao, W.~Xing, H.~Yuan, and P.~Shi, ``A collaborative control framework with
  multi-leaders for auvs based on unscented particle filter,'' \emph{Journal of
  the Franklin Institute}, vol. 353, no.~3, pp. 657--669, 2016.

\bibitem{8327867}
Q.~{Li}, Y.~{Ben}, S.~M. {Naqvi}, J.~A. {Neasham}, and J.~A. {Chambers},
  ``Robust student’s $t$ -based cooperative navigation for autonomous
  underwater vehicles,'' \emph{IEEE Transactions on Instrumentation and
  Measurement}, vol.~67, no.~8, pp. 1762--1777, 2018.

\bibitem{xu2019cooperative}
B.~Xu, S.~Li, A.~A. Razzaqi, and J.~Zhang, ``Cooperative localization in harsh
  underwater environment based on the mc-anfis,'' \emph{IEEE Access}, vol.~7,
  pp. 55\,407--55\,421, 2019.

\bibitem{li2020improved}
S.~Li, B.~Xu, L.~Wang, and A.~A. Razzaqi, ``Improved maximum correntropy
  cubature kalman filter for cooperative localization,'' \emph{IEEE Sensors
  Journal}, vol.~20, no.~22, pp. 13\,585--13\,595, 2020.

\bibitem{xu2021novel}
B.~Xu, S.~Li, A.~A. Razzaqi, Y.~Guo, and L.~Wang, ``A novel measurement
  information anomaly detection method for cooperative localization,''
  \emph{IEEE Transactions on Instrumentation and Measurement}, vol.~70, pp.
  1--18, 2021.

\bibitem{xu2020novel}
B.~Xu, Y.~Guo, L.~Wang, and J.~Zhang, ``A novel robust gaussian approximate
  smoother based on em for cooperative localization with sensor fault and
  outliers,'' \emph{IEEE Transactions on Instrumentation and Measurement},
  vol.~70, pp. 1--14, 2020.

\bibitem{rui2010cooperative}
G.~Rui and M.~Chitre, ``Cooperative positioning using range-only measurements
  between two auvs,'' in \emph{OCEANS'10 IEEE SYDNEY}.\hskip 1em plus 0.5em
  minus 0.4em\relax IEEE, 2010, pp. 1--6.

\bibitem{bahr2012dynamic}
A.~Bahr, J.~J. Leonard, and A.~Martinoli, ``Dynamic positioning of beacon
  vehicles for cooperative underwater navigation,'' in \emph{2012 IEEE/RSJ
  International Conference on Intelligent Robots and Systems}.\hskip 1em plus
  0.5em minus 0.4em\relax IEEE, 2012, pp. 3760--3767.

\bibitem{7353681}
J.~M. {Walls}, S.~M. {Chaves}, E.~{Galceran}, and R.~M. {Eustice}, ``Belief
  space planning for underwater cooperative localization,'' in \emph{2015
  IEEE/RSJ International Conference on Intelligent Robots and Systems (IROS)},
  2015, pp. 2264--2271.

\bibitem{5658626}
{Mingyong Liu}, {Wenbai Li}, {Bingxian Mu}, and {Fuqiang Liu}, ``Cooperative
  navigation for multiple auvs based on relative range measurements with a
  single leader,'' in \emph{2010 IEEE International Conference on Intelligent
  Computing and Intelligent Systems}, vol.~2, 2010, pp. 762--766.

\bibitem{4449404}
D.~K. {Maczka}, A.~S. {Gadre}, and D.~J. {Stilwell}, ``Implementation of a
  cooperative navigation algorithm on a platoon of autonomous underwater
  vehicles,'' in \emph{OCEANS 2007}, 2007, pp. 1--6.

\bibitem{6847925}
J.~{Liu}, Z.~{Wang}, Z.~{Peng}, J.~{Cui}, and L.~{Fiondella}, ``Suave: Swarm
  underwater autonomous vehicle localization,'' in \emph{IEEE INFOCOM 2014 -
  IEEE Conference on Computer Communications}, 2014, pp. 64--72.

\bibitem{8084652}
J.~S. {Willners}, P.~{Patron}, and Y.~R. {Pettilot}, ``Moving baseline
  localization for multi-vehicle maritime operations,'' in \emph{OCEANS 2017 -
  Aberdeen}, 2017, pp. 1--6.

\bibitem{allison2020resilient}
M.~Allison, ``A resilient cooperative localization strategy for autonomous
  underwater vehicles in swarms,'' in \emph{2020 10th Annual Computing and
  Communication Workshop and Conference (CCWC)}.\hskip 1em plus 0.5em minus
  0.4em\relax IEEE, 2020, pp. 0150--0156.

\bibitem{matsuda2012performance}
T.~Matsuda, T.~Maki, T.~Sakamaki, and T.~Ura, ``Performance analysis on a
  navigation method of multiple auvs for wide area survey,'' \emph{Marine
  Technology Society Journal}, vol.~46, no.~2, pp. 45--55, 2012.

\bibitem{6964386}
T.~{Matsuda}, T.~{Maki}, Y.~{Sato}, and T.~{Sakamaki}, ``Cooperative navigation
  method of multiple autonomous underwater vehicles for wide seafloor survey
  — sea experiment with two auvs,'' in \emph{OCEANS 2014 - TAIPEI}, 2014, pp.
  1--9.

\bibitem{6405084}
T.~{Matsuda}, T.~{Maki}, T.~{Sakamaki}, and T.~{Ura}, ``State estimation of
  multiple auvs with limited communication traffic,'' in \emph{2012 Oceans},
  2012, pp. 1--10.

\bibitem{Matsuda_2018}
T.~{Matsuda}, T.~{Maki}, Y.~{Sato}, and T.~{Sakamaki}, ``Experimental
  evaluation of accuracy and efficiency of alternating landmark navigation by
  multiple auvs,'' \emph{IEEE Journal of Oceanic Engineering}, vol.~43, no.~2,
  pp. 288--310, 2018.

\bibitem{5152859}
A.~{Bahr}, M.~R. {Walter}, and J.~J. {Leonard}, ``Consistent cooperative
  localization,'' in \emph{2009 IEEE International Conference on Robotics and
  Automation}, 2009, pp. 3415--3422.

\bibitem{liu_convex_2010}
\BIBentryALTinterwordspacing
M.-Y. Liu, W.-B. Li, and X.~Pei, ``Convex {Optimization} {Algorithms} for
  {Cooperative} {Localization} in {Autonomous} {Underwater} {Vehicles},''
  \emph{Acta Automatica Sinica}, vol.~36, no.~5, pp. 704--710, May 2010.
  [Online]. Available:
  \url{https://linkinghub.elsevier.com/retrieve/pii/S1874102909600318}
\BIBentrySTDinterwordspacing

\bibitem{chen2013minimizing}
B.~Chen and D.~Pompili, ``Minimizing position uncertainty for under-ice
  autonomous underwater vehicles,'' \emph{Computer Networks}, vol.~57, no.~18,
  pp. 3840--3854, 2013.

\bibitem{7017127}
V.~{Ludovico}, J.~{Gomes}, J.~{Alves}, and T.~C. {Furfaro}, ``Joint
  localization of underwater vehicle formations based on range difference
  measurements,'' in \emph{2014 Underwater Communications and Networking
  (UComms)}, 2014, pp. 1--5.

\bibitem{parlangeli2012relative}
G.~Parlangeli, P.~Pedone, and G.~Indiveri, ``Relative pose observability
  analysis for 3d nonholonomic vehicles based on range measurements only,''
  \emph{IFAC Proceedings Volumes}, vol.~45, no.~27, pp. 182--187, 2012.

\bibitem{viegas2015distributed}
D.~Viegas, P.~Batista, P.~Oliveira, C.~Silvestre, and C.~P. Chen, ``Distributed
  state estimation for linear multi-agent systems with time-varying measurement
  topology,'' \emph{Automatica}, vol.~54, pp. 72--79, 2015.

\bibitem{viegas2016decentralized}
D.~Viegas, P.~Batista, P.~Oliveira, and C.~Silvestre, ``Decentralized state
  observers for range-based position and velocity estimation in acyclic
  formations with fixed topologies,'' \emph{International Journal of Robust and
  Nonlinear Control}, vol.~26, no.~5, pp. 963--994, 2016.

\bibitem{viegas2018discrete}
------, ``Discrete-time distributed kalman filter design for formations of
  autonomous vehicles,'' \emph{Control Engineering Practice}, vol.~75, pp.
  55--68, 2018.

\bibitem{7778673}
G.~{Rui} and M.~{Chitre}, ``Cooperative multi-auv localization using
  distributed extended information filter,'' in \emph{2016 IEEE/OES Autonomous
  Underwater Vehicles (AUV)}, 2016, pp. 206--212.

\bibitem{8206528}
Z.~{Song} and K.~{Mohseni}, ``Facon: A flow-aided cooperative navigation
  scheme,'' in \emph{2017 IEEE/RSJ International Conference on Intelligent
  Robots and Systems (IROS)}, 2017, pp. 6251--6256.

\bibitem{8604531}
L.~{Zhang}, X.~{Tao}, and H.~{Liang}, ``Multi auvs cooperative navigation based
  on information entropy,'' in \emph{OCEANS 2018 MTS/IEEE Charleston}, 2018,
  pp. 1--10.

\bibitem{sabra2020fuzzy}
A.~Sabra and W.-K. Fung, ``A fuzzy cooperative localisation framework for
  underwater robotic swarms,'' \emph{Sensors}, vol.~20, no.~19, p. 5496, 2020.

\bibitem{5278138}
R.~{Engel} and J.~{Kalwa}, ``Relative positioning of multiple underwater
  vehicles in the grex project,'' in \emph{OCEANS 2009-EUROPE}, 2009, pp. 1--7.

\bibitem{5509573}
G.~{Antonelli}, F.~{Arrichiello}, S.~{Chiaverini}, and G.~S. {Sukhatme},
  ``Observability analysis of relative localization for auvs based on ranging
  and depth measurements,'' in \emph{2010 IEEE International Conference on
  Robotics and Automation}, 2010, pp. 4276--4281.

\bibitem{6094466}
F.~{Arrichiello}, G.~{Antonelli}, A.~P. {Aguiar}, and A.~{Pascoal},
  ``Observability metric for the relative localization of auvs based on range
  and depth measurements: Theory and experiments,'' in \emph{2011 IEEE/RSJ
  International Conference on Intelligent Robots and Systems}, 2011, pp.
  3166--3171.

\bibitem{arrichiello2013observability}
F.~Arrichiello, G.~Antonelli, A.~P. Aguiar, and A.~Pascoal, ``An observability
  metric for underwater vehicle localization using range measurements,''
  \emph{Sensors}, vol.~13, no.~12, pp. 16\,191--16\,215, 2013.

\bibitem{parlangeli2015single}
G.~Parlangeli and G.~Indiveri, ``Single range observability for cooperative
  underactuated underwater vehicles,'' \emph{Annual Reviews in Control},
  vol.~40, pp. 129--141, 2015.

\bibitem{allotta2016cooperative}
B.~Allotta, A.~Caiti, R.~Costanzi, F.~Di~Corato, D.~Fenucci, N.~Monni, L.~Pugi,
  and A.~Ridolfi, ``Cooperative navigation of auvs via acoustic communication
  networking: field experience with the typhoon vehicles,'' \emph{Autonomous
  Robots}, vol.~40, no.~7, pp. 1229--1244, 2016.

\bibitem{tan2016cooperative}
Y.~T. Tan, M.~Chitre, and F.~S. Hover, ``Cooperative bathymetry-based
  localization using low-cost autonomous underwater vehicles,''
  \emph{Autonomous Robots}, vol.~40, no.~7, pp. 1187--1205, 2016.

\bibitem{wiktor2020collaborative}
A.~Wiktor and S.~Rock, ``Collaborative multi-robot localization in natural
  terrain,'' in \emph{2020 IEEE International Conference on Robotics and
  Automation (ICRA)}.\hskip 1em plus 0.5em minus 0.4em\relax IEEE, 2020, pp.
  4529--4535.

\bibitem{7747236}
C.~{Cadena}, L.~{Carlone}, H.~{Carrillo}, Y.~{Latif}, D.~{Scaramuzza},
  J.~{Neira}, I.~{Reid}, and J.~J. {Leonard}, ``Past, present, and future of
  simultaneous localization and mapping: Toward the robust-perception age,''
  \emph{IEEE Transactions on Robotics}, vol.~32, no.~6, pp. 1309--1332, 2016.

\bibitem{5509869}
M.~F. {Fallon}, G.~{Papadopoulos}, and J.~J. {Leonard}, ``A measurement
  distribution framework for cooperative navigation using multiple auvs,'' in
  \emph{2010 IEEE International Conference on Robotics and Automation}, 2010,
  pp. 4256--4263.

\bibitem{6107061}
J.~M. {Walls} and R.~M. {Eustice}, ``Experimental comparison of
  synchronous-clock cooperative acoustic navigation algorithms,'' in
  \emph{OCEANS'11 MTS/IEEE KONA}, 2011, pp. 1--7.

\bibitem{mirza2015real}
D.~Mirza, P.~Naughton, C.~Schurgers, and R.~Kastner, ``Real-time collaborative
  tracking for underwater networked systems,'' \emph{Ad Hoc Networks}, vol.~34,
  pp. 196--210, 2015.

\bibitem{rego2021cooperative}
F.~Rego and A.~Pascoal, ``Cooperative single-beacon multiple auv navigation
  under stringent communication bandwidth constraints,''
  \emph{IFAC-PapersOnLine}, vol.~54, no.~16, pp. 216--223, 2021.

\bibitem{7271663}
A.~{Munafò}, J.~{Sliwka}, and J.~{Alves}, ``Dynamic placement of a
  constellation of surface buoys for enhanced underwater positioning,'' in
  \emph{OCEANS 2015 - Genova}, 2015, pp. 1--6.

\bibitem{8546733}
S.~{Sun}, S.~{Yu}, Z.~{Shi}, J.~{Fu}, and C.~{Zhao}, ``A novel single-beacon
  navigation method for group auvs based on simo model,'' \emph{IEEE Access},
  vol.~6, pp. 75\,155--75\,168, 2018.

\end{thebibliography}
%








\end{document}